\documentclass[a4paper, 11pt]{book}
\pdfoutput=1

\usepackage[margin=2.5cm]{geometry}
\usepackage{latexsym}
\usepackage{graphics}
\usepackage{epsfig}
\def\be{\begin{equation}}
\def\ee{\end{equation}}

\def\bea{\begin{eqnarray}}
\def\eea{\end{eqnarray}}

\begin{document}

\markboth{Alfredo Iorio}{Curved Spacetimes and Curved Graphene}

%
%

\title{CURVED SPACETIMES AND CURVED GRAPHENE: \\ A STATUS REPORT OF THE WEYL-SYMMETRY APPROACH}
\author{Alfredo Iorio\thanks{E-mail: alfredo.iorio@mff.cuni.cz} \\
Faculty of Mathematics and Physics \\ Charles University in Prague \\ V Hole\v{s}ovi\v{c}k\'ach 2, 18000 Prague 8, Czech Republic
\\
alfredo.iorio@mff.cuni.cz}

\date{\today}

\maketitle




\tableofcontents

\chapter{Introduction}

The research reported about here \cite{iorio, ioriolambiase,ioriotolor,ioriodice12,prd2014,iorioMaterials,reviewslectures} moves at the crossroad of condensed matter and high energy theory. It is carried out with the main goal of constructing, with graphene, a real table-top system where quantum field in a curved spacetime, and local Weyl symmetry \cite{lor}, are at work. This is a multifaceted enterprize, with many steps. Some steps have been done \cite{iorio,ioriolambiase,prd2014}, some are in progress \cite{pablo1,ioVacuum}, some are in the list of to-dos, some are just foreseeable future developments. We offer here an overview of the issues, that will help have the matters all in one place.

Physics is an experimental science. Nonetheless, in the last decades, it proved very difficult (if not impossible) to reconcile theoretical investigations of the fundamental laws of nature with the necessary experimental tests. The divarication between theory and experiments is the central problem of contemporary fundamental physics that, to date, is still unable to furnish a consistent quantum theory of gravity, or to obtain experimental evidences of milestones theories like supersymmetry (see, e.g., the whole Issue \cite{thooftstring} on the status of string theory).

A widespread view of this problem is that experimental observations of these type of phenomena can only be achieved at energies out of the reach of our laboratories (e.g., the Planck energy). In our view, due to unprecedented behaviors of certain novel materials, nowadays indirect tests should be considered as a viable alternative to direct observations.

This field of research is not a novelty. It usually goes under the generic name of ``analogue gravity''. See \cite{visser} for a review, and \cite{stringari} and \cite{cks}, for recent results on the Bose-Einstein condensates and on hadronization, respectively. Perhaps, to call it ``bottom-up'' approach does it more justice: analogue experimental settings on the bottom, fundamental theories of nature on the top. Nonetheless, for various reasons, it has been seen as little more than a curiosity. An amusing and mysterious series of coincidences, that never (up to our knowledge) were taken as tests of those aspects of the fundamental theories that they are reproducing. Neither are taken as experimental tests of the fundamental theories, the mysterious and amusing coincidences of the ``top-bottom'' approach of the AdS/CFT correspondence (where theoretical constructions of the fundamental world,  are used to describe experimental results at our energy scale \cite{maldacena}).

With graphene we have a quantum relativistic-like Dirac massless field available on a nearly perfectly two-dimensional sheet of carbon atoms. While this special-relativistic-like behavior of a condensed matter system is quite unusual (and it came as a surprise at the time of its discovery\cite{geimnovoselovFIRST}), it is, by now, a well established experimental fact\cite{geim, pacoreview2009}. A promising new direction, that is the one we shall review here, is the emergence of gravity-like phenomena on graphene\cite{iorio,ioriolambiase,ioriotolor,ioriodice12,prd2014,iorioMaterials,reviewslectures,pablo1,ioVacuum}.

One important feature here is the embedding nature of  these spacetimes that are in real terrestrial laboratories, i.e., spatially, $\mathbf{R}^3$. For the sake of extracting measurable effects, such as a Hawking-Unruh effect, a crucial role is played by surfaces of constant negative Gaussian curvature. As is well known, those surfaces can only be embedded into $\mathbf{R}^3$ at the price of non-removable singularities, as proven by a theorem of Hilbert \cite{eisenhart, hilbert, ovchinnikov}. We shall explain this in detail here, alongside the role of coordinates here, that have a more important status than in the customary proper relativistic setting.

Important are also the quantum vacua, and the related (unitarily) inequivalent quantization schemes \cite{habilitation,uirs,birrellanddavies,takagi,israel}. That is at the core of the Hawking-Unruh effect: an observer in $A$ sees the quantum vacuum in the frame $B$ as a condensate of $A$-particles. The matter is under investigation, and outlined here \cite{siddhartha, ioVacuum, kastrup,derco}.

Local Weyl symmetry will play an important role in this work. This is a symmetry of the massless Dirac action under transformations that, in (2+1) dimensions, take the following form
\begin{equation}
g_{\mu \nu} (x) \to \phi^2 (x) g_{\mu \nu} (x) \quad {\rm and} \quad \psi (x) \to \phi^{-1} (x) \psi (x) \label{firstweyltrnsf}\;,
\end{equation}
note that all the quantities are computed at the same point $x$ in spacetime.

Here our notations: $\phi (x)$ is a scalar field (conformal factor), $\mu , \nu = 0, 1, ..., n-1$ are Einstein indices, responding to diffeomorphisms, $a, b = 0, 1, ..., n-1$ are flat indices, responding to local Lorentz transformations, while $\alpha,\beta$ are spin indices. The covariant derivative is
\be
\nabla_\mu \psi_\alpha = \partial_\mu \psi_\alpha + {\Omega_\mu}_\alpha^{\; \beta} \psi_\beta \;,
\ee
with
\be
{\Omega_\mu}_\alpha^{\; \beta} =  \frac{1}{2} \omega_\mu^{a b} (J_{a b})_\alpha^{\; \beta} \;,
\ee
where $(J_{a b})_\alpha^{\; \beta}$ are the Lorentz generators in spinor space, and
\be
{\omega_\mu}^a_{\; b} = e^a_\lambda (\delta^\lambda_\nu \partial_\mu + \Gamma_{\mu \nu}^\lambda) E^\nu_b \;,
\ee
is the spin connection, whose relation to the Christoffel connection comes from the metricity condition $\nabla_\mu e^a_\nu = \partial_\mu e^a_\nu - \Gamma_{\mu \nu}^\lambda e^a_\lambda +  \omega_{\mu \; b}^a e^b_\nu = 0$. We also introduced the Vielbein $e^a_\mu$ (and its inverse $E_a^\mu$), satisfying $\eta_{a b} e^a_\mu e^b_\nu = g_{\mu \nu}$, $e^a_\mu E_a^\nu = \delta_\mu^\nu$, $e^a_\mu E_b^\mu = \delta_b^a$, where $\eta_{a b} = {\rm diag} (1, -1, ...)$. The Weyl dimension of the Dirac field $\psi$ in $n$ dimensions is $d_\psi = (1 - n)/2$. Here $n=3$, and we can move one dimension up (embedding), or down (boundary). More notation can be found in \cite{iorio}.

To appreciate the physical meaning of this symmetry, it should be considered that it relates the physics in a given spacetime (metric $g_{\mu \nu}$), to the physics in a {\it different} spacetime (metric $\phi^2 g_{\mu \nu}$). For instance, when the background spacetime $g_{\mu \nu}$ is curved but conformally flat, since we can take advantage from the privileged link with the flat spacetime counterpart, Weyl symmetry might allow for exact nonperturbative results in the computation of the Green functions \cite{iorio}, a very difficult task to accomplish by other means \cite{birrellanddavies}. When we are dealing with a conformally invariant field in a conformally flat spacetime this is sometimes referred to as {\it conformal triviality} \cite{birrellanddavies}, a name that emphasizes the simplest possible case of QFT in curved space, but, perhaps, does not justice to the fact that the key features are indeed at work. If the spacetime is only curved in a conformally flat fashion, the effects of curvature are null on the \textit{classical physics} of a massless Dirac field. To spot the effects of curvature we need to move to the {\it quantum regime}.

A further reason for considering the conformally flat cases, lies with the gravitational analogues of these settings. For instance, there are interesting configurations, such as the (2+1)-dimensional  black hole of Ba\~{n}ados, Teitelboim, and Zanelli (BTZ) \cite{btz,BHTZ,carlipreview}
(see \cite{zermelo} for a study of graphene and the BTZ black hole) and the gravitational kink of \cite{cs3}, that are conformally flat configurations. For all these reasons, in this review we shall almost entirely focus on the conformally flat cases. Nonetheless, many of the arguments presented here apply to the general case.

\chapter{Deformed graphene: the large wavelength/small energy regime}\label{fieldgraphene}

Graphene is an allotrope of carbon. It is one-atom-thick, hence it is the closest in nature to a 2-dimensional object. It was first theoretically speculated about \cite{wallace,semenoff}, and, decades later, experimentally found \cite{geimnovoselovFIRST}.

\begin{figure}
\begin{center}
 \includegraphics[height= .4\textheight]{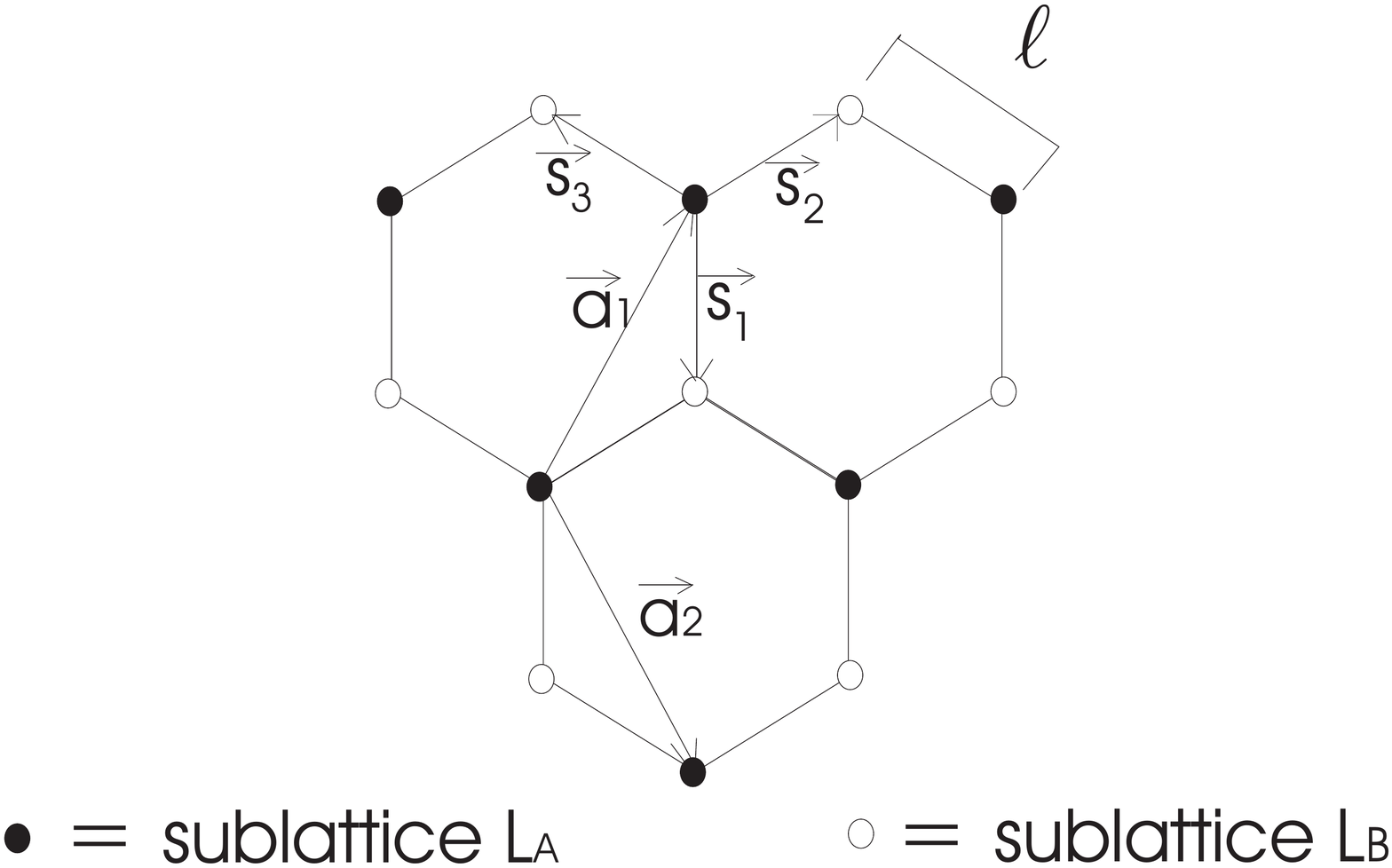}
\end{center}
  \caption{The honeycomb lattice of graphene, and its two triangular sublattices. The choice of the basis vectors, $(\vec{a}_1, \vec{a}_2)$ and
  $(\vec{s}_1, \vec{s}_2, \vec{s}_3)$, is, of course, not unique. Here we indicate our own.}
\label{honeycombpaper}
\end{figure}

The honeycomb lattice of graphene is made of two intertwined triangular sub-lattices, see Fig.~\ref{honeycombpaper}. Of the carbon's four electrons available to form covalent bonds, three are put in common with the three nearest neighbors (one each), forming what are known as $\sigma$-bonds (the molecular-level merging of the atomic $2s$-orbitals). These bonds are the responsible for the {\it elastic properties} of the membrane. The fourth electron also forms a covalent bond, called $\pi$-bond, but only with one of the three neighbors. Furthermore, being the $\pi$-orbitals the molecular-level merging of the atomic $2p$-orbitals, it has nodes on the membrane, and the electrons there are much more free to ``hop''. Thus, the latter bond ($\pi$) is of a much weaker kind than the former ($\sigma$). The {\it electronic properties} of graphene are due to the electrons belonging to the $\pi$-orbitals.

Here we deal with the electrons of the $\pi$-orbitals. Hence, although the effects of the deformations of the membrane will be duly taken into account, the elastic properties are not our direct concern. We shall propose various morphologies, for reasons that are on the theoretical side, but we shall not prove from the theoretical point of view whether those shapes are elastically permitted, or feasible from the practical/engineering point-of-view. Nonetheless, ongoing work shows that certain shapes, crucial for our purposes, are indeed possible to engineer \cite{simorugg}. On the other hand, as well known, suspended graphene's samples come with corrugations and ripples \cite{ripples1}, and many deformations have been induced to study the effects of curvature and strain on the electronic properties, a central issue in the ongoing studies of graphene on the condensed matter side, see, e.g., \cite{conference2011}.

\section{The Dirac Hamiltonian scenario for planar and unstrained graphene}

As is by now well known, graphene's lattice structure is behind a natural description of its electronic properties in terms of massless, (2+1)-dimensional, Dirac (hence, relativistic-like) pseudoparticles. Below we give one derivation of this result, and more is in \cite{pacoreview2009}. In the honeycomb lattice, there are two inequivalent sites per unit cell, the white and black spots in Fig.~\ref{honeycombpaper}, that do not refer to different atoms (they all are carbons), but to their topological inequivalence. Contrary to a square lattice, the basis vectors, $(\vec{a}_1, \vec{a}_2)$, are not enough to reach all the points (black and white), and an extra set of vectors, $(\vec{s}_1, \vec{s}_2, \vec{s}_3)$, is needed. That is how the two-component Dirac spinor emerges (see later). Hence, the Dirac description is resistent to changes of the lattice that preserve this aspect of the structure. See, e.g., the developments in \cite{graphyne}, where the authors discuss \textit{graphyne}, by departing from the hexagonal structure, but retaining the two inequivalent sites per unit cell, hence the Dirac description. Very resistent to strain and deformations is also the masslessness (zero gap) of these excitations. That is why it is impossible to open-up a gap by mechanically deforming the material \cite{gapSOA}.

\begin{figure}
 \centering
  \includegraphics[height=.5\textheight]{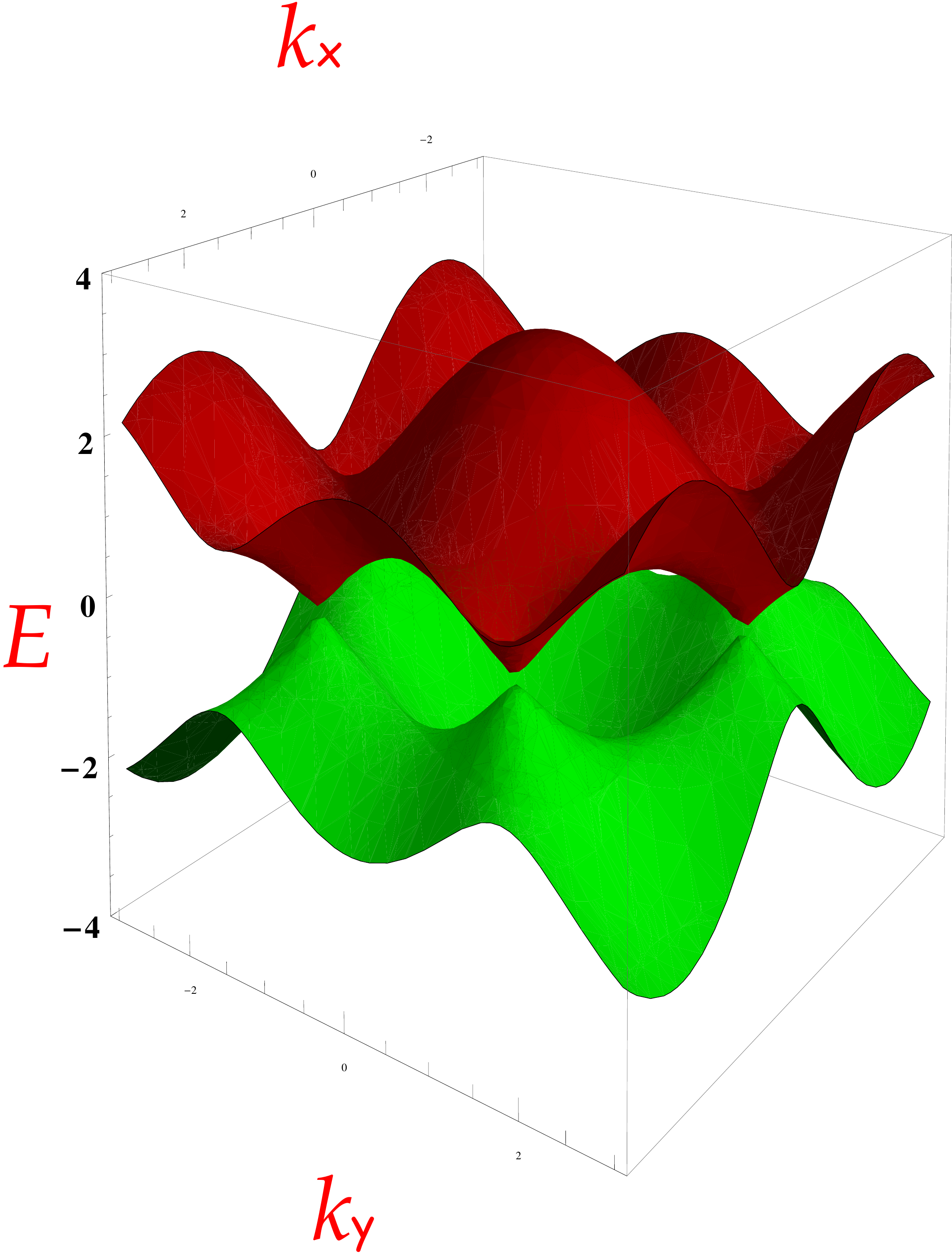}
  \caption{The dispersion relations of graphene, $E(\vec{k})$, in the approximation where the contribution coming from the next-to-nearest hopping, $\eta' \sim 0.1 \eta$, is not considered. The energy $E$ is measured in units of the hopping energy, $\eta \simeq 2.8$~eV. The conductivity band (red) and the valence band (green) touch at six {\it points}, reflecting, in the Brillouin zone, the hexagonal symmetry of the lattice. At those points the energy is zero. Only two of the six points are inequivalent. In the figure the two bands do not actually touch, due to graphic limitations of the plot, but in reality they do (see also Fig.~\ref{lindisprel}).}
\label{fulldisprel}
\end{figure}

The electronic properties of graphene, in the tight-binding low-energy approximation, are customarily described by the Hamiltonian (we use $\hbar = 1$)
\begin{eqnarray}
    H = - \eta \sum_{\vec{r} \in L_A} \sum_{i =1}^3 \left( a^\dagger (\vec{r}) b(\vec{r} +\vec{s}_i)
    + b^\dagger (\vec{r} +\vec{s}_i) a (\vec{r}) \right) \label{primaH} \;,
\end{eqnarray}
where the nearest-neighbor hopping energy is $\eta \simeq 2.8$~eV, and $a, a^\dagger$ ($b, b^\dagger$) are the anti-commuting annihilation and creation operators, respectively, for the planar electrons in the sub-lattice $L_A$ ($L_B$), see Fig.~\ref{honeycombpaper}. All the vectors are bi-dimensional,  $\vec{r} = (x,y)$, and, for the choice of basis vectors made in Fig.~\ref{honeycombpaper}
\be
\vec{a}_1 = \frac{\ell}{2} (\sqrt{3}, 3) \;, \; \vec{a}_2 = \frac{\ell}{2} (\sqrt{3}, - 3)
\ee
we have
\be
\vec{s}_1 = \ell (0, - 1) \;, \; \vec{s}_2 = \frac{\ell}{2} (\sqrt{3}, 1)\ ;, \; \vec{s}_3 = \frac{\ell}{2} (-\sqrt{3}, 1) \;,
\ee
with $\ell \simeq 1.4${\AA} the carbon-to-carbon distance, that we call the lattice {\it length} (the lattice {\it spacing} is, as usual, the length of the basis vectors, $|\vec{a}_1| = |\vec{a}_2| = \sqrt{3} \ell \simeq 2.5$\AA).

If we Fourier transform, $a(\vec{r}) = \sum_{\vec{k}} a(\vec{k}) e^{i \vec{k} \cdot \vec{r}}$, etc,  then $H = \sum_{\vec{k}} ( f(\vec{k}) a^\dagger (\vec{k}) b(\vec{k}) + {\rm h.c.})$, with
\begin{equation}
  f(\vec{k}) = - \eta \; e^{-i \ell k_y} \left( 1 + 2 \; e^{i 3 \ell k_y / 2} \cos(\sqrt{3} \ell k_x / 2) \right) \;.
\end{equation}
Solving $E(\vec{k}) = \pm |f (\vec{k})| \equiv 0$ tells us where, in the Brillouin zone, conductivity and valence bands touch (if they do). Indeed, for graphene, this happens, pointing to a gapless spectrum, for which we expect massless excitations to emerge. Furthermore, the solution is not a Fermi \textit{line} (the $(2+1)$-dimensional version of the Fermi surface of the $(3+1)$ dimensions), but rather they are two Fermi \textit{points},
\be
\vec{k}^D_\pm = \left( \pm \frac{4 \pi}{3 \sqrt{3} \ell}, 0 \right) \;.
\ee
There are actually six such points, but only the indicated two are inequivalent. The above is illustrated in Fig.~\ref{fulldisprel}. Notice that
\be
\vec{k}^D_+ = - \vec{k}^D_- \;,
\ee
i.e., the two momenta, with this choice basis vectors, point in opposite directions.

\begin{figure}
 \centering
  \includegraphics[height=.4\textheight]{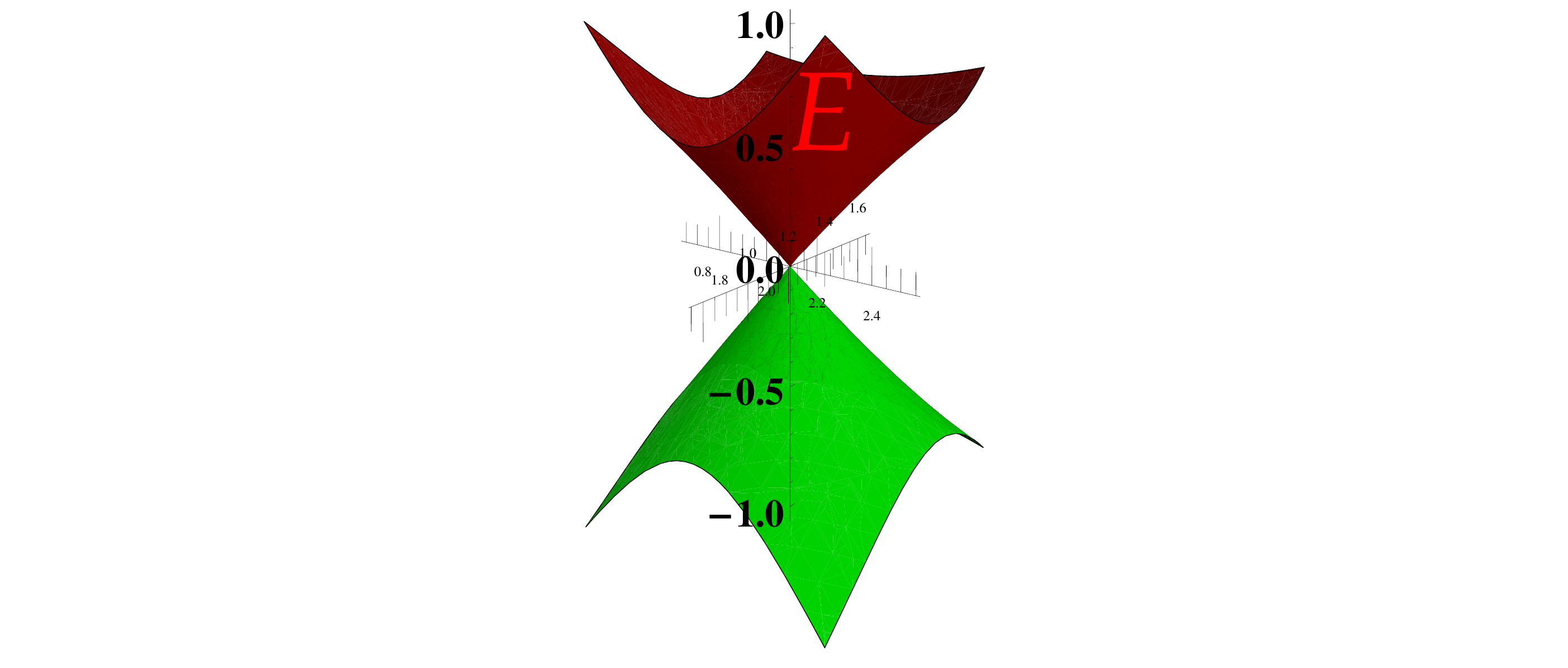}
  \caption{The linear dispersion relations near one of the Dirac points, showing the typical behavior of a relativistic-like system (the ``$v_F$-light-cone'' in $k$-space). Note that we are below the threshold $E_\ell \simeq 4.2$eV, as it must be, and the energy range is of the order of one unit of the hopping energy $\eta$, i.e., here $E \in [-2.8{\rm eV},2.8{\rm eV}]$. Within this range, the continuum field approximation is tenable.}
\label{lindisprel}
\end{figure}

The label ``$D$'' on the Fermi points stands for ``Dirac''. That refers to the all-important fact that, near those points  the spectrum is {\it linear}, as can be seen from Fig.~\ref{lindisprel}, $E_{\pm} \simeq \pm v_F |\vec{k}|$, where $v_F = 3 \eta \ell / 2 \sim c/300$ is the Fermi velocity. This behavior is expected in a relativistic theory, whereas, in a non-relativistic system, the dispersion relations are usually quadratic.

The physical setting we are focusing on is that of a full valence band, and of an empty conductivity band, a regime referred to as ``half filling'' in the condensed matter literature. The electronic properties of the material are described by the behavior a couple of electronvolts (eV) away from $E=0$, both above and below, and due to the linear behavior there, see Fig.~\ref{lindisprel}, it makes perfect sense to consider only the linear contributions to the Hamiltonian. If one linearizes around $\vec{k}^D_\pm$, $\vec{k}_\pm \simeq \vec{k}^D_\pm + \vec{p}$:
\be
f_+ (\vec{p}) \equiv f (\vec{k}_+) = v_F (p_x + i p_y) \;, \quad f_- (\vec{p}) \equiv f (\vec{k}_-) = - v_F (p_x - i p_y)\;,
\ee
and $a_\pm (\vec{p}) \equiv a (\vec{k}_\pm)$,  $b_\pm (\vec{p}) \equiv b (\vec{k}_\pm)$, then the Hamiltonian (\ref{primaH}) becomes
\begin{eqnarray}
    H|_{\vec{k}_\pm} & \simeq & \sum_{\vec{p}} \left[ f_+ a_+^\dagger b_+ +
f_- a_-^\dagger b_- + f^*_+ b_+^\dagger a_+ + f^*_- b_-^\dagger a_- \right] (\vec{p}) \nonumber \\
    & = & v_F  \sum_{\vec{p}} \Bigg[ (b_+^\dagger , a^\dagger_+) \left(\begin{array}{cc} 0 & 1 \\ 1 & 0 \\ \end{array} \right) p_x
\left(\begin{array}{c} b_+ \\ a_+ \end{array}\right)
+ (b_+^\dagger , a^\dagger_+) \left(\begin{array}{cc} 0 & - i \\ i & 0 \\ \end{array} \right) p_y
\left(\begin{array}{c} b_+ \\ a_+ \end{array}\right) \nonumber \\
& - & (b_-^\dagger , a^\dagger_-) \left(\begin{array}{cc} 0 & 1 \\ 1 & 0 \\ \end{array} \right) p_x
\left(\begin{array}{c} a_- \\ b_- \end{array}\right)
+ (b_-^\dagger , a^\dagger_-) \left(\begin{array}{cc} 0 & -i \\ i & 0 \\ \end{array} \right) p_y
\left(\begin{array}{c} a_- \\ b_- \end{array}\right) \Bigg] \nonumber \\
& = &  v_F \sum_{\vec{p}} \left(\psi_+^\dagger \vec{\sigma} \cdot \vec{p} \; \psi_+
    - \psi_-^\dagger \vec{\sigma}^* \cdot \vec{p} \; \psi_- \right) \label{hmeta}
\end{eqnarray}
where
\be\label{stripepsi}
\psi_\pm \equiv \left(\begin{array}{c} b_\pm \\ a_\pm \end{array}\right)
\ee
are two-component Dirac spinors, and $\vec{\sigma} \equiv (\sigma_1, \sigma_2)$, $\vec{\sigma}^* \equiv (\sigma_1, - \sigma_2)$, with $\sigma_i$ the Pauli matrices.

\begin{figure}
 \centering
  \includegraphics[height=.5\textheight]{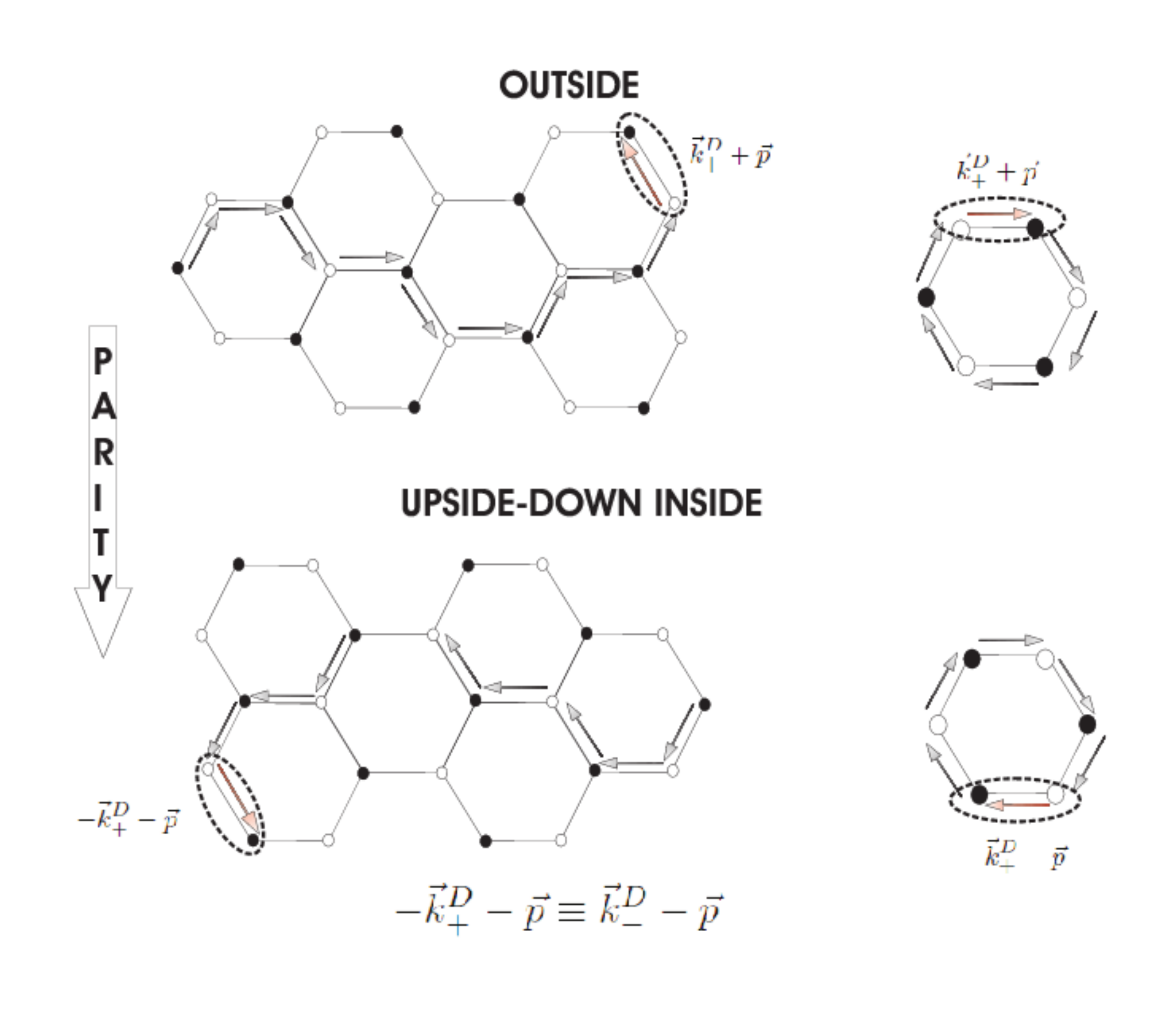}
  \caption{Schematic view of the outside $\to$ upside-down inside transformation. For a given electron, near one Dirac point, say $\vec{k}^D_+$, to a jump $\circ \to \bullet$ of momentum $\vec{k}^D_+ + \vec{p}$, corresponds a jump $\circ \to \bullet$ of opposite momentum $- \vec{k}^D_+ - \vec{p} = \vec{k}^D_- - \vec{p}$. The indicated circled jumps correspond to each other via this transformation. That is the {\it same jump} seen on the two sides of the membrane.}
\label{stripe}
\end{figure}

The two-spinors are connected by the inversion of the full momentum
\be
\vec{k}^D_+ + \vec{p} \to - \vec{k}^D_+ - \vec{p} \equiv \vec{k}^D_- - \vec{p} \;,
\ee
and this is consistent with the following picture, illustrated in Fig.~\ref{stripe}: If near a given Dirac point, say $\vec{k}^D_+$, the physics is described by a spinor $\psi_+$ that describes hoppings $\bullet \to \circ \to \bullet$ {\it on the outside} of the membrane, then the very same hoppings, seen from {\it the upside-down inside} of the membrane are described by the spinor $\psi_-$, with opposite momentum. Thus, we can think of graphene as a two-sided membrane: on one side lives $\psi_+$, on the other $\psi_-$.

If nothing mixes the two sides, then one spinor $\psi_+$, or $\psi_-$, describes the whole physics of the $\pi$ electrons in the low energy regime. This is indeed the case when there is no torsion, as we shall discuss later, when we shall face the issue of topological defects in this context.

From now on we shall only consider the linear/relativistic-like regime. Let us now establish, within that regime, what we shall call ``high'' and ``low'' energies, i.e. let us clear-up the picture of the range of validity of the approximations we want to make. The first reference energy is
\be
E_\ell \sim v_F / \ell \sim 4.2 \; {\rm eV}. \label{Escale}
\ee
Notice that $E_\ell \sim 1.5 \eta$, and that the associated wavelength, $\lambda = 2 \pi / |\vec{p}| \simeq 2 \pi v_F / E$, is $2 \pi \ell$. The electrons' wavelength, at energies below $E_\ell$, is large compared to the lattice length, $\lambda > 2 \pi \ell$. Those electrons see the graphene sheet as a continuum, hence, over the whole linear regime, the following Hamiltonian well captures the physics of undeformed (planar and unstrained) graphene
\begin{equation}
    H    = - i v_F \int d^2 x \; \psi^\dagger \vec{\sigma} \cdot \vec{\partial} \; \psi \;, \label{HGrapheneBpaper}
\end{equation}
where the two component spinor is $\psi \equiv \psi_+$, we moved back to configuration space, $\vec{p} \to - i \vec{\partial}$, and sums turned into integrals because of the continuum limit.

\section{Local Lorentz group structures}

Eventually, we shall come back to the two-component spinor form of the Hamiltonian (\ref{HGrapheneBpaper}), as that will be the starting point to implement Weyl symmetry in the most straightforward way. Nonetheless, this comes at a price, and we shall be forced to do certain approximations. To understand that in some detail, let us consider a more general expression, that involves both Dirac points, hence {\it four component} spinors
\begin{equation}\label{diracpsi4dappendix}
   \Psi \equiv \left( \begin{array}{c} \psi_+ \\ \psi_- \\ \end{array} \right)
= \left( \begin{array}{c} b_+ \\ a_+ \\ b_- \\ a_- \\ \end{array} \right) \;,
\end{equation}
and $4 \times 4$ Dirac matrices
\begin{equation}\label{alphabetaappendix}
\alpha^i = \left(\begin{array}{cc} \sigma^i & 0 \\ 0 & - {\sigma^*}^i \\ \end{array} \right) \;, \;
\beta = \left(\begin{array}{cc} \sigma^3 & 0 \\ 0 & \sigma^3 \\ \end{array} \right)
\end{equation}
$i = 1, 2$. These matrices satisfy the standard properties: $\alpha^i \alpha^j + \alpha^j \alpha^i = 2 \delta^{i j} \mathbf{1}_4$, $\alpha^i \beta + \beta \alpha^i = 0$, $\beta^2 = \mathbf{1}_4$, $(\alpha^i)^\dag = \alpha^i$, $\beta^\dag = \beta$, ${\rm Tr} \alpha^i = {\rm Tr} \beta = 0$.

The gamma matrices are, as usual $\gamma^0 = \beta$, $\gamma^i = \beta \alpha^i$, that is
\be
\gamma^0 = \left(\begin{array}{cc} \sigma^3 & 0 \\ 0 & \sigma^3 \\ \end{array} \right) \;,
\gamma^1 = i \left(\begin{array}{cc} \sigma^2 & 0 \\ 0 & - \sigma^2 \\ \end{array} \right) \;,
\gamma^2 = - i \left(\begin{array}{cc} \sigma^1 & 0 \\ 0 & \sigma^1 \\ \end{array} \right) \label{gamma4times4}\;.
\ee
These obey all standard properties, as derived from the properties of the $\alpha^i$ and $\beta$, e.g.
\be
\gamma^a \gamma^b + \gamma^b \gamma^a = 2 \eta^{a b} \mathbf{1}_4 \;,
\ee
where $a, b= 0, 1, 2$.

From these, one also obtains
\be
i \gamma^0 \gamma^1 \gamma^2 = \left(\begin{array}{cc} \mathbf{1}_2 & 0 \\ 0 & - \mathbf{1}_2 \\ \end{array} \right) \;,
\ee
that, whatever choice one makes for $\gamma^3$, is the same as $\gamma^5 \gamma_3$ (use $\gamma^3 \gamma_3 = + \mathbf{1}_4$), with standard $\gamma^5 \equiv i \gamma^0 \gamma^1 \gamma^2 \gamma^3$.

With these, the Hamiltonian can then be written as
\begin{eqnarray}
H & = & - i v_F \int d^2 x \left( \psi_+^\dagger \vec{\sigma} \cdot \vec{\partial} \; \psi_+
    - \psi_-^\dagger \vec{\sigma}^* \cdot \vec{\partial} \; \psi_- \right) \nonumber \\
&=&  - i v_F \int d^2 x \; \Psi^\dagger \vec{\alpha} \cdot \vec{\partial} \; \Psi
= - i v_F \int d^2 x \; \bar{\Psi} \vec{\gamma} \cdot \vec{\partial} \; \Psi \;,  \label{Hdiracgraphene1}
\end{eqnarray}
where
\begin{equation}
\bar{\Psi} \equiv \Psi^\dag \gamma^0 = ( \bar{\psi}_+, \bar{\psi}_- ) = (\psi_+^\dag \sigma^3, \psi_-^\dag \sigma^3) \;.
\end{equation}

The $4 \times 4$ $\gamma$s are reducible to two sets of $2 \times 2$ $\gamma$s, that are essentially the Pauli matrices suitably multiplied by imaginary factors to give the right signature of $\eta_{a b}$
\be
\gamma^a = \left(\begin{array}{cc} \gamma_+^a & 0 \\ 0 & \gamma_-^a \\ \end{array} \right) \;,
\ee
with (see (\ref{gamma4times4}))
\begin{eqnarray}
\gamma_+^0 & = & \sigma^3 = \left(\begin{array}{cc} 1 & 0 \\ 0 & -1 \\ \end{array} \right)  \;, \;
\gamma_+^1 = i \sigma^2 = \left(\begin{array}{cc} 0 & 1 \\ - 1 & 0 \\ \end{array} \right)   \;, \;
\gamma_+^2 = - i \sigma^1 = \left(\begin{array}{cc} 0 & -i \\ -i & 0 \\ \end{array} \right) \;, \; \\
\gamma_-^0 & = & \sigma^3 =  \left(\begin{array}{cc} 1 & 0 \\ 0 & -1 \\ \end{array} \right) \;, \;
\gamma_-^1 = - i \sigma^2 = \left(\begin{array}{cc} 0 & - 1 \\ 1 & 0 \\ \end{array} \right) \;, \;
\gamma_-^2 = - i \sigma^1 = \left(\begin{array}{cc} 0 & -i \\ -i & 0 \\ \end{array} \right)  \;.
\end{eqnarray}
The Lorentz generators in spinor space are
\begin{equation}\label{generatorA}
J^{a b} =  \frac{1}{4} [\gamma^a, \gamma^b] \;.
\end{equation}
Using the 3D property that, for any antisymmetric tensor, $T_{a b} = \epsilon_{a b c} T^c$, then $T_a = \frac{1}{2} \epsilon_{a b c} T^{b c}$, we have
\begin{equation}\label{3dgenerators}
J_a = \frac{1}{4} \epsilon_{a b c} \gamma^b \gamma^c \;,
\end{equation}
with $J_{a b} = \epsilon_{a b c} J^c$. This can be decomposed in the $\pm$ components
\begin{equation}\label{3dgeneratoridentity}
j^+_a = \frac{1}{4} \epsilon_{a b c} \gamma_+^b \gamma_+^c \;, \;\; j^-_a = \frac{1}{4} \epsilon_{a b c} \gamma_-^b \gamma_-^c \;.
\end{equation}
Let us show now that the choice for the $\gamma_+$s and $\gamma_-$s consistently gives two copies of the local Lorentz group, SO(1,2) in terms of $j^+_a$ and $j^-_a$: one copy for the $(1/2, 0)$ representation, one copy for the conjugate representation $(0,1/2)$. By exploiting the properties of the Pauli matrices
\be
\sigma_I \sigma_J = \delta_{I J} + i \epsilon_{I J K} \sigma_K \;, \label{sigma}
\ee
here $I, J = 1,2,3$, we have
\begin{equation} \label{3dgeneratorsgammas}
j^+_0 =  \frac{i}{2} \gamma^+_0 \;, \; \;
j^+_1 =  \frac{i}{2} \gamma^+_1 \;, \; \;
j^+_2 =  \frac{i}{2} \gamma^+_2 \;,
\end{equation}
from which
\begin{equation}\label{explicitJSO12}
[j^+_0, j^+_1] = - j^+_2 \;, \; \; [j^+_1, j^+_2] = + j^+_0 \;, \; \; [j^+_2, j^+_0] = - j^+_1 \;,
\end{equation}
together
\begin{equation}
[j^+_a, j^+_b] = \epsilon_{a b c} j_+^c \; \; {\rm or} \; \; [{\cal J}^+_a, {\cal J}^+_b] = i \epsilon_{a b c} {\cal J}_+^c \;,
\end{equation}
i.e. SO(1,2), as it should be\footnote{This is not SO(3) because the indices are lowered/raised with $\eta$. See the explicit expressions (\ref{explicitJSO12})}. Here ${\cal J}^+_a \equiv i j^+_a$.

Following the same road, for the other generators we obtain
\begin{equation}\label{Jstar}
j^-_0 = - j^+_0 \;, \; \; j^-_1 = + j^+_1 \;, \; \; j^-_2 = - j^+_2 \;,
\end{equation}
which gives the same algebra of $j^+_a$
\begin{equation}
[j^-_a, j^-_b] = \epsilon_{a b c} j_-^c \; \; {\rm or} \; \; [{\cal J}^-_a, {\cal J}^-_b] = i \epsilon_{a b c} {\cal J}_-^c \;,
\end{equation}
i.e. a second copy (representation) of SO(1,2). Here ${\cal J}^-_a \equiv i j^-_a$.

By writing, e.g.,
\begin{equation}
R(\alpha) = e^{\alpha {\cal J}^+_1} \simeq I_2 + \frac{\alpha}{2} \left(\begin{array}{cc} 0 &  - 1 \\  1 & 0 \\ \end{array} \right)
\simeq \left(\begin{array}{cc} \cos(\alpha/2) & - \sin(\alpha/2) \\ \sin(\alpha/2) & \cos(\alpha/2) \\ \end{array} \right) \;,
\end{equation}
where $R^\dagger(\alpha) = R^{-1} (\alpha) = R(-\alpha)$, we see that ${\cal J}^+_1$ corresponds to the only rotation, hence ${\cal J}^+_{0, 2}$ are the two boosts, hence, by looking at (\ref{Jstar}), this means that $\psi_+$ and $\psi_-$ are one the time reversal transformed of the other. Furthermore, they transform under conjugate representations of the Lorentz group because, in any dimension, one moves from one representation to the conjugate one, precisely by changing the sign of the boosts, but keeping the same sign for the rotations (see, e.g., \cite{schweber}).

\section{The diffeomorphic Dirac Lagrangian scenario for deformed graphene}

Long before the advent of graphene, a field theoretical Dirac approach has also been successfully put forward on ``inflated'' \textit{buckyball fullerenes} \cite{vozmediano2} (carbon structures that can be thought of as graphene sheets, warped to make spheres). Later, this approach was extended to {\it curved graphene}, see, e.g. \cite{vozmediano}, where spatial curvatures in this Dirac field theoretical model were taken into account. Those were the pioneering steps towards the use of techniques of QFT in curved spacetimes with the goal of describing the electronic properties of graphene.
In \cite{iorio} the goal was of the opposite kind. Namely, to identify the conditions for which graphene might get as close as possible to a full-power QFT in curved spacetime. Therefore, key issues had to be faced, such as the proper inclusion of the time variable in a relativistic-like description, and the role of the nontrivial vacua and their relation to different quantization schemes for different observers. All of this finds its synthesis in the Unruh or the Hawking effects, the clearest and unmistakable signatures of QFT in curved spacetime (see, e.g., \cite{penrose}). Therefore, starting from \cite{iorio}, this road was pursued in \cite{ioriolambiase, prd2014}. Let us explain here the main issues an the approximations made there.

Besides the scale (\ref{Escale}), when we introduce curvature, we also have a second scale. When this happens, $E_\ell$ is our ``high energy regime''. This is so because we ask the curvature to be small compared to a limiting maximal curvature, $1/\ell^2$, otherwise: i) it would make no sense to consider a smooth metric, and ii) $r < \ell$ (where $1/r^2$ measures the intrinsic curvature), means that we should bend the very strong $\sigma$-bonds, an instance that does not occur. Therefore, our second scale is
\be
E_r \sim v_F / r  \;,
\ee
with $E_r =  \ell / r \; E_\ell  < E_\ell$. To have a quantitative handle on these scales, let us take, e.g., $r \simeq 10 \ell$ as a small radius of curvature (high intrinsic curvature). To this corresponds an energy $E_r \sim 0.4$eV, whereas, to $r \sim 1 {\rm mm} \sim 10^6 \ell$, corresponds $E_r \sim 0.6 \mu$eV. The ``high energy'' to compare with is $E_\ell \sim 4$eV.

When energies are within $E_r$ (wavelengths comparable to $2 \pi r$) the electrons experience the global effects of curvature. That is to say that, at those wavelengths, they can distinguish between a flat and a curved surface, and between, e.g., a sphere and a pseudosphere. Therefore, whichever curvature $r < \ell$ we consider, the effects of curvature are felt until the wavelength becomes comparable to $2 \pi \ell$. The formalism we shall use, though, are taking into account all deformations of the geometric kind, with the exception of torsion. Hence, this includes intrinsic curvature, and elastic strain of the membrane (on this see \cite{pablo1}), but our predicting power stops before $E_\ell$, because there local effects (such as the actual structure of the defects) play a role that must be taken into account.

Let us recall the various kinds of possible deformation, and how they can be encoded within the Dirac field formalism. First of all, we do not consider here impurities, vacancies, wrinkles, Coulomb scatterers, resonant scatterers (see, e.g., \cite{peres}). The only defects are disclinations, necessary for the intrinsic curvature \cite{Kleinert} \cite{corvoz}: that is squares, pentagons, heptagons, octagons, etc, within the hexagonal lattice. We can then safely assume that the mobility of the electrons is not highly affected by strong scatterers, such as the Coulomb and resonant ones, for which the scattering cross section diverges at long wavelengths \cite{peres}. In fact, in the latter case the defects are such that their local effects are so strong (think, e.g., of a $\delta$-function type of potential), that even electrons with very large wavelength are affected on length scales of the order of $\ell$, hence the lattice/discrete structure of the graphene membrane needs to be taken into account. We are, instead, in a weak scatterers regime\cite{corvoz}, where the Born approximation is valid. That said, there are then three kinds of deformation that can still be at work \cite{pacoreview2009}: intrinsic and extrinsic curvature, torsion, and strain.

The intrinsic curvature is taken here as produced by the mentioned disclination defects, that are customarily described in elasticity theory (see, e.g., \cite{Kleinert,KatVol92}), by the (smooth) derivative of the (non-continuous) SO(2)-valued rotational angle $\partial_i {\omega} \equiv {\omega_i}$, where $i=1,2$ is a curved spatial index (see the Introduction for notation on indices etc). The corresponding (spatial) Riemann curvature tensor is easily obtained
\begin{equation}\label{a11}
    {R^{i j}}_{k l} =
    \epsilon^{i j} \epsilon_{k l} \epsilon^{m n} \partial_{m} \omega_{n} =
    \epsilon^{i j} \epsilon_{l k} 2 {\cal K}.
\end{equation}
where $\cal K$ is the Gaussian (intrinsic) curvature of the surface. In our approach we include time, although the metric we shall adopt is
\begin{equation}\label{mainmetric}
g^{\rm graphene}_{\mu \nu}  = \left(\begin{array}{cc} 1 & 0  \quad 0 \\ \begin{array}{c} 0 \\ 0 \end{array} & g_{i j} \\ \end{array} \right)\;,
\end{equation}
i.e., the curvature is all in the spatial part, and $\partial_t g_{i j}= 0$. Since the time dimension is included, the SO(2)-valued (abelian) disclination field has to be lifted-up to a SO(1,2)-valued (non-abelian) disclination field\footnote{Recall that in three dimensions $\omega_{\mu \; a b} = \epsilon_{a b c} \,\omega_\mu^{\;\; c}$.}, ${\omega_\mu}^a$, $a=0,1,2$, with $\omega_\mu^{\; a} = e^b_\mu \omega_b^{\; a}$ and the expression
\begin{equation}\label{a9}
\omega_a^{\; d}  = \frac{1}{2} \epsilon^{b c d} \left( e_{\mu a} \partial_b E_c^\mu + e_{\mu b} \partial_a E_c^\mu
+ e_{\mu c} \partial_b E_a^\mu \right) \;,
\end{equation}
gives the relation between the disclination field and the metric (dreibein). All the information about intrinsic curvature does not change. For instance, the Riemann curvature tensor, ${R^\lambda}_{\mu \nu \rho}$, has only one independent component, proportional to $\cal K$, just like in (\ref{a11}) (see \cite{iorio}). What we gain here is the possibility to use a full relativistic approach, where, for instance, a change of frame (or the inclusion of an external potential, mimicking the gravitational potential) might change $g_{00}$ in (\ref{mainmetric}). In the next Sections we shall exploit these features.

\begin{figure}
 \centering
  \includegraphics[height=.4\textheight]{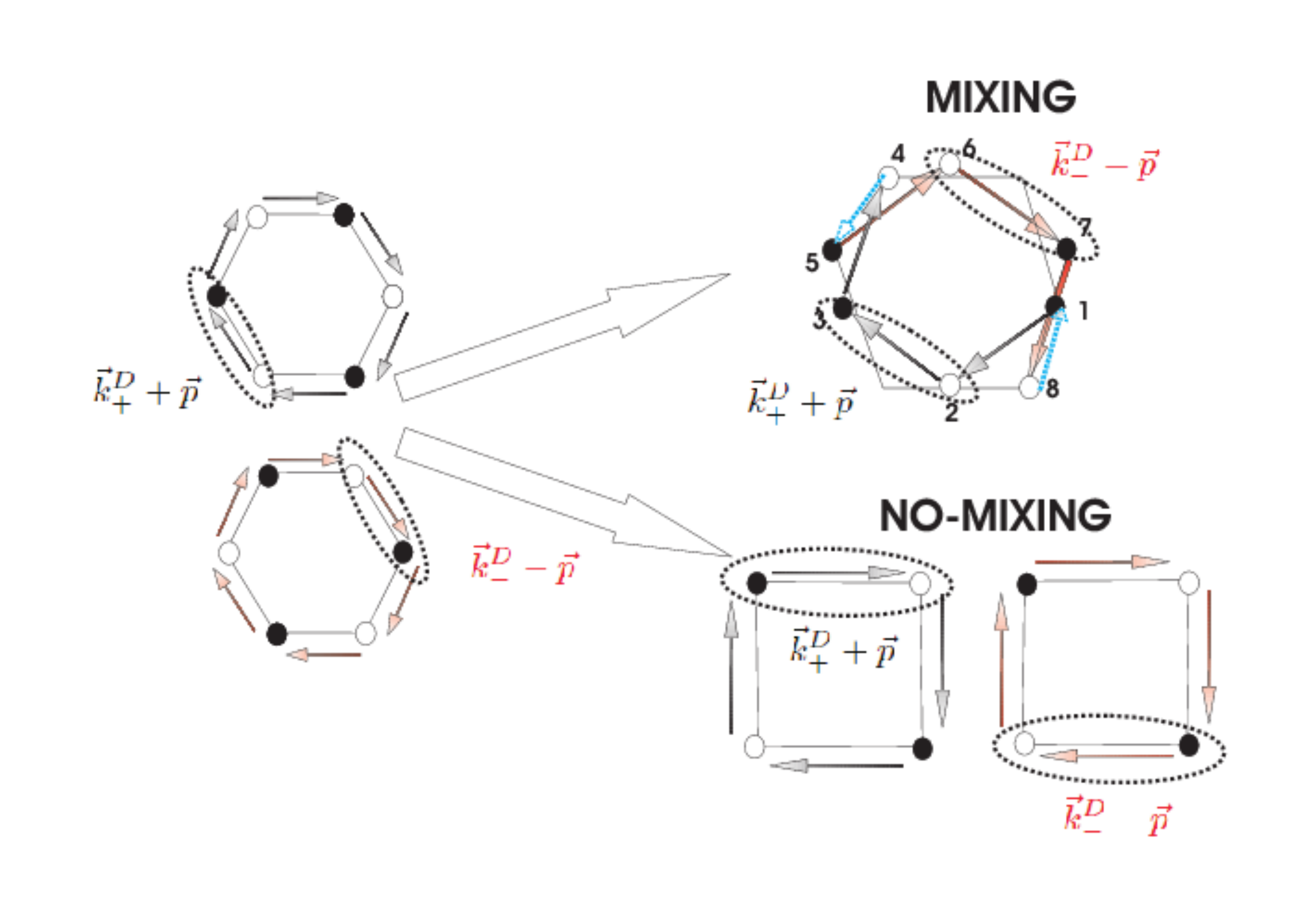}
  \caption{When a disclination defect is odd-sided the two descriptions, $\psi_+$ and $\psi_-$ are necessarily mixed. This happens because, if one starts at point $1$, say, and wants to make a full turn around the defect, to keep the rule that after a $\bullet$ there is a $\circ$ (so that one retains the doublet $\psi$ Dirac description), at some stage (that is point $4$ here) one is forced to move to the defect as seen from the other side of the membrane (following the same parity correspondence depicted in Fig.~\ref{stripe}, and indicated here by the circled jumps). Then, keeping going there (that is the route $5 \to 8$), the same issue applies in reverse, and one needs to consider again the defect in the starting picture. This, on the one hand, clearly shows how one needs both picture at once, and on the other hand, shows that odd-sided defects here are introducing torsion. Indeed, following the same rules, one sees that even-sided defects do not have this behavior, hence one Dirac point only is enough, just like in the flat case. But curvature is present for both defects! So what is happening for the even-folded defects is that the geometry is the same as that of a stripe with an even number of  ``Moebius'' turns, that is continuously deformable into a stripe with no turns at all \cite{reviewslectures}.}
\label{defects}
\end{figure}

These defects come in two categories, odd-folded (pentagons, heptagons), and even-folded (squares, octagons). When the odd-folded are present, the two Fermi points mix, whereas this does not happen for even-folded defects. This is explained in Fig.~\ref{defects}, for the examples of a pentagon and a square, both positive curvature defects. With this in mind, let us consider the action for the Hamiltonian (\ref{Hdiracgraphene1})
\begin{equation} \label{actionDirac}
    A =  v_F \int d^3 x (i \Psi^\dagger \; \dot{\Psi} - {\cal H}) = i v_F \int d^3 x \bar{\Psi} \; \gamma^a \partial_a \; \Psi
\end{equation}
where, $x^0 \equiv v_F t$, with the $4 \times 4$ $\gamma$ matrices given by (\ref{gamma4times4}). It is easy to convince oneself that, no matter the choice of the two-component spinors $\psi_\pm$, and no matter the choice of the $\gamma$s, the action
\begin{equation}\label{Acurved}
{\cal A} = i \int d^3 x \sqrt{g} \; \bar{\Psi} \gamma^a E^\mu_a (\partial_\mu + \Omega_\mu) \Psi
\;,
\end{equation}
with
\be
\Omega_\mu \equiv {\omega_\mu}^a J_a \;,
\ee
and $J_a$ the generators of SO(1,2), the local Lorentz transformations (see above), can never take into account mixing of the $\psi_\pm$.

This is so, because the generators $J_a$ are of the form
\be
J^a = \left(\begin{array}{cc} j^a_+ & 0 \\ 0 & j^a_- \\ \end{array} \right) \;,
\ee
whereas, what is necessary are generators of the form
\be
K^a = \left(\begin{array}{cc} 0 & k^a_+ \\  k^a_- & 0 \\ \end{array} \right) \;.
\ee
From Fig.~\ref{defects}, and from the picture of graphene as a two-sided membrane (see Fig.~\ref{stripe}), it is clear that the mixing is happening when the membrane has torsion. Indeed, when outside and inside-up-side-down parts of a membrane need be considered together, it means that we are in the presence of a Moebius stripe, otherwise, a continuous transformation brings back to a simple stripe \cite{reviewslectures}. Therefore, the most general gauge field, that takes into account strain, curvature (intrinsic and extrinsic) and torsion is of the form $A_\mu = \Omega_\mu + {\cal K}_\mu$.

Work is in progress \cite{pablo1} to fully take into account all these features, and put them in contact with known structures of QFT in curved spacetime \cite{witten3dgravity}. Nonetheless, the mixing terms $\bar{\Psi} \Psi \sim \psi^\dag_+ \psi_- + \psi^\dag_- \psi_+$, appear to spoil Weyl invariance as $\bar{\Psi} \Psi$ is not invariant. The matter might be faced by invoking extra couplings, as done, e.g., in \cite{mauriciosusy}, but we shall not do that here. In fact, we shall take a shortcut that allows for curvature but avoid torsion, and that boils down to the request that only even-folded defects are present.

When this happens, not only we retain Weyl invariance in the most straightforward way, but we also avoid mixing of the two Fermi points. Henceforth, from now on, we shall work with one Dirac point only, hence we go back to Hamiltonian (\ref{HGrapheneBpaper}), and Legendre transform that to
\begin{equation} \label{actionfirst}
    A =  i v_F \int d^3 x \bar{\psi} \; \gamma^a \partial_a \; \psi \;,
\end{equation}
and from here we have that the long wavelength/small energy electronic properties of graphene, within the approximations that keep the Weyl invariance, are well described by the following action
\begin{equation}\label{actionAcurvedpaper}
{\cal A} = i  v_F \int d^3 x \sqrt{g} \; \bar{\psi} \gamma^\mu (\partial_\mu + \Omega_\mu) \psi \;,
\end{equation}
that we shall use from now on.

\chapter{Use of Weyl symmetry to probe QFT in curved graphene spacetimes}

Bearing in mind the previous discussion, to extract experimental predictions from the hypothesis that graphene conductivity electrons realize a quantum field on a curved background that enjoys local Weyl symmetry, hence described by the action (\ref{actionAcurvedpaper}), we proceed as follows.

First of all, we focus on surfaces of constant $\cal K$. As recalled at the end of the Introduction, and as explained in \cite{iorio}, to make the most of the Weyl symmetry of (\ref{actionAcurvedpaper}), we better focus on conformally flat metrics. The simplest metric to obtain in a laboratory is of the kind (\ref{mainmetric}). For this metric the Ricci tensor is ${R_\mu}^\nu = {\rm diag}(0, {\cal K}, {\cal K})$. This gives as the only nonzero components of the Cotton tensor, $C^{\mu \nu} = \epsilon^{\mu \sigma \rho} \nabla_\sigma {R_\rho}^\nu + \mu \leftrightarrow \nu$, the result $C^{0 x} = - \partial_y {\cal K} = C^{x 0} $ and $C^{0 y} = \partial_x {\cal K} = C^{y 0}$. Since conformal flatness in (2+1) dimensions amounts to $C^{\mu \nu} = 0$, this shows that all surfaces of constant $\cal K$ give raise in (\ref{mainmetric}) to conformally flat (2+1)-dimensional spacetimes (note that the result holds for $(+,-,-)$ and for $(+,+,+)$). This means that we focus on {\it surfaces of constant Gaussian curvature}.

The result $C^{\mu \nu} = 0$ is intrinsic (it is a tensorial equation, true in any frame), but to exploit Weyl symmetry to extract non-perturbative exact results, we need to find the coordinate frame, say it $Q^\mu \equiv (T,X,Y)$, where
\begin{equation}\label{genexplicitconfflat}
g^{\rm graphene}_{\mu \nu}  (Q) = \phi^2(Q) g^{\rm flat}_{\mu \nu} (Q) \;.
\end{equation}
Here, besides the technical problem of finding these coordinates, the issue is: \textit{what is the physical meaning of the coordinates $Q^\mu$, and their practical feasibility}.

Tightly related to the previous point is the issue of a conformal factor that makes the model {\it globally predictive, over the whole surface/spacetime}. The simplest possible solution would be a single-valued, and time independent $\phi(q)$, already in the original coordinates frame, $q^\mu \equiv (t,u,v)$, where $t$ is the laboratory time, and, e.g., $u, v$ the meridian and parallel coordinates of the surface.

Here we are dealing with a spacetime that is embedded into the flat (3+1)-dimensional Minkowski. Although, as said, we shall focus on intrinsic curvature effects, just like in a general relativistic context, issues related to the embedding, even just for the spatial part, are important. For instance, when the surface has negative curvature, we need to move from {\it the abstract objects of non-Euclidean geometry} (say the coordinates of the upper-half plane  model of Lobachevsky geometry), {\it to objects measurable in a Euclidean real laboratory}. This will involve the last issue above about global predictability, and, in the case of negative curvature, will necessarily lead to singular boundaries for the surfaces, as proved in a theorem by Hilbert, see, e.g., \cite{penrose, ovchinnikov, mclachlan}. Even the latter fact is, once more, a coordinates effect, due to our insisting in embedding in $\mathbf{R}^3$, and clarifies the hybrid nature of these pseudo-relativistic settings.

We then need to find the {\it quantum vacuum of the field}, to properly take into account: (a) {\it the measurements}, as for any QFT on a curved spacetime, and (b) {\it the graphene hybrid situation}. As well known, in QFT, in general, we have choices of the ground states that are not equivalent, i.e. they are not connected through a unitary transformation \cite{habilitation, uirs}. This instance becomes particularly important in QFT in curved spacetimes, where those inequivalent vacua are related to different observers, see, e.g., \cite{birrellanddavies, takagi, israel}.

Having in mind that the most clear prediction of QFT on a curved spacetime is the \textit{Hawking effect}, if we want to prove beyond doubt that graphene realizes such a system, we shall have to face the challenge to reproduce on graphene the conditions for this effect to take place. Thus, one of the main challenges is to realize the conditions for which an event horizon appears. Having confined ourselves to metrics of the kind (\ref{mainmetric}) the task is indeed a difficult one, and, since we shall focus on surfaces of constant negative $\cal K$, we have to face the fact that the surface might end before the horizon is reached, see \cite{zermelo}. In \cite{ioriolambiase,prd2014}, for one specific case, and in the following, for more cases, we show that these problems can be solved.

Of course, we do not argue that this procedure is the only one leading to an experimental test of the validity of the QFT in curved spacetime description of the physics of graphene's $\pi$ bonds. One could depart from the beginning from the metric (\ref{mainmetric}), for instance by applying external electromagnetic fields, or imagining more or less exotic situations where the $g_{00} \neq 1$. Nonetheless, acting as explained above merges two goals: to be experiments-friendly, and to keep on board as many as possible of the crucial aspects of QFT in curved spacetimes.

Let us now face all the issues of this list.

\chapter{Embedding surfaces in $\mathbf{R}^3$, obtaining conformally flat (2+1)d spacetimes}

Our focus will be on the surfaces of constant Gaussian curvature, one of the subject matters of the classic studies of differential geometry \cite{spivak, eisenhart}. Let us recall here the main facts about them that we shall need in the following.

\begin{figure}
 \centering
  \includegraphics[height=.4\textheight]{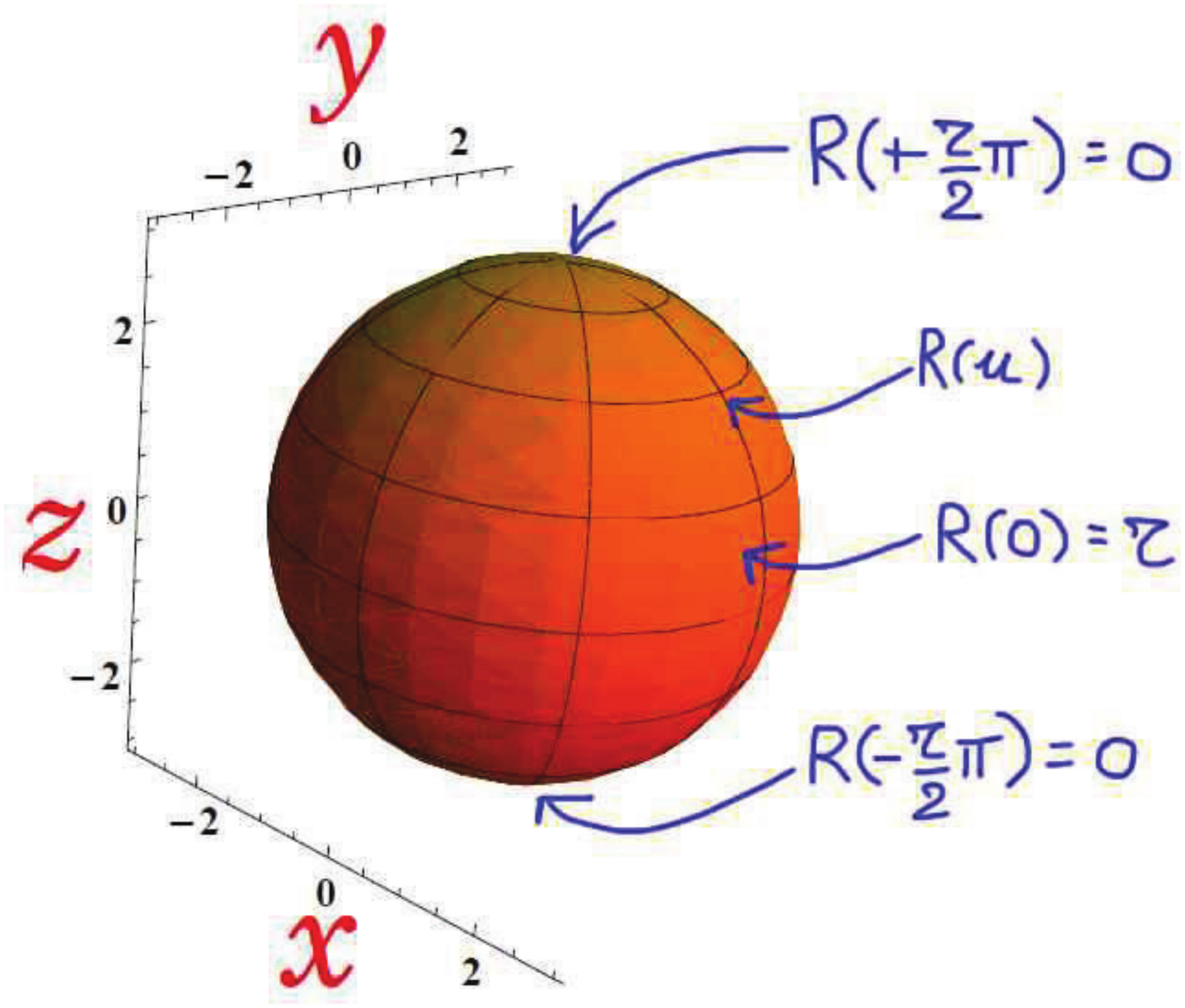}
  \caption{We follow here the general procedure of the successive circular ($v\in [0,2\pi]$) slicing of varying radius $R(u)$, for any given $u$ in the range, in the example of the sphere. $R(u) = r \cos u/r$, and  $u \in [- \frac{r}{2} \pi, + \frac{r}{2} \pi]$. Thus $R=0$ at the south pole $u = - \frac{r}{2} \pi$,  $R = r$ at the equator $u = 0$, and again $R=0$ at the north pole $u = + \frac{r}{2} \pi$. This clarifies that the ``radius'' $R(u)$ is actually a function related to the profile curve, $z(u)$. In the plot $r=2$.}
\label{sphere}
\end{figure}

In general, there is no single parametrization good for all surfaces. In fact, for the surfaces of revolution, there is one such parameterizations, sometimes called ``canonical'', that we now introduce. Surfaces of revolution are the surfaces swapped by a (profile) curve, say in the plane $(x,z)$, rotated of a full angle around the $z$-axis. All such surfaces (both of constant and nonconstant $\cal K$) can be parameterized in $\mathbf{R}^3$ as
\begin{equation} \label{canonicalpar}
x(u,v) = R(u) \cos v \;, \; y(u,v) = R(u) \sin v \; , \; z(u) = \pm \int^u \sqrt{1 - {R'}^2(\bar{u})} d \bar{u} \;,
\end{equation}
where prime denotes derivative with respect to the argument, $v\in [0, 2 \pi]$ is the parallel coordinate (angle), and $u$ is the meridian coordinate whose range is fixed by the knowledge of $R(u)$, i.e. of the type of surface, through the request that $z (u) \in {\mathbf R}$. The relation between $z$ and $R$ comes from the constraint ${R'}^2 (u) + {z'}^2 (u) = 1$, that amounts to a choice (that we are free to make) for the coefficients of Gauss's first fundamental form given by \cite{spivak} $E = 1, F = 0, G = R (u)$. A direct proof of this last statement is obtained by considering the embedded line element descending from (\ref{canonicalpar})
\begin{equation}\label{linelsurfrev}
d l^2 \equiv dx^2 + dy^2 + dz^2 = du^2 + {R}^2(u) dv^2  \;.
\end{equation}
The expression on the far right side above is the typical line element of a surface of revolution. We shall always deal with such type of line element, for the spatial part. The way the surface of revolution can be plotted via the line element (\ref{linelsurfrev}) is by drawing successive circular ($v\in [0,2\pi]$) slices of varying radii $R(u)$, an example is explicitly shown in Fig.~\ref{sphere}.

The Gaussian curvature is given by the simple expression \cite{spivak}
\begin{equation}\label{gaussian}
{\cal K} = - \frac{R''(u)}{R(u)} \;.
\end{equation}
Thus, the knowledge of $R(u)$ amounts to the knowledge of the surface of revolution. When ${\cal K}$ is constant, (\ref{gaussian}) is an easy equation to solve
\begin{equation}\label{solgaussian1}
R(u) = c \cos (u/r + b) \quad {\rm for} \quad {\cal K} = \frac{1}{r^2} \;,
\end{equation}
\begin{equation}\label{solgaussian2}
R(u) = c_1 \sinh (u/r) + c_2 \cosh (u/r )\quad {\rm for} \quad {\cal K} = - \frac{1}{r^2} \;,
\end{equation}
where $r \in {\mathbf R}$ is constant, and $c, b, c_1, c_2$ are also real constants, that determine the type of surface, and/or set the zero and scale of the coordinates.

When ${\cal K} = 1/r^2$, one first chooses the zero of $u$ in such a way that $b$ in (\ref{solgaussian1}) is zero, then distinguishes three cases,
$c = r$, $c > r$, $c < r$. The first case is the sphere of radius $r$, the other two surfaces are applicable to the sphere through a redefinition of the meridian coordinate $v \to (c/r) v$. With these, $z(u) = \int^u \sqrt{1 - (c^2 / r^2) \sin^2(\bar{u}/r)} d\bar{u}$, and the range of $u$ changes according to the relation between $c$ and $r$, being $z(u) = r \sin (u/r)$, with $u/r \in [-\pi/2 , + \pi/2]$ for the sphere.

When ${\cal K} = - 1/r^2$, all the surfaces described by (\ref{solgaussian2}) can be applied to one of the following three cases: $c_1 = c_2 \equiv c$, giving
\be \label{Rbeltrami}
R(u) = c \; e^{u/r} \;,
\ee
or $c_1 = 0$, $c_2 \equiv c$, giving
\be \label{Rhyperbolic}
R(u) = c \; \cosh(u/r) \;,
\ee
or $c_2 = 0$, $c_1 \equiv c$, giving
\be \label{Relliptic}
R(u) = c \; \sinh (u/r) \;.
\ee
They are called the \textit{Beltrami}, the \textit{hyperbolic}, and the \textit{elliptic} pseudospheres, respectively, and the corresponding expressions for $z(u)$ are given by substituting $R(u)$ in the integral in (\ref{canonicalpar}). Very importantly for us, {\it all surfaces} of constant negative $\cal K$, not only the surfaces of revolution, are applicable to either the Beltrami, or the hyperbolic, or the elliptic pseudospheres, see, e.g., \cite{eisenhart}.

The condition $z \in {\mathbf R}$ gives the range of $R$ and $u$ in the various cases
\bea
R(u) \in [0, r] \Leftrightarrow u & \in & [- \infty, r \ln(r/c)] \label{raggiobel} \\
R(u) \in [c, \sqrt{r^2 + c^2}] \Leftrightarrow u & \in & [- {\rm arccosh} (\sqrt{1 + r^2 /c^2}) , + {\rm arccosh} (\sqrt{1 + r^2/c^2})]  \label{raggiohyp} \\
R(u) \in [0, r \cos \vartheta] \Leftrightarrow u & \in & [0 , {\rm arcsinh} \cot \vartheta] \label{raggioell}
\eea
where, in the first two cases, $c$ is only bound to be a real positive number, while in the last case $0 < c = r \sin \vartheta < r$. Furthermore, in the second case, $R(u)$ is an even function of $u$, hence, in the symmetric interval, reaches the maximum twice. Notice also that the only case where the range of $R$ is independent from $c$ is for the Beltrami surface. More details of these surfaces are in the captions of the relative figure, see Figs. \ref{Beltrami}, \ref{hyperbolic} and \ref{elliptic}.

On the mathematics side, our goal is to find the coordinate frame $Q^\mu \equiv (T,X,Y)$ where the metric (\ref{mainmetric}) is explicitly conformally flat. On the physics side, we have to understand what are the conditions that need to be realized on graphene to correspond to this frame, and how feasible this is.

One problem to solve, on the spatial part, is to combine the canonical parametrization (\ref{canonicalpar}), for which it is immediate to plot the surface, with the spatial isothermal coordinates, $(\tilde{x}, \tilde{y})$, where $dl^2 = \varphi^2 (\tilde{x}, \tilde{y}) (d\tilde{x}^2 + d\tilde{y}^2)$, where the task to find the coordinate frame $Q^\mu$ is easier. Indeed,
\begin{equation}\label{metricconf2conf3}
g^{\rm graphene}_{\mu \nu}  = {\rm diag} \left( 1, - \varphi^2 (\tilde{x}, \tilde{y}), - \varphi^2 (\tilde{x}, \tilde{y}) \right)
= \phi^2 (T,X,Y) \; {\rm diag} (1, -1, -1) \;,
\end{equation}
hence, using the standard $g_{\mu \nu} (Q) = (\partial Q_\mu / \partial q_\lambda) \; (\partial Q_\nu / \partial q_\kappa) \; g_{\lambda \kappa} (q)$,  the system of partial differential equations to solve simplifies to
\be \label{changecoord1}
\phi^2 \left( T^2_t - X^2_t - Y^2_t \right) =  1 \;, \;
\phi^2 \left( T^2_{\tilde{y}} - X^2_{\tilde{y}} - Y^2_{\tilde{y}} \right) = - \varphi^2  =  \phi^2 \left( T^2_{\tilde{x}} - X^2_{\tilde{x}} - Y^2_{\tilde{x}} \right)
\ee
\be \label{changecoord6}
T_t T_{\tilde{x}} - X_t X_{\tilde{x}} - Y_t Y_{\tilde{x}} =
T_t T_{\tilde{y}} - X_t X_{\tilde{y}} - Y_t Y_{\tilde{y}} =
T_{\tilde{x}} T_{\tilde{y}} - X_{\tilde{x}} X_{\tilde{y}} - Y_{\tilde{x}} Y_{\tilde{y}} = 0 \;,
\ee
where $T_t \equiv \partial_t T(t, \tilde{x}, \tilde{y})$, etc..

Isothermal coordinates can always be found for surfaces of revolution, by using the meridian and parallel parametrization. To see it, use the following re-parametrization of (\ref{canonicalpar})
\begin{equation} \label{isothermpar}
x(\tilde{R},v) = \tilde{R} \cos v \;, \; y(\tilde{R},v) = \tilde{R} \sin v \; , \; z(\tilde{R}) = f(\tilde{R}) \;,
\end{equation}
so that $d l^2 \equiv dx^2 + dy^2 + dz^2 = (1 + {f'}^2 (\tilde{R})) d{\tilde{R}}^2 + {\tilde{R}}^2 dv^2$,
then use $\tilde{u} \equiv \int \sqrt{1 + {f'}^2 (\tilde{R})} / \tilde{R} \; d \tilde{R}$, that gives
$dl^2 = {\tilde{R}}^2 (\tilde{u}) \left( d{\tilde{u}}^2 + dv^2 \right)$ with ${\tilde{R}} (\tilde{u})$ obtained by inverting the definition of $\tilde u$. Note that $\tilde{u} (u)$, and ${\tilde{R}} (\tilde{u}(u)) = R(u)$. This means that for surfaces of revolution\footnote{For a generic surface, i.e. not a surface of revolution, this is not the case. If we could have the general procedure to go from a parametrization of the surface where the visualization is easy (the canonical parametrization (\ref{canonicalpar}) being one example), to the isothermal coordinates, that give the conformal factor $\varphi^2$, we would immediately know the profiles that graphene should have in order to correspond to important algebraic structures, such as the Virasoro algebras (for ${\cal K} = 0$) and the Liouville structures (for ${\cal K} \neq 0$), which naturally emerge  here in terms of the conformal factor $\varphi^2$, see  \cite{iorio}. Among the latter, for instance, particularly rich are the vortex solutions of Liouville equation found in \cite{liouville4}.}
\be\label{virasoroliouville}
\varphi (\tilde{x}, \tilde{y}) = {\tilde{R}} (\tilde{u}) \;.
\ee
Thus focusing on the surfaces of revolution (and of constant $\cal K$), we are moving in the right direction, but this does not guarantee that we can always succeed to find the coordinates $Q^\mu$ this way. For instance, for the sphere, on the one hand, it is easy to find the isothermal coordinates
\be \label{firstabstactcoord}
\tilde{x} = v \;, \quad \tilde{y} = \ln \left(1 + \frac{2}{\cot(u/2r) - 1} \right) \;,
\ee
for which\footnote{To see the full match with (\ref{virasoroliouville}) let us consider, for simplicity, a unit sphere, $r=1$, centered at the origin, and let us use the standard parametrization  $x = \sin \theta \cos \chi$, $y = \sin \theta \sin \chi$, $z = \cos \theta$. For this, $\tilde{R} (\tilde{u} (\theta)) = \sin \theta$. Hence, from $x^2 + y^2 + z^2 \equiv {\tilde{R}}^2 + z^2 = 1$, one gets $z = \sqrt{1 - {\tilde{R}}^2} = f({\tilde{R}}) = \cos \theta$. Applying the procedure above,
$\tilde{u} = \int 1/{\tilde{R}} \sqrt{1 + {\tilde{R}}^2 / (1 - {\tilde{R}}^2)} = \ln \tan(\theta/2)$, or $\theta = 2 \arctan e^{\tilde{u}}$, that gives ${\tilde{R}} = \sin \theta = 1/ \cosh \tilde{u}$ (that is also a nice formula to relate trigonometric and hyperbolic functions without resorting to complex numbers). Then, defining $\tilde{u} = \tilde{y}$, and $v = \tilde{x}$, one gets the line element (\ref{isothermsphere}).}
\be\label{isothermsphere}
dl^2 = du^2 + r^2 \cos^2 \frac{u}{r} dv^2 = \frac{r^2}{\cosh^2 \tilde{y}} \left( d \tilde{x}^2 + d \tilde{y}^2 \right) \;,
\ee
hence\footnote{We used here that, when $u = 2 r \; {\rm arccot} (\frac{2}{e^{\tilde{y}} - 1} + 1)$,  $\cos (u/r) = 1/\cosh \tilde{y}$} $\varphi^2 (\tilde{y}) = r^2 /\cosh^2 \tilde{y} = r^2 \cos^2 \frac{u}{r} = \varphi^2(u)$. One can also check that the Liouville equation is satisfied\footnote{In the isothermal coordinates $(\tilde{x}, \tilde{y})$, Liouville equation is ${\cal K} = - \frac{1}{2 \varphi} (\ln \varphi)''$. With $\varphi = r^2 / \cosh^2 \tilde{y}$, one immediately finds the result ${\cal K} = 1/r^2$.}. On the other hand, the difficult task is to fit that into the space{\it time}, by solving the system (\ref{changecoord1})-(\ref{changecoord6}), to find the conformal factor for the (2+1)-dimensional spacetime, $\phi^2$, and then understand the physical meaning of the frame where this happens.

Nonetheless, although it would be interesting to use our approach also on this sort of Einstein static universe (ESU) \cite{wald}, later we shall show  that there is no horizon in this case, hence no Hawking phenomenon takes place. This makes the sphere a case less apt for the emergence of unmistakable signatures of QFT in curved spacetime. All of this makes us focus on the cases of constant negative curvature.

\begin{figure}
 \centering
  \includegraphics[height=.4\textheight]{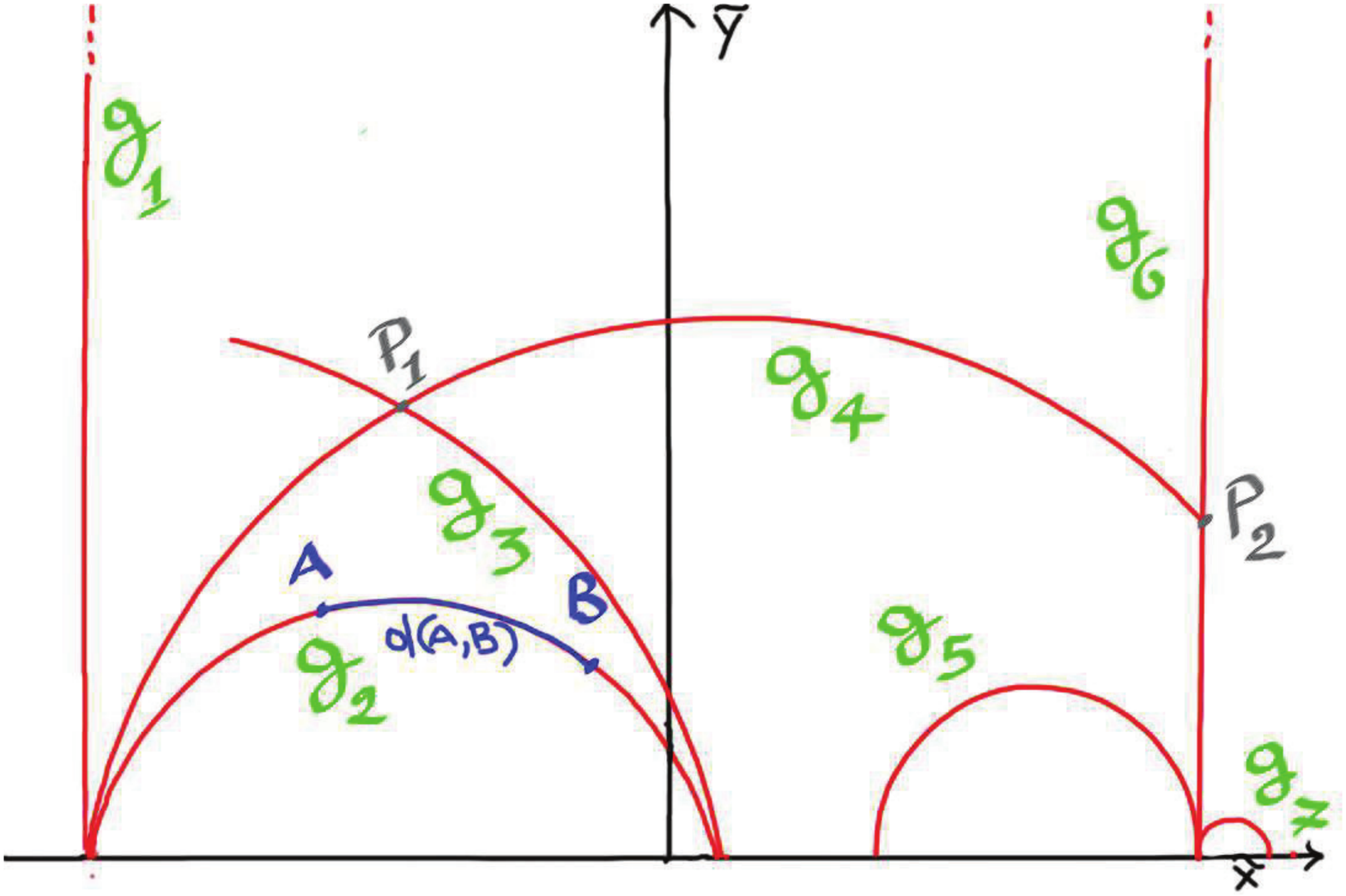}
  \caption{The upper-half plane model, $\{ (\tilde{x}, \tilde{y}) | \tilde{y} > 0 \}$, with metric $d l^2 = \frac{r^2}{{\tilde y}^2}(d{\tilde x}^2+d{\tilde y}^2)$, represents Lobachevsky geometry, both globally and locally. The geodesics, indicated with $g$, are half-circles, starting and ending on the absolute, the $\tilde{x}$-axis. These include the half-lines, e.g.,  $g_1$ and $g_6$ in the figure, as the limiting case of infinite radius. The distance between two points, $A$ and $B$, is
  $d(A,B) = {\rm arcosh} \left( 1 + \frac{(\tilde{x}_B - \tilde{x}_A)^2 + (\tilde{y}_B - \tilde{y}_A)^2}{2 \tilde{y}_A \tilde{y}_B}\right)$. Notice that the points on the absolute $\tilde{y} = 0$ are points at infinity, as measured with the above distance, so they are the points at $\tilde{y} = \infty$. Two geodesics are parallel if they share one point at infinity. In the figure, $g_1 || g_2 || g_4$, as well as $g_5 || g_6 || g_7$, besides, of course $g_1 || g_6$. In the figure there are also two illustrations of the way Lobachevsky geometry overcomes the fifth postulate of Euclidean geometry. Through point $P_1$ go {\it two} parallels to the geodesic $g_2$, namely $g_3$ and $g_4$ (notice that $g_3$ and $g_4$, though, are not parallel to each other). Through point $P_2$ go {\it two} parallels to the geodesic $g_1$, namely $g_4$ and $g_6$ (notice that $g_4$ and $g_6$, though, are not parallel to each other). These are examples of the general statement that, in Lobachevsky geometry, given a straight-line and a point, there are two parallels to the straight-line going through that point.}
\label{circles}
\end{figure}

Before moving to those cases, let us add here that the formulae for the sphere we have re-obtained above are well known, see, e.g., \cite{eisenhart}. The reason for showing them here is that they illustrate, in a very familiar case, that having found the isothermal coordinates, $(\tilde{x}, \tilde{y})$, their link with Euclidean  coordinates (those measurable in a lab) needs to be made explicit. The expressions (\ref{firstabstactcoord}) are one example. This issue, for the sphere, has been solved over the centuries by map-makers\footnote{For instance, the Mercator projection is precisely the $\tilde{x} = v$, $\tilde{y} = \ln \tan(\theta/2)$ discussed in the footnote.}. Less usual is to find solutions for the surfaces of our interest, of which we shall discuss next.

\chapter{Surfaces of constant negative Gaussian curvature}

For these surfaces, the spatial part of the metric of graphene can be written, in isothermal coordinates, as
\begin{equation}
d l^2 = \frac{r^2}{{\tilde y}^2}(d{\tilde x}^2+d{\tilde y}^2) \label{lobsurface}\;,
\end{equation}
where ${\tilde x}, {\tilde y}$ are the \textit{abstract coordinates} of the Lobachevsky geometry in the upper half-plane (${\tilde y}>0$) model. One then immediately realizes that our goal is nearer. Indeed, the full line element is
\begin{equation}
ds^2_{\rm graphene}=\frac{r^2}{{\tilde y}^2}\left[\frac{{\tilde y}^2}{r^2}dt^2-d{\tilde x}^2-d{\tilde y}^2\right] \label{lobspacetime}\;,
\end{equation}
where the line element in square brackets is {\it flat}. This apparently solves our problem: the coordinates $Q^\mu$ appear to be $(t, {\tilde x}, {\tilde y})$, as there we shall always have the explicit conformal factor $\varphi^2 (\tilde{y}) = r^2/{\tilde y}^2$ to implement the Weyl symmetry. Furthermore, the line element in square brackets is of the Rindler kind, see, e.g., \cite{birrellanddavies}, hence it is pointing towards a Unruh kind of effect available for all surfaces of this family. Nonetheless, this is just an important indication but we cannot conclude for any Unruh-Hawking kind of effect hidden in the line element (\ref{lobspacetime}) until we make contact with what can be seen in a real laboratory (not in a ``Lobachevsky laboratory'', so to speak).

As said, we need to refer to coordinates measurable in the Euclidean space ${\mathbf R}^3$ of the laboratory, hence we have to specify ${\tilde x}$ and ${\tilde y}$ in terms of coordinates measurable using the Euclidean distance (embedding), say the $(u,v)$ coordinates. If we are lucky, the result will be globally valid already in the frame $(t,u,v)$. Otherwise, we need to change coordinates again, and this means, in general, that we have to abandon the lab time $t$.

\begin{figure}
\centering
\includegraphics[height=.4\textheight]{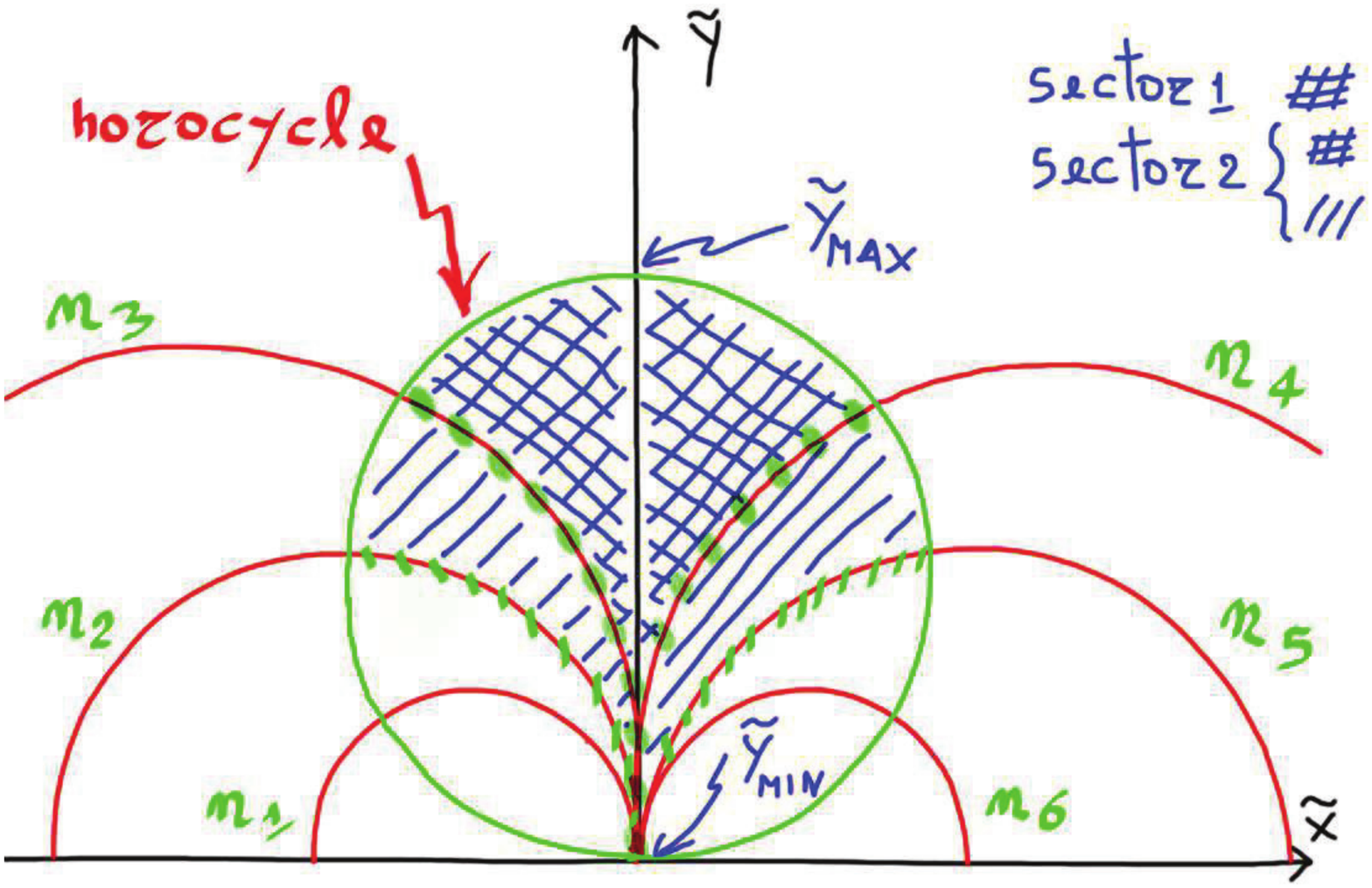}
  \caption{Two horocyclic sectors, in the upper-half model. The horocycle is a curve whose normals ($n_1, ..., n_6$ in the figure), all converge to the same point ($(\tilde{x}, \tilde{y}) = (0,0)$ in the figure). Here it is a full circle, tangent to the absolute. In the next example, see Fig.~\ref{horocycleandBeltrami}, it is a line parallel to the absolute. Here we give two examples of horocyclic sectors. Sector 2 captures a bigger portion of the Lobachevsky geometry than sector 1. Nonetheless, none of them can ever capture the whole upper-half plane. The surface is obtained by identifying the boundaries of the given sector, in the picture, the indicated portion of normals $n_3$ and $n_4$, for sector 1, and the indicated portion of normals $n_2$ and $n_5$, for sector 2.}
\label{horocycle1}
\end{figure}

This is the right place to recall the basic facts of Lobachevsky geometry, the discovery of which put to an end the efforts of generations of mathematicians to prove the fifth postulate of Euclidean geometry as a consequence of the first four. In particular, we focus on the subtle points of embedding this geometry into the Euclidean space. The upper-half plane, $\{ (\tilde{x}, \tilde{y}) | \tilde{y} > 0 \}$, equipped with the metric (\ref{lobsurface}), represents Lobachevsky geometry, both locally and globally. All other realizations (the Poincar\`e disc, and the Minkowski model, being the other most used two) are related to it. The geodesics for (\ref{lobsurface}) are semi-circles\footnote{For the Poincar\`e model the absolute is itself a full circle, and the geodesics are arcs of circle, starting and ending on the absolute. This has been represented in various famous drawings of M.~C.~Escher, one example being ``Circle Limit'', see, e.g., pgs form 34 on of \cite{penrose}.}, starting and ending on the ``absolute'', the boundary of the space, namely the $\tilde{x}$-axis, as shown in Fig.~\ref{circles}.

Every surface of constant negative Gaussian curvature is locally isometric to the upper-half plane with metric (\ref{lobsurface}). This means that we can work with this metric, but have to remember its abstract nature. If we explicitly have $\tilde{x}$ or  $\tilde{y}$ in a formula, when it is time to measure, we have to express them in terms of Euclidean coordinates. Since, from (\ref{lobspacetime}) we already have the explicit conformal factor we wanted, that is $r^2/\tilde{y}^2$, we have to focus on the $\tilde{y}$ coordinate. Indeed, we might run into troubles when we write the specific $\tilde{y}$ for the given surface. The reason is that non Euclidean geometry objects are ``intruders'' in a Euclidean world.

In the literature, when a spacetime of negative curvature in $n$ dimensions (such as the anti-de Sitter, $AdS_n$), is considered, the embedding is done in the un-physical higher dimensional flat spacetime with signature, e.g., $(+, -, \cdots, -, +)$, see, e.g., \cite{btz, BHTZ}. This could be described, for instance, by a spacetime with all real coordinates, say $t, x, \cdots, y \in {\mathbf R}$, but one spatial coordinate, that necessarily is imaginary, say $z = i \zeta, \zeta \in {\mathbf R}$. Here, as we want to realize such spacetimes in a real lab, we shall always embed in the physical spacetime $(+, -, -, -)$, i.e. $t, x, y, z \in {\mathbf R}$.  On this we shall elaborate more in the following Section.

\section{Beltrami pseudosphere}

\begin{figure}
 \centering
  \includegraphics[height=.45\textheight]{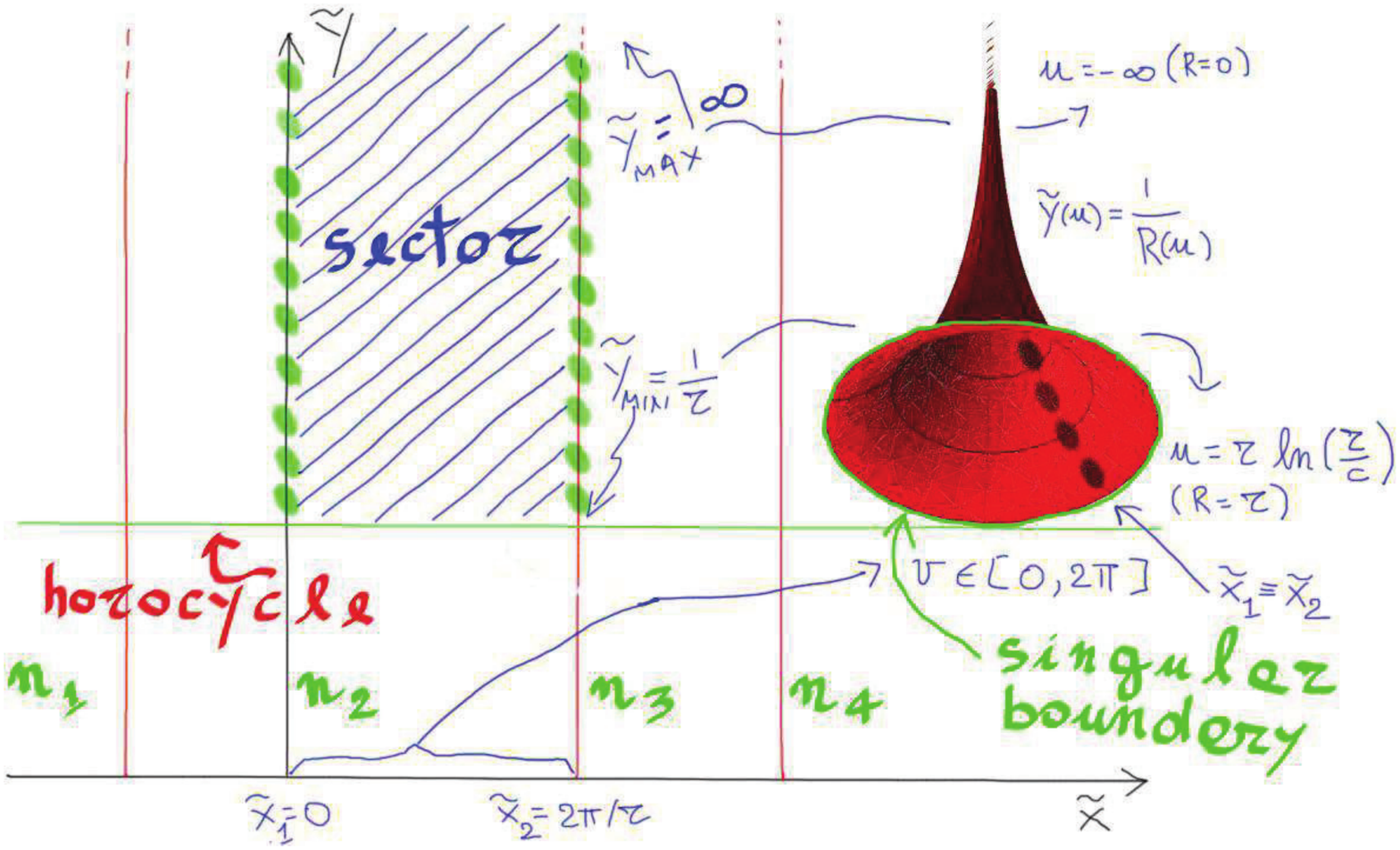}
  \caption{Construction of the Beltrami pseudosphere from an horocyclic sector. The boundaries to be identified are the indicated portions of the normals $n_2$ and $n_3$. The point at infinity here is $\tilde{y}_{max} = \infty$, and it would be the same for any choice of horocycles of this kind. The other end of the surface, corresponding to $\tilde{y}_{min} = 1/r$, is a singular boundary, as predicted by the Hilbert theorem. The range of the meridian coordinate is, of course, $v \in [0,2\pi]$, while the range of the parallel coordinate $u$ is obtained through the equation $\tilde{y} (u) = 1/R(u)$, and $R(u) = c e^{u/r}$, see Eqs.~\ref{explicitxybeltrami}. This figure is taken from \cite{prd2014}.}
\label{horocycleandBeltrami}
\end{figure}

\begin{figure}
\centering
\includegraphics[height=.6\textheight]{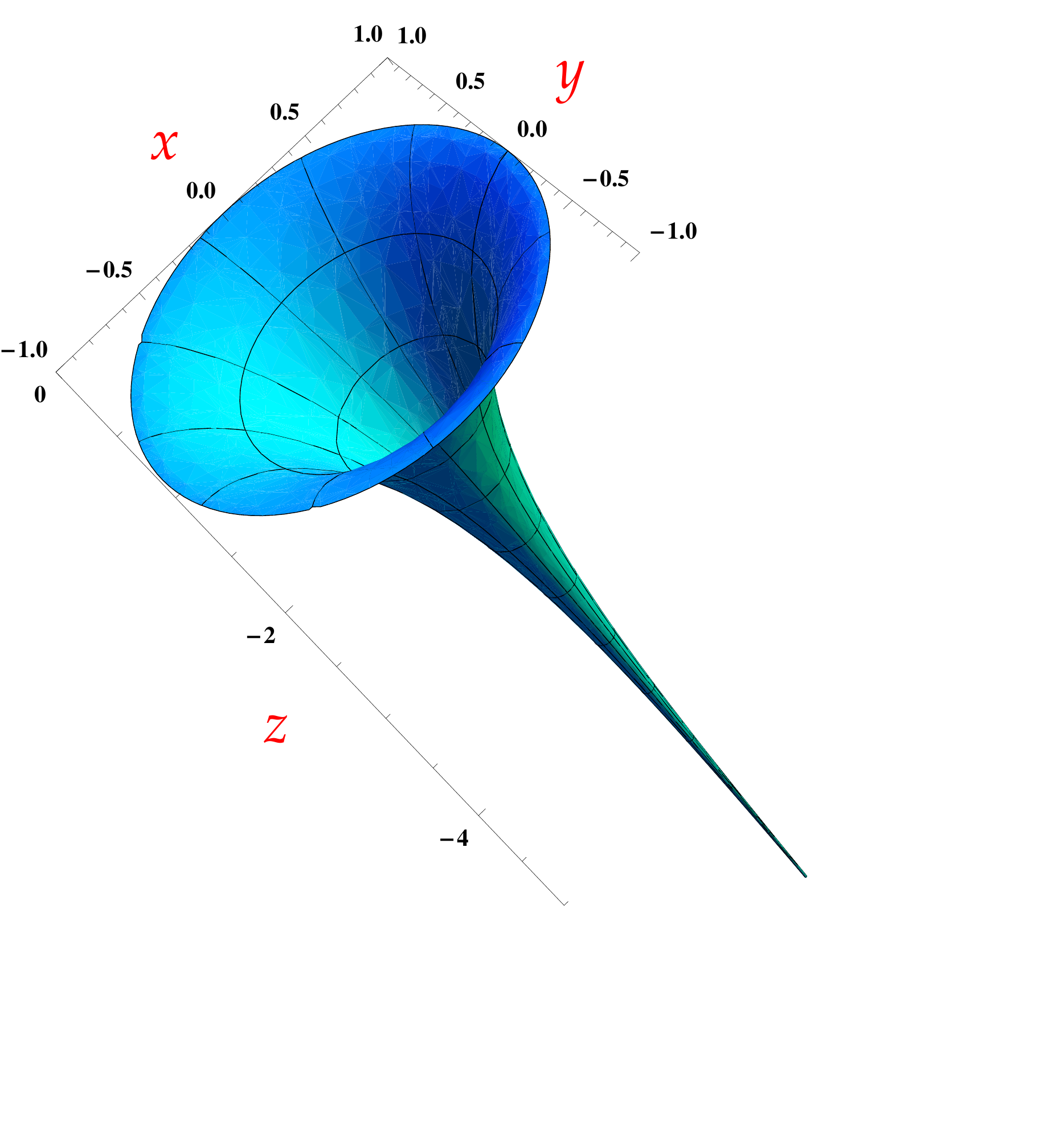}
\caption{Beltrami pseudosphere embedded in $\mathbf{R}^3$: $x(u,v) = c \, e^{u/r} \cos v$, $y(u,v) = c \, e^{u/r} \sin v$, $z(u) = r (\sqrt{1 - (c^2/r^2) e^{2 u/r}} - {\rm arctanh}\sqrt{1 - (c^2/r^2) e^{2 u/r}})$, hence $R(u) =  c \, e^{u/r}$, with $c$ any positive real constant, and $r = \sqrt{-{\cal K}}$. The range of $R$ is $R(u) \in [0,r]$ as $u \in [-\infty, r \ln(r/c)]$, therefore it is an infinite surface, with boundary at $R=r$, where $u=r \ln(r/c)$. Beyond that ($R>r$) $z(u)$ becomes imaginary. The range of $v$ is $[0, 2 \pi]$. In the plot $r=1=c$ and $u \in [-6, 0]$, $v\in [0, 2\pi]$.} \label{Beltrami}
\end{figure}

Let us now see how all the above comes about in practice. We focus on the Beltrami pseudosphere\footnote{Historically, this pseudosphere has been the first example of a surface of constant negative curvature. It convinced the mathematicians of the 1890s, that the exotic scenarios of the Lobachevsky geometry were possible to make real. It is said in many places, see, e.g., \cite{penrose}, that this surface does not deserve much admiration as, after all, it only conveys a little portion of the full intricacies of the Lobachevsky geometry. This latter instance is true, of course. Nonetheless, this surface (and the many others that came afterwards) is a true piece of wander. It is like the intrusion into this (Euclidean, as for space) reality of a different reality, the border between them being the singular boundary (that we call a ``Hilbert horizon'', see later), beyond which the surface ceases to exist {\it here} ($z$ becomes imaginary), and continues {\it there}.}. If we take $\tilde{y}$ in (\ref{lobsurface}) such that
\be
\ln \tilde{y} = - (u/r + \ln c) \;,
\ee
then, $d\tilde{y} / \tilde{y} = - du / r$, or $du^2 = ( r^2 / \tilde{y}^2) d\tilde{y}^2$, so the choice
\be \label{explicitxybeltrami}
\tilde{x} = \frac{v}{r} \quad \;, \quad \tilde{y} = \frac{1}{c} e^{- u / r} \;,
\ee
in (\ref{lobsurface}), gives the line element of the Beltrami pseudosphere
\be
dl^2 = du^2 + c^2 e^{2 u/r} dv^2 \;.
\ee
The equations above connect ``unmeasurable'' objects, $\tilde{x}$,  $\tilde{y}$, to measurable ones, $u$ and $v$, the meridian and the parallel coordinates. Up to now, it does not look such a different situation as the one depicted before for the sphere, see (\ref{firstabstactcoord}). In fact, the big difference is that we can only represent a tiny sector of the Lobachevsky plane into ${\mathbf R}^3$, namely what is called a ``horocyclic sector'', see Fig.~\ref{horocycle1} and Fig.~\ref{horocycleandBeltrami}. This fact is governed by a deep result of Hilbert that says\cite{ovchinnikov, hitchin}: {\it There exists no analytic complete surface of constant negative Gaussian curvature in the Euclidean three-dimensional space.} By ``complete'' it is meant a surface that does not exhibit singularities.

The horocycles, in the upper-half plane model, are curves whose normals all converge asymptotically. Since the geodesics here are: (a) semicircles starting and ending on the $\tilde{x}$-axis, and (b) half-lines starting on the $\tilde{x}$-axis (that are just the limiting case of the former case), we have two kinds of horocycles: full circles tangent to the absolute for (a), and lines parallel to the absolute for (b). Sectors of horocycles are, essentially, the Lobachevsky version of stripes (see, especially Fig.~\ref{horocycleandBeltrami}). Once one realizes that it is impossible to represent the whole of the Lobachevsky geometry on a real surface, the next most natural thing to try is to see whether at least a stripe can be represented. This is, essentially, what Eugenio Beltrami discovered. The details about the actual construction of this pseudosphere from the Lobachevsky plane are in Fig.~\ref{horocycleandBeltrami}, the details on the parametric expression in ${\mathbf R}^3$ are in Fig.~\ref{Beltrami}.

One important point for us is that, by looking at the expressions (\ref{explicitxybeltrami}), we see that $\tilde{y}$ is a smooth, well-behaved, single-valued function, and, to take a full turn on a parallel, $\tilde{x} \to \tilde{x} + 2 \pi / r$, has no effects on $\tilde{y}$. For this coordinate, the only thing we need to care about is that the surface ends abruptly at $\tilde{y}_{min} = 1/r$, corresponding to the maximal circle, $R(u= r \ln \frac{r}{c}) = r$. This maximal circle is what we call ``Hilbert horizon'', to recall that it is an effect of the Hilbert theorem on the embedding in ${\mathbf R}^3$, and that there the Beltrami spacetime ends. This notion of horizon will be put in contact with that of a proper event horizon in the following. Let us notice here that, not always the singular boundaries one has to expect for a generic surface of this family, are so clean. In general, they could be: discrete set of points, open/closed curves, self-intersecting open/closed curves, or a combination of these. This depends on the particular embedding, that can be quite involved (see, e.g., Fig.~\ref{genericpseudosphere} here, and \cite{mclachlan}). Thus, not always it is such an easy task to identify a Hilbert horizon.

For the Beltrami  surface, all the geometrical problems are solved at ones: we have an explicit conformal factor that, through the realization (\ref{explicitxybeltrami}), is well defined all over the non-singular part of the surface (and the singular boundary is a circle), already in the frame of reference $q^\mu = (t,u,v)$, where the time is exactly the lab time. For this surface, then, using the Weyl symmetry, we can extract sensible predictions based on the line element in square brackets in (\ref{lobspacetime}). The latter line element, for this pseudosphere, coincides with a proper Rindler line element, modulo some differences, see \cite{ioriolambiase,prd2014}, and next Section. We shall discuss all of this points in the next Section. Before that, let us have a closer look at the other surfaces of this family.

\section{Hyperbolic, Elliptic, and other pseudospherical surfaces}

\begin{figure}
\centering
\includegraphics[height=.34\textheight]{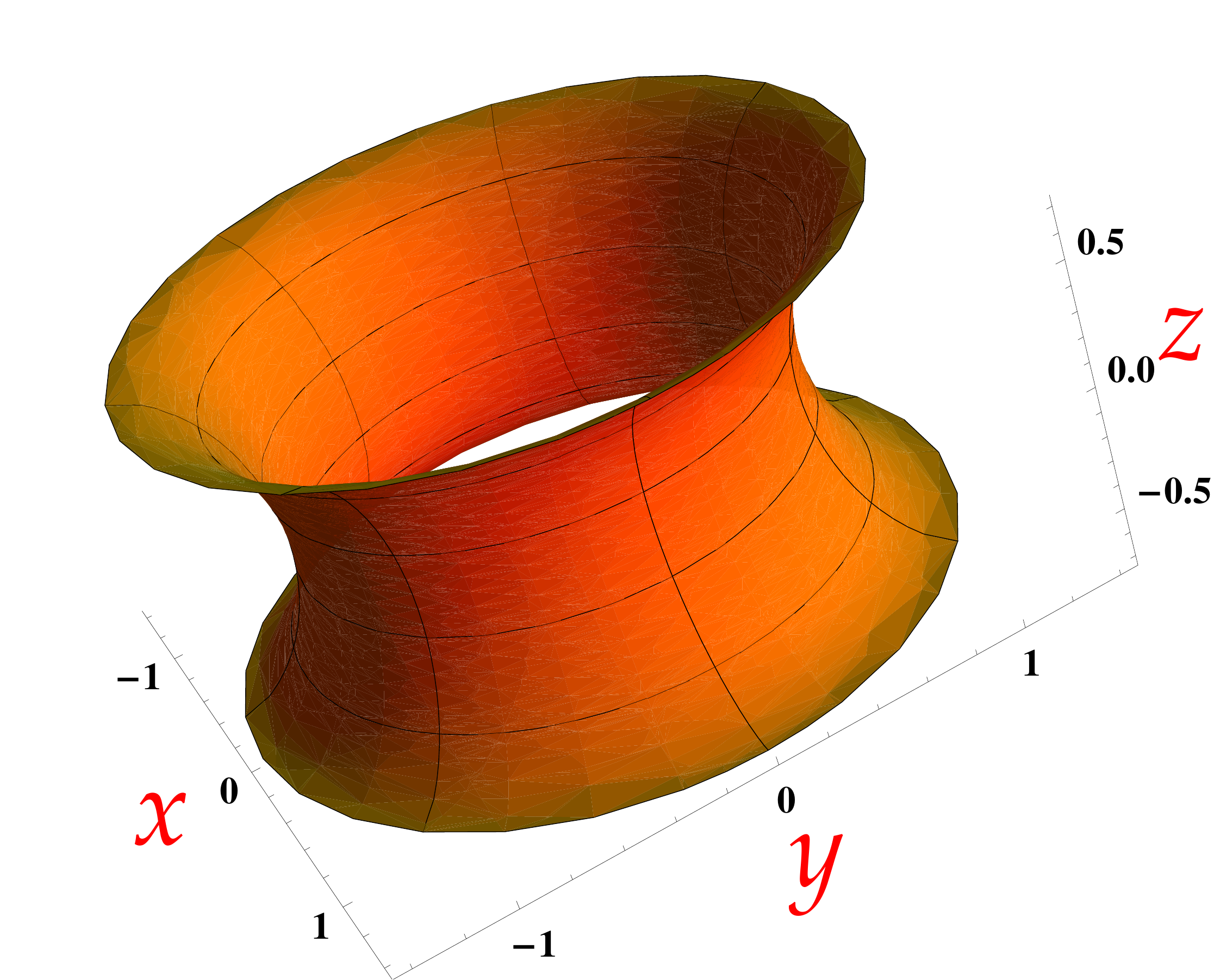}
\caption{The hyperbolic pseudosphere is, in general, a finite surface. It is identified by $R(u) = c \cosh(u/r)$, with $c$ any positive real constant, and $r = \sqrt{-{\cal K}}$. The range of $R$ is, $R \in [c, \sqrt{r^2 + c^2}]$. According to the general results, the ${\mathbf R}^3$ coordinates are \cite{eisenhart} $x(u,v) = c \cosh(u/r) \cos v$, $y(u,v) = \cosh(u/r) \sin v$, $z(u) = -i E\int [i (u/r) , - (c/r)^2]$, where the last symbol is the elliptic integral. In the plot $r=1 = c$, and $u \in [- {\rm arc}\cosh \sqrt{2}, + {\rm arc}\cosh \sqrt{2} ]$, $v\in [0, 2\pi]$ and the range of $u$ is dictated by the condition that $z\in {\mathbf R}$ in terms of the elliptic integral. The singular boundaries in this case are the two extremal circles $R = \sqrt{r^2 + c^2} = \sqrt{2}$, where $u = \pm  {\rm arc} \cosh \sqrt{2}$.} \label{hyperbolic}
\end{figure}

For the hyperbolic pseudosphere, we need to solve
\begin{equation}
d l^2 = \frac{r^2}{{\tilde y}^2}(d{\tilde x}^2+d{\tilde y}^2) \equiv du^2 + c^2 \cosh^2 \frac{u}{r} \; dv^2 \;,
\label{lobhyperbpseudosphere}
\end{equation}
that gives
\be\label{explicithyperbolic}
\tilde{x} = e^{c \; v / r} \tanh(u / r) \;, \quad \tilde{y} = e^{c \; v / r} \frac{1}{\cosh(u/r)} \;.
\ee
From here it is evident that, in the frame $q^\mu = (t,u,v)$, contrary to the Beltrami pseudosphere, we do not have a globally predictive power. The conformal factor, $r^2/ \tilde{y}^2$, at a fixed value of the meridian, $\bar{u}$,  will jump of $e^{- 4 \pi c  / r} r^2 \cosh^2 (\bar{u}/r)$ after a complete turn from 0 to $2 \pi$, see Fig.~\ref{hyperbolic}.

This does not mean that we cannot do anything with this pseudosphere. We still have three roads to follow: (a) we make {\it locally} valid predictions in the coordinates $q^\mu$; (b) we find a new coordinate system, where Weyl symmetry gives a globally well defined conformal factor, but points to a curved spacetime, rather than to a flat spacetime; (c) we find coordinates $Q^\mu$ for which we have both things at work, Weyl symmetry linking this spacetime to the flat one, \textit{and} a global predictive conformal factor. Later we shall use a mixture of the options (a) and (b), but let us illustrate the strategy (c) at work for yet another pseudosphere, the elliptic.

\begin{figure}
\centering
\includegraphics[height=.4\textheight]{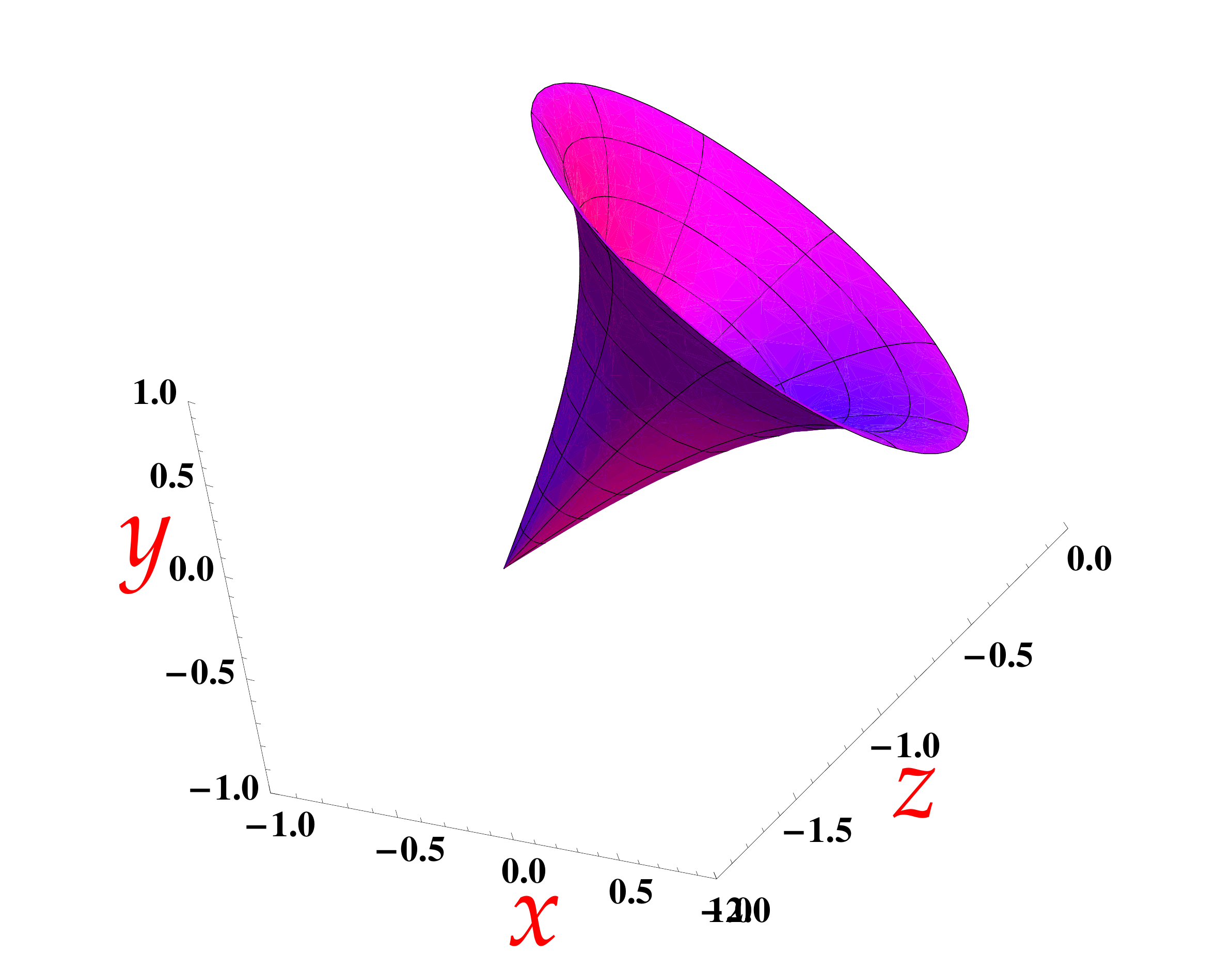}
\caption{The elliptic pseudosphere is, in general, a finite surface. It is identified by $R(u) = c \sinh(u/r)$, with $c < r$, and $r = \sqrt{-{\cal K}}$. A good parametrization of the constants, due to Ricci\cite{eisenhart}, is $c = r \sin \vartheta$ (note that $c=r$ is a degenerate case, giving a circle), hence the range of $R$ is
$[0, r \cos \vartheta]$. The ${\mathbf R}^3$ coordinates can be obtained from the general results, $x(u,v) = r \sin \vartheta \sinh(u/r) \cos v$, $y(u,v) = r \sin \vartheta \sinh(u/r) \sin v$, with $z(u)$ given in terms of elliptic integrals through $z = \pm \int \sqrt{1 - {R'}^2(u)}$.  In the plot $r=1$, $\vartheta = \pi /11$ and $u \in [0, 2]$, $v\in [0, 2\pi]$ and the range of $u$ is dictated by the condition that $z\in {\mathbf R}$. The singular boundaries in this case are:  the point $R=0$, corresponding to $u = 0$, and the maximal circle of radius $R= r \cos \vartheta$, corresponding to $u_{\rm max}$.} \label{elliptic}
\end{figure}

To solve
\begin{equation}
d l^2 = \frac{r^2}{{\tilde y}^2}(d{\tilde x}^2+d{\tilde y}^2) \equiv du^2 + c^2 \sinh^2 \frac{u}{r} \; dv^2 \;,
\label{lobellipticpseudosphere}
\end{equation}
is cumbersome, because the most natural model for this pseudosphere is the Poincar\'e disc model, rather than the upper half-plane. The results for $\tilde{x} (u,v)$ and $\tilde{y} (u,v)$ are too messy to show here, and, in any case, as for the previous pseudosphere, they also exhibit multivaluedness of the $\tilde{y}$ coordinate. On the other hand, by solving (a system related to) (\ref{changecoord1})-(\ref{changecoord6}), we find that the following coordinates
\be\label{newcoord}
T = r \, e^{t /r} \cosh\left(\frac{u}{r}\right) \;, \; X =  r \, e^{t / r} \sinh \left(\frac{u}{r}\right) \cos \left( c \, \frac{v}{r}\right) \;,
\;  Y =  r \, e^{t / r} \sinh \left(\frac{u}{r}\right) \sin \left( c \, \frac{v}{r}\right) \;,
\ee
with $0 < c < r$, give a perfectly well defined conformal factor $\phi$ for the (2+1)-dimensional metric (see last expression in (\ref{metricconf2conf3}))
\be\label{newconffact}
\phi^2 = r^2 / \left( T^2 - X^2 - Y^2 \right) = e^{- 2 t /r} \;.
\ee
One can surely imagine a physical situation where those coordinates make physical sense, but we shall not follow this road, because its experimental realization does not look easy. In fact, there is the same alternative road than for the previous pseudosphere, involving another curved spacetime and the Beltrami spacetime, that gives sensible results

\begin{figure}
\centering
\includegraphics[height=.6\textheight]{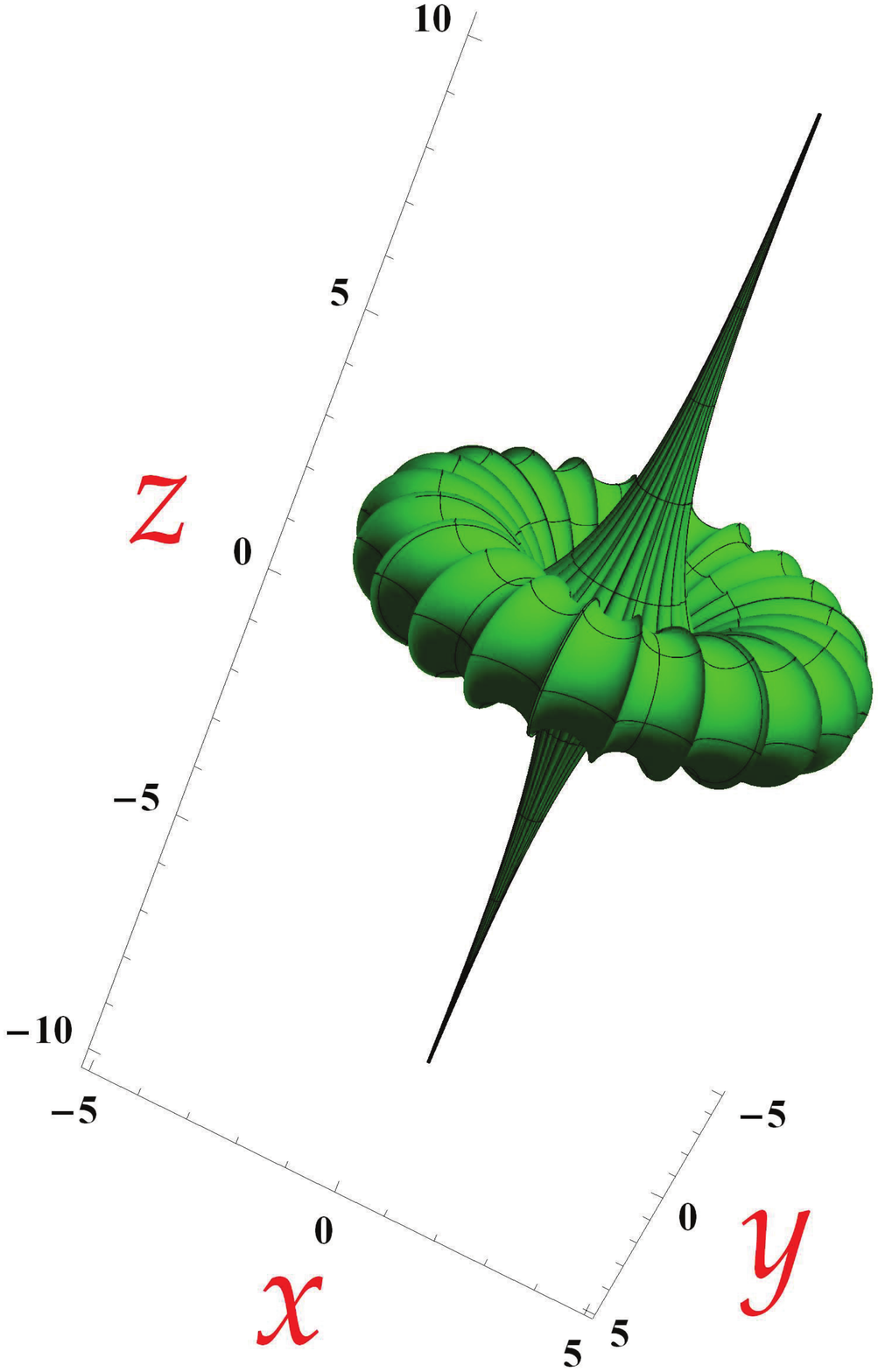}
\caption{This is a surface of constant negative Gaussian curvature, ${\cal K} = -1$. It is called ``breather''. There is a correspondence between these surfaces, and the (solitonic) solutions of the sin-Gordon equation, easy to see when the Tchebychev net parametrization of the surface is used, see, e.g., \cite{mclachlan}. Its parametric equations are $x(u,v) = [- 2 \sqrt{1-c^2} \cosh(c u) (\sqrt{1-c^2} \cos(v) \cos(\sqrt{1-c^2} v)) + \sin(v) \sin(\sqrt{1-c^2} v)]/ D$, $y(u,v) = 2 \sqrt{1-c^2} \cosh(c u) [-\sqrt{1-c^2} \sin(v) \cos(\sqrt{1-c^2} v) + \cos(v) \sin(\sqrt{1-c^2} v)]/D$, $z(u,v) = -u + 2 (1 - c^2) \cosh(c u) \sinh(c u) / D$, where $D \equiv c [(1-c^2)\cosh^2(cu) + c^2 \sin^2(\sqrt{1-c^2}v)]$, with $c\in[0,1]$. In the plot $c=0.4$, $u\in [-15,15]$, and $v\in [-37.4,37.4]$.}
\label{genericpseudosphere}
\end{figure}

To have a flavor of the other pseudospherical surfaces, infinite in number, in Fig.~\ref{genericpseudosphere} we have plotted one exotic example, the so-called ``breather''. Many more can be found in the literature, see, e.g. \cite{ovchinnikov,mclachlan}. From the parametric equations of the breather, one immediately realizes that the tasks we have accomplished in such a clean way with the Beltrami spacetime, are, in general, much more difficult. For instance, it seems a hard problem to identify the $\tilde{x} (u,v)$ and $\tilde{y} (u,v)$, and study the behavior of the conformal factor $r^2 / \tilde{y}^2$.

\section{A unifying limit. The special status of Beltrami.}

Nonetheless, all those surfaces are described by the line element (\ref{lobsurface}), hence the general result of geometry recalled earlier, comes in hand \cite{eisenhart}: {\it The line element of any surface of constant negative Gaussian curvature, not necessarily a surface of revolution, is reducible to the line element of either the Beltrami, or the hyperbolic, or else the elliptic pseudospheres}, given here by (\ref{linelsurfrev}) with (\ref{Rbeltrami}), (\ref{Rhyperbolic}), and (\ref{Relliptic}), respectively. This is achieved by having the geodesics system of the given surface coincide (though a mapping) with the geodesics system of the particular pseudosphere \cite{eisenhart}. To this well known result, that points towards three surfaces only, we want to add the following considerations that merge the three into one: the Beltrami. This shows that, to consider the Beltrami means to consider all the surfaces of the family.

\begin{figure}
\begin{tabular}{ccc}
\epsfig{file=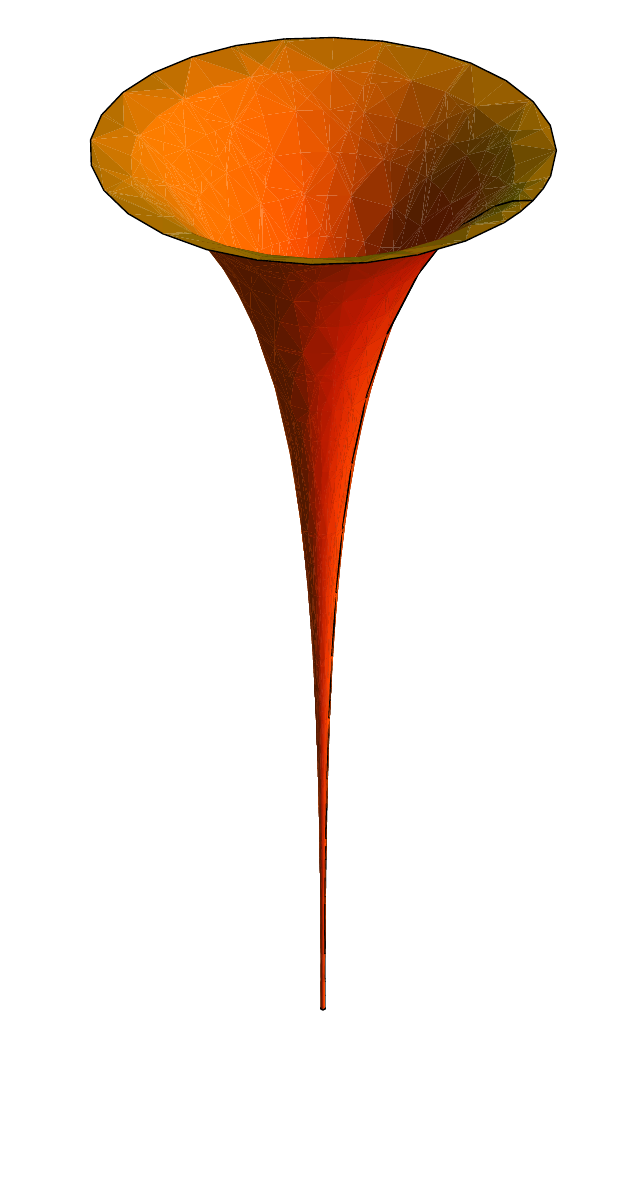, width=0.3\linewidth, clip=} &
\epsfig{file=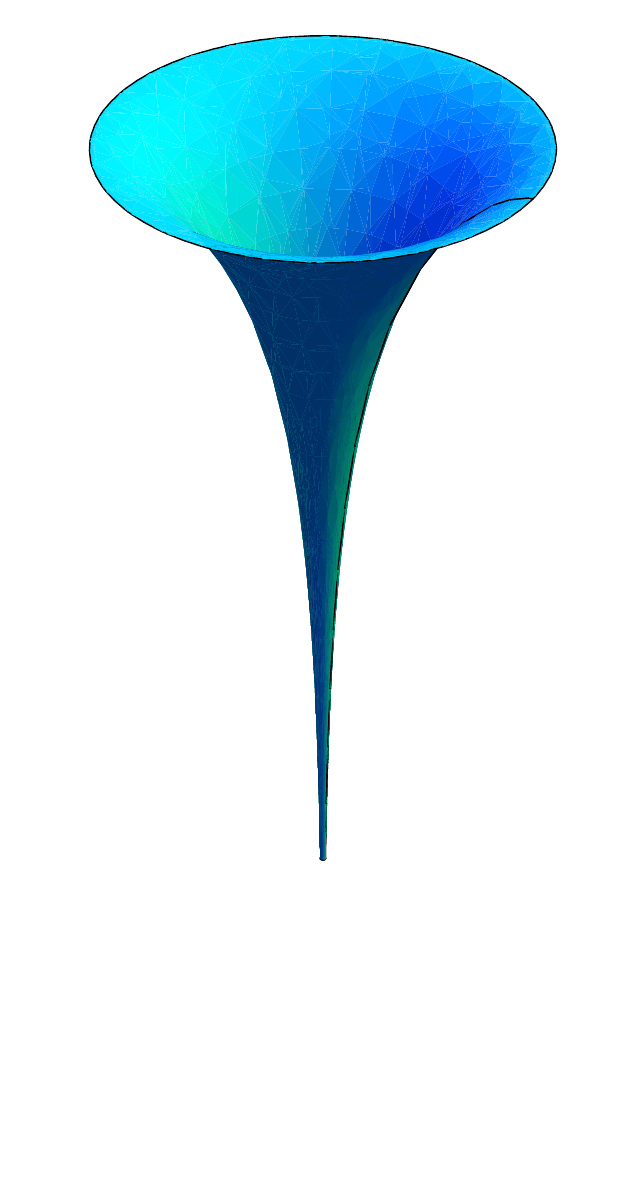, width=0.3\linewidth, clip=} &
\epsfig{file=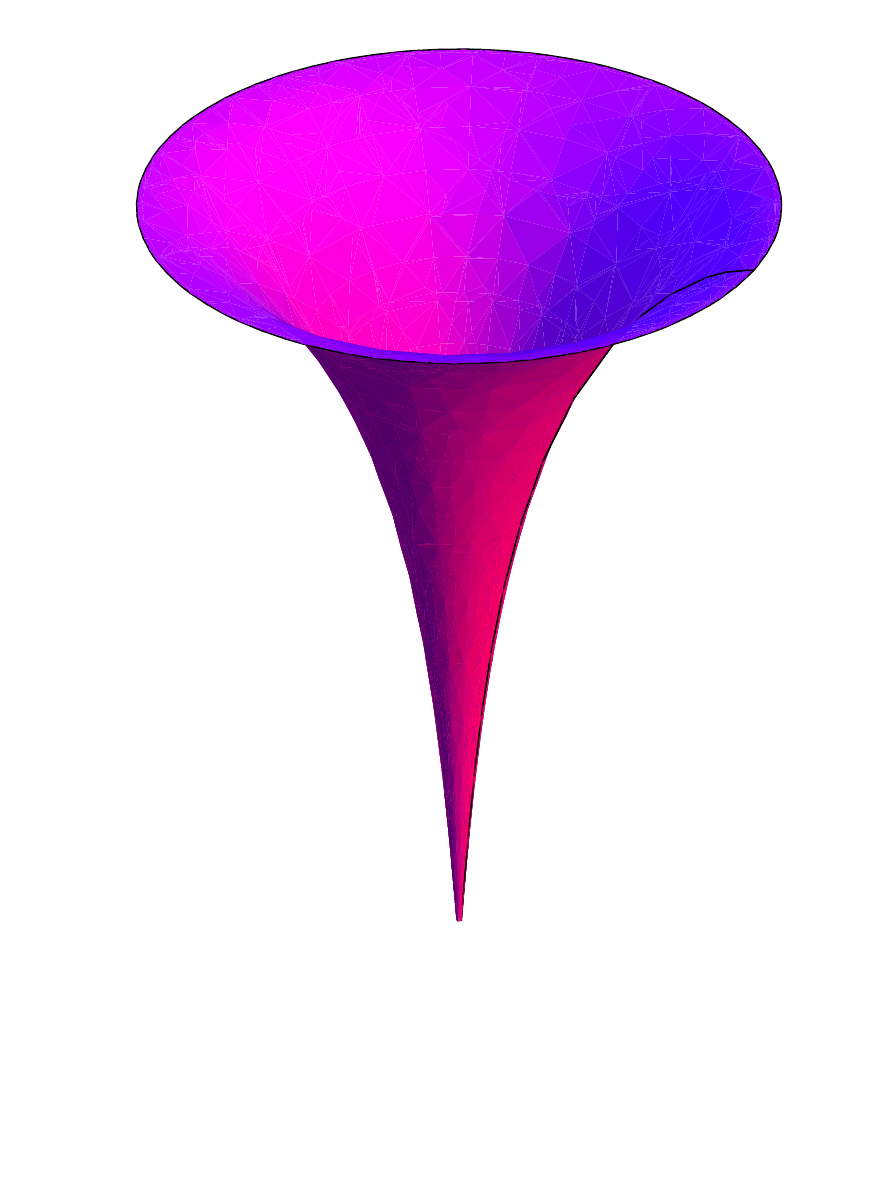, width=0.4\linewidth, clip=} \\
\end{tabular}
\caption{The three pseudospheres, for small values of $c/r$. In the middle the Beltrami. On its left, one half of the hyperbolic pseudosphere, on its right the full elliptic pseudosphere. For the first figure (half hyperbolic pseudosphere), the plot is for $r =1$ and $c=0.01$. For the third figure (elliptic pseudosphere), the plot is for $r=1$, and $\vartheta = \pi/50 \simeq 0.06 \simeq c$.}
\label{comparison}
\end{figure}

In general, the three pseudospheres (\ref{Rbeltrami})--(\ref{Relliptic}) differ importantly from each other, and the previous theorem is one example of this. Indeed, besides the differences just discussed about their natural coordinate systems in a (2+1)-dimensional spacetime, the Beltrami surface is infinite, while the other two are not. Furthermore, the Beltrami surface has one singular boundary only, at $R=r$, while the other two pseudospheres have two singular boundaries: the hyperbolic pseudosphere when $R = \sqrt{r^2 + c^2}$, the elliptic at $R=0$, and at $R = r \cos \vartheta$. Nonetheless, \textit{in the limit of very small} $c/r$, the three surfaces have very similar behavior. This can be seen by inspection of the expressions in (\ref{Rbeltrami})--(\ref{raggioell}), and is depicted in the plots of Fig.~\ref{comparison} that have to be compared with the plot in Fig.~\ref{Beltrami}. In that limit the range of $u$ is infinite for all, the range of $R$ is $[0,r]$ for all, and in the positive $u$ sector, they all approach the same form (see (\ref{Rbeltrami})--(\ref{Relliptic}))
\be
R(u) \sim c \; e^{u/r} \in [0, r] \quad {\rm when} \quad u \in [0,+ \infty] \;,
\ee
(we shall be more precise in the following Section about the actual limits of the range of $u$, that crucially depend upon the physics of the application to graphene).

With this in mind, we shall mostly focus on the Beltrami spacetime, as the results are the cleanest there, and can serve as a guide for the other (infinite number of) cases too.

\chapter{The conformal Killing horizon from three perspectives}

There are various kinds of horizon in general relativity, sometimes differing for very subtle reasons, see, e.g., \cite{wald,dinverno,booth}. We do not have yet at our disposal the gravity/geometrical theory that describes the dynamics of the elastic membrane of graphene (the effective description of the dynamics of the $\sigma$-bonds), thus we cannot embark in such subtle distinctions, as yet. One distinction we can attempt to make, though, is between the singular boundary of the surfaces of constant negative curvature that we call Hilbert horizon, (when it is possible to identify it as the end of the spacetime, like, e.g., the circle $R=r$ of the Beltrami surface), and a standard event horizon (e.g., the horizon of Rindler spacetime). These two types of horizon, for a generic surface of constant negative Gaussian curvature, are in principle different. But, below, shall show that, in the case of a Beltrami spacetime, in the physically appropriate limit of small $c/r$, the two horizons coincide within very reasonable experimental errors. Furthermore, in that limit, the same Hilbert horizon, $R=r$, is (Weyl) related to three standard event horizons: the Rindler kind, the cosmological kind, and the black-hole kind (although, in this latter case, in the limit of vanishing black-hole mass).

\section{Hilbert theorem standing on the way to the horizon}

Let us start by finding some general results, valid for the infinite number of cases of constant negative Gaussian curvature. We need to consider the line element (\ref{lobspacetime})
\begin{eqnarray}
ds^2_{\rm graphene} =  \frac{r^2}{{\tilde y}^2}\left[\frac{{\tilde y}^2}{r^2}dt^2 - d{\tilde y}^2 - d{\tilde x}^2\right] \;, \nonumber
\end{eqnarray}
and, either study separately the Rindler-like spacetime and the conformal factor, or study directly the full line element
$dt^2- (r/{\tilde y})^2 (d{\tilde x}^2 + d{\tilde y}^2)$ (the results are, of course, the same). The null geodesics\footnote{Our considerations refer to the pseudoparticles of graphene, that are massless (Dirac) excitations.} of both spacetimes are of this form
\be
\tilde{x} (t) = {\rm constant} \quad {\rm and} \quad \tilde{y} (t) = \tilde{y}_0 \; e^{\pm v_F t /r} \label{geodesics} \;,
\ee
where we have, momentarily, reintroduced $v_F$ (our ``speed of light''), and $+$ ($-$) is for the outgoing (ingoing) trajectories (for antiparticles signs swap). The actual \textit{Euclidean length} can be obtained only when the Lobachevsky coordinate $\tilde{y}$ has been expressed in terms of Euclidean measurable spatial coordinates, hence when the surface has been actually specified. On the other hand, to make general statements we look at the \textit{Lobachevsky length}. Equations (\ref{geodesics}) identify a straight line of the degenerate type in the Lobachevsky plane (see geodesics $g_1$ and $g_6$ in Fig.~\ref{circles}), and the Lobachevsky distance between two points is
\be
d(\tilde{y}(t_2),\tilde{y}(t_1)) = {\rm arccosh} \left( 1 + \frac{(\tilde{y}(t_2) - \tilde{y}(t_1))^2}{2 \tilde{y}(t_2) \tilde{y}(t_1)}\right)
= \frac{v_F}{r} |t_2 - t_1| \;,
\ee
with $\tilde{y}(t)$ in (\ref{geodesics}).

The above discussion means that the pseudoparticles, on a generic graphene negatively curved surface, see
\[
\tilde{y} = 0 \;,
\]
as an event horizon: (a) the metric elements are singular there, and (b) it can only be reached asymptotically at future null infinity (it can never be crossed). Thus, one thing we learn from the above discussion, is that, even when only spatial curvature is present, and all metric variables are time independent, an event horizon is indeed possible. The issue here is: does the curve\footnote{By ``curve'' we mean the set of points \textit{in the Euclidean coordinates} that are solution of $\tilde{y}=0$, for the given surface/spacetime. Strictly speaking, there is no such solution.} $\tilde{y}=0$ belong to the spacetime?

First of all, by the very definition of the Lobachevsky plane, strictly speaking, $\tilde{y} = 0$ is excluded from the manifold. It is the absolute\footnote{This makes clear that the choice of other models for Lobachevsky geometry would introduce no difference, in this important respect.} of the upper-half plane model, hence all considerations about having it into the spacetime have to be about limiting processes. Nonetheless, this situation is common to standard event horizons, when the coordinates are such that the inner region beyond the horizon is out of reach. A well known example is the event horizon for a spherically symmetric black hole in the Schwarzschild coordinates.

In this latter case, the coordinates can be changed, for instance to the Eddington-Finkelstein coordinates, and the singular behaviors of infinite geodesic distance, and infinite metric elements (that we just used to identify $\tilde{y} = 0$ as an event horizon) go away, changing the properties of an horizon into those of a one-way membrane (or one-brane, for this (2+1)d case): ingoing particles can cross the horizon, but outgoing cannot.

In our case there is no equivalent of the Eddington-Finkelstein coordinates. The horizon can never be crossed, no matter the coordinates. The coordinates, though, are crucial, because only when we specify the surface (i.e. when we find the Euclidean coordinates to realize a portion of the Lobachevsky plane in the  $\mathbf{R}^3$ of the laboratory) we can face the question on whether the Hilbert horizon, when it is a well defined object, is close enough to the event horizon. This is the other issue here.

In all cases, the Hilbert horizon is located where its smaller $\tilde{y}$ coordinate, say it $\tilde{y}_{H h}$, is strictly bigger than that of the event horizon: $\tilde{y}_{H h} > 0$, and, in general, we cannot say whether the Hilbert horizon is close or far from the event horizon. Indeed, this depends on the fine details of the given surface. Each of the infinite surfaces has its own structure of singularities. It might well be that there is more than one Hilbert horizon (see, e.g., the hyperbolic pseudosphere in Fig.~\ref{hyperbolic}), or it might even happen that it is not easy to identify a reasonable Hilbert horizon, i.e. a curve where the spacetime ends (see, e.g., the breather in Fig.~\ref{genericpseudosphere}). What we shall see now is that: (a) for the Beltrami spacetime, in the limit of small $c/r$, the Hilbert horizon is a clean object, and Weyl-related to a very reasonable \textit{Rindler event horizon}; (b) for the elliptic pseudosphere, in the same limit, the Hilbert horizon is a clean object, and Weyl-related to a reasonable \textit{cosmological (de Sitter) event horizon}; (c) for the hyperbolic pseudosphere, in the same limit, the Hilbert horizon is a clean object, and Weyl-related to a \textit{black hole (BTZ) event horizon, although in the limiting case of vanishing mass}. In the last two cases, as already shown, in the limit of small $c/r$, the spacetime tends to the Beltrami spacetime\footnote{There are, though, some global differences: the elliptic pseudosphere is singular also at the tip of the tail ($R=0$), while the hyperbolic pseudosphere tends to two Beltramis joined at $R=0$, see Fig.~\ref{hyperbolicwormhole}, hence evoking a wormhole-type of spacetime with two Hilbert/event horizons.}, hence the surface and the Hilbert horizon, in all cases, are the same ones.

Since all surfaces of constant negative Gaussian curvature are locally isometric to the Beltrami pseudosphere, these results also are an empirical proof that, when on one surface of the family the conditions for an horizon are reached, the results about thermal Green functions found for the Beltrami spacetime \cite{ioriolambiase,prd2014} can be used, although their validity might be confined to a small neighbor. On this latter point we shall come back later. We want to prove now the previous statements about Rindler, de Sitter, and BTZ horizons.

\section{The Beltrami spacetime: conformal to Rindler}

By using the expressions (\ref{explicitxybeltrami}) in (\ref{lobspacetime}), and by introducing the correct dimensionality for the coordinates, so that the conformal factor is dimensionless, the line element of the Beltrami spacetime is
\be\label{rightchoice}
ds_B^2 =  \frac{c^2}{r^2} \; e^{2 u/r} \left[ \frac{r^2}{c^2} \; e^{-2 u/r} \left( dt^2 - du^2 \right) - r^2 dv^2 \right]
\equiv \varphi^2 (u) \; ds^2_R \;,
\ee
with $\varphi (u) \equiv  c/ r \; e^{u/r}$, and
\be\label{firstRindlerds}
ds^2_R \equiv \frac{r^2}{c^2} \; e^{-2 u/r} \left( dt^2 - du^2 \right) - r^2 dv^2 \;,
\ee
where the subscript ``$R$'' stands for Rindler, and from now on we take
\be
c < r \;,
\ee
so that $\ln (r/c)$ is always greater than zero.

This amounts to introduce Lobachevsky coordinates of dimension of [length], $\tilde{x} \to r^2 \tilde{x}$, and $\tilde{y} \to r^2 \tilde{y}$. Having done that, though, there is a more straightforward choice than (\ref{rightchoice}) to have a dimensionless conformal factor multiplying a line element of the right dimensions of $[{\rm length}]^2$, namely $ds_B^2 = e^{2 u/r} [e^{- 2 u/r}(dt^2 - du^2) - c^2 dv^2]$. For the latter choice, the conformal factor, evaluated at the Hilbert horizon, diverges for $c \to 0$, $\varphi(u= r \ln (r/c)) = r/c \to \infty$. This is as it must be for a proper event horizon, and it can also be taken as a piece of evidence of the coincidence of Hilbert and event horizons in the limit for $c \to 0$. Nonetheless, the metric elements (see the angular part $c^2 dv^2$) make no sense in the limit $c \to 0$, no matter whether one is at the horizon or not. Thus we prefer the choice (\ref{rightchoice}), along with a redefinition of the coordinates ($t \to (r/c) \, t$, $u \to (r/c) \, u$, see later), so that the divergent behavior in the $c \to 0$ limit is passed on from the metric elements to the range of the coordinates. The choice (\ref{rightchoice}) gives a conformal factor that, at the horizon, is always finite, $ \varphi(u= r \ln (r/c)) = 1$. No matter the choice, though, the indication of how close the Hilbert horizon of the Beltrami spacetime is to the event horizon of a Rindler spacetime is always given by how small is $c/r$, as will be clear from the following.

Just like in the standard case, the line element $ds^2_R$ in (\ref{firstRindlerds}) describes both the left {\it and} the right Rindler wedges. Indeed,
\be
\eta \equiv \frac{r}{c} \; t \in [- \infty, + \infty] \; {\rm and} \; v \in [0,2 \pi]
\ee
but, for $a \equiv c / r^2 > 0$
\be\label{xirigth}
\xi \equiv - \frac{r}{c} \; u \in [- (r^2 /c) \, \ln (r/c), + \infty] \;,
\ee
while, for $a \equiv - c / r^2 < 0$
\be\label{xileft}
\xi \equiv \frac{r}{c} \; u \in [- \infty, + (r^2 /c) \, \ln (r/c)] \;,
\ee
so that
\be\label{ytwosectors}
\tilde{y} = \frac{1}{c} \; e^{- u/r} = \left\{ \begin{array}{cc}
+ \displaystyle{\frac{1}{a \, r^2}} \; e^{a \xi} \; > 0 & {\rm right} \; {\rm wedge} \;, \\
 {\rm or}  &  \\
- \displaystyle{\frac{1}{a \, r^2}} \; e^{a \xi} \; > 0 & {\rm left} \; {\rm wedge} \;.
\end{array}
\right.
\ee
Here $a$ is the value of the proper acceleration $\alpha (\xi)$ evaluated at the origin of the Rindler spatial coordinate, $\xi = 0$, and the proper acceleration is defined in accordance to the line element
\be\label{dsrindler}
ds^2_R = e^{2 a \xi} (d\eta^2 - d \xi^2) - r^2 dv^2 \;,
\ee
so that $e^{- a \xi}$ is the appropriate Tolman factor
\be
\alpha (\xi) = a \; e^{- a \xi} \;.
\ee

Now we focus on the line element (\ref{dsrindler}), knowing that this spacetime differs from a standard Rindler spacetime only with respect to the range of the $\xi$ coordinate (and for the fact that the ``speed of light'' here is $v_F$, on this see next Section). In the standard Rindler spacetime, the event horizon is identified \textit{spacewise} by
\be
\xi_{E h} = - \infty \quad {\rm right} \; {\rm wedge} \quad {\rm and} \quad
\xi_{E h} = + \infty \quad {\rm left} \; {\rm wedge} \;,
\ee
and \textit{timewise} by
\be
\eta = + \infty \;.
\ee
In \cite{ioriolambiase} the focused was on the latter. In \cite{prd2014} the focus is on the former. Let us recall here the conditions such that the event horizon is within the reach of the Beltrami spacetime, and its relationship to the Hilbert horizon, located at
\be
\xi_{H h} = - \frac{r^2}{c} \ln (r/c) \quad {\rm right} \; {\rm wedge} \quad {\rm and} \quad
\xi_{H h} = + \frac{r^2}{c} \ln (r/c) \quad {\rm left} \; {\rm wedge} \;.
\ee
Clearly, $\xi_{H h} \to \xi_{E h}$ for $c \to 0$. Let us now investigate the physics of this limit. For definitiveness, we shall consider {\it the right wedge} for the rest of this Section.

In the standard Rindler case, the inertial observer is the one for which the proper acceleration is zero, i.e. $\xi_{max} = + \infty$. Thus, the range of $\xi$ is dictated by the two conditions: (i) its minimum corresponds to the event horizon, and there the acceleration reaches its maximum; (ii) its maximum corresponds to the inertial observer ($\alpha = 0$)
\be\label{standardRW}
\xi \in [- \infty, \cdots, 0, \cdots, + \infty] \Rightarrow \alpha(\xi) \in [+ \infty,\cdots, a, \cdots, 0] \;,
\ee
where we included the middle range value, important for us, and wrote the range of $\alpha(\xi)$ so that it corresponds to the range of $\xi$.
An observer at standard Rindler space coordinate $\xi$ is constantly at a distance
\be\label{dstandrind}
d(\xi) \equiv 1/\alpha(\xi) - 1 / \alpha_{max} = 1 / \alpha(\xi) \;,
\ee
from the horizon. The inertial observer is infinitely far away, $d(\xi_{max} = + \infty) = \infty$, and $d(\xi_{Eh} \equiv \xi_{min} = -\infty) = 0$. For our spacetime, the ranges in (\ref{standardRW}), for finite $c < r$, become
\be
\xi \in [- \frac{r^2}{c} \ln (r/c),\cdots, 0, \cdots, + \infty] \Rightarrow \alpha(\xi) \in [1/r, \cdots, a = c/r^2, \cdots, 0] \;.
\ee
If we now consider the mathematical limit $c \to 0$, with $r$ finite, we see two things. First $a \to 0$, hence it is $\xi = 0$ (corresponding to $\alpha = a \to 0$) the coordinate corresponding to the inertial observer. Second, the lower bound of the range, corresponding to the Hilbert horizon, is
$ - (r^2 / c ) \ln (r / c) \to - \infty$, there $\alpha = 1/r$. So, the range of $\xi$ gets halved, and the maximal acceleration is finite and related to the curvature
\be
\xi \in [- \infty, 0] \Rightarrow \alpha(\xi) \in [1/r, 0] \quad {\rm when} \quad c \to 0 \;.
\ee
In the limit $c \to 0$, $\xi_{H h} \to \xi_{E h}$, and an observer with space coordinate $\xi$ is constantly at a distance
\be\label{dbeltrind}
d(\xi) = 1 / \alpha(\xi) - 1 / \alpha_{max} = 1 / \alpha(\xi) - r
\ee
from the horizon. Thus, the inertial observer is infinitely far away, $d(\xi_{max} = 0) = \infty$, while $d(\xi_{H h} \equiv \xi_{min} = - \infty)= 0$.

It is important to notice that, even in the limit $c \to 0$, we are not changing the location of the Hilbert horizon in terms of the Lobachevsky coordinate, as this is once an for all given by $\tilde{y}_{H h} \equiv \tilde{y} (u_{max} = r \ln(r/c)) = 1/r$. We are changing its location in terms of the coordinate $u$.

It is crucial to implement the limit $c \to 0$ \textit{physically}, i.e. to have $c$ small compared to the only physical scale we have used, that is $r$, thus, the crucial parameter is $c/r$, rather than $c$. In a moment we shall identify the physical and geometrical meaning of $c$ for the graphene membrane, and shall fix this length. Thus, we shall have that $\xi_{H h} \to - \infty = \xi_{E h}$ only approximately. Furthermore, to make $c/r$ small we have to make $r$ big.

To understand the physical and geometrical role of $c$ for the Beltrami spacetime, one recalls that $R (u) = c \; e^{u/r}$, hence we see that $c$ fixes the origin of the $u$ coordinate
\be
c = R (u = 0) \;,
\ee
and this explains, from the point of view of the geometry of the pseudosphere, why, when $c \to 0$, the range of $u$ gets halved: in that limit, the value $R = 0$ is reached already at $u=0$. On the other side of the range there is $r$, and
\be
r = R (u_{max} = r \ln(r/c)) \;.
\ee
Thus the pace at which one reaches the end of the surface, starting from the origin of the Euclidean measurable coordinate $u$, is fixed by $c$: the smaller $c$, the farther away is the end of the surface, i.e. the more $u$-steps are necessary to reach there. In the limit $c \to 0$ the number of steps is infinite. The most natural choice for $c$, then, is to link it to the natural pace of the graphene membrane, that is the lattice spacing. Thus, we choose
\be
c = \ell \;.
\ee
There is also another reason, perhaps clearer, to fix $c = \ell$, that comes from the geometry of the hyperbolic pseudosphere. There, $c = R (u = 0)$ always corresponds to the minimum value of $R$, see Fig.~\ref{hyperbolic}. Hence, one cannot think of going below $R = \ell$ for the real membrane. Since, as shown, in the limit of small $c/r$ the two surfaces (and the elliptic pseudosphere) become, in a way, the same surface, that argument can be imported here too.

This choice of $c$ fits very well our requests of small curvatures ($r > \ell$), necessary for the approach based on the action (\ref{actionAcurvedpaper}) to work.

Of course, even the choice $c = \ell$ is an idealization, and it must serve only as a guide for the real situation. In fact, our approximations on the dynamics of the conductivity electrons of graphene cannot hold down to such small radii of the pseudosphere. For instance, the distance between different sides of the membrane will become too small to ignore out-of-surface interactions, not to mention the distortions in the lattice structure to make a section of radius\cite{ioriodice12} $R = \ell$. Nonetheless, as can be easily read off from the Table~\ref{tablebeltrami}, the approximations within which the Hilbert horizon and the event horizon coincide are so good that, even a much larger $c$, would not change our conclusions. That is why we prefer to present the values for the choice $c = \ell$, that can easily be adapted to a realistic engineering of the graphene membrane, rather than present values for a choice $c = \alpha \ell$, with $\alpha$ a number that, at the present level of experimental and theoretical knowledge on the manipulation of graphene in order to induce different shapes and patterns, it is simply a number that one has to guess.

\begin{table}[h]
\caption{Quantification of how good is to approximate the Hilbert horizon of the Beltrami spacetime, $R=r$, with a Rindler event horizon. The closer $\zeta_B \equiv  - (\ell / r)^2 / \ln (r / \ell)$ is to zero, the better is the approximation. In the table we indicate three values of $r$, the corresponding values of $\zeta_B$, and we also explicitly indicate the corresponding values of $\ell / r$ (recall that $\ell \simeq 2$\AA.). This latter parameter is also a measure of how close to zero is $\tilde{y}_{H h} = 1/r$, in units of the lattice spacing $\ell$: $1/(r/\ell)$. The values are all approximate.}
\centering
\begin{tabular}{|l|c|l|}
  \hline
  $r$ & $ \zeta_B $ & $ \ell / r $ \\ \hline
  20${\AA}$ & $ - 4 \times 10^{-3}$ & $ 0.1$ \\
  1 $\mu$m & $ - 5 \times 10^{-9}$ &  $ 2 \times 10^{-4}$ \\
  1 mm & $ - 3 \times 10^{-15}$ & $ 2 \times 10^{-7}$ \\
    \hline
\end{tabular}
\label{tablebeltrami}
\end{table}

Since the event horizon of a Rindler spacetime is at $1 / \xi_{E h} = 0$, a way to quantify how good are our approximations for the Beltrami spacetime, is to see how close to zero, in units of $\ell$, is $1 / \xi_{H h}$, that is the dimensionless parameter $\zeta_B \equiv  - (\ell / r)^2 / \ln (r / \ell)$. Some values are shown in Table~\ref{tablebeltrami}.

Notice that the bigger $r$, the closer is the horizon. This could have guessed immediately from $\tilde{y}_{H h} = 1/r$. Nonetheless, we have to measure in terms of the $u$ coordinate, so, even a finite and not too small $1/r$, can still give a very good approximation. For instance, already at $r = 20 \AA$, that gives a large value 0.1 for $\ell / r$, the error in identifying $R = r$ as the Rindler event horizon is of a more reasonable four parts per one thousand. On the other hand, at a more realistic value of $r = 1 \mu$m, that error becomes a reassuring five parts per one billion. For experimental detections of Hawking phenomena associated to the existence of this horizon, we need to compromise between a large enough $r$ for a good horizon, and a small enough $r$ for a detectable Hawking temperature, ${\cal T} \sim 1/r$. As seen in \cite{ioriolambiase,prd2014}, $r$s in the range of $1 \mu$m--1mm are good for the latter purpose, and are shown in Table~\ref{tablebeltrami} to be good for the former purpose too.

\section{The elliptic pseudosphere spacetime: conformal to de Sitter}\label{deser}

In this Subsection, and in the next one, we should evoke two spacetimes, de Sitter, important for cosmology, and BTZ, that is a black-hole spacetime, respectively. Our main scope is to illustrate how the same Hilbert horizon of the Beltrami spacetime, besides being Weyl-related to the Rindler-like event horizon, it is also Weyl-related to the event horizon of those two spacetimes. Thus, although we shall establish links between important physical quantities on both sides (the de Sitter/BTZ side, and the graphene side), we shall present here only a kinematical starting point for a much deserved study that probes full power into those analogies.

De Sitter (dS) spacetime, in (2+1) dimensions, can be described by the following line element
\begin{equation}\label{dS1}
ds_{{\rm dS}_3}^2 = \left( 1 - {\cal R}^2/ r^2 \right) d t^2 - \left( 1 - {\cal R}^2/ r^2 \right)^{-1} d {\cal R}^2 - {\cal R}^2 d v^2 \;,
\end{equation}
where $t$ and $v$ are the time, and angular variables, respectively, and ${\cal R}$ is the radial coordinate. The positive quantity $r$ is related to the cosmological constant through $\Lambda = 1 /r^2 > 0$, hence, through the relation ${\rm Ricci} = 6 \Lambda$, valid in (2+1) dimensions, the Ricci scalar curvature is $+ 6 /r^2$. Clearly, this spacetime has an event horizon at ${\cal R}_{E h} = r$. After the discovery of the positive (but tiny) value of the cosmological constant, this spacetime became of great importance for nowadays cosmology. Its horizon is often referred to as ``cosmological horizon'', i.e., the horizon that limits what we can observe of the expanding universe, due to the finiteness of the speed of light (see, e.g., \cite{cosmoriz}). We shall not probe into this here.

On the other hand, Anti de Sitter (AdS) spacetime, in (2+1) dimensions, can be described by substituting $r \to i r$ in (\ref{dS1})
\begin{equation}\label{AdS1}
ds_{{\rm AdS}_3}^2 = \left( 1 + {\cal R}^2/ r^2 \right) d t^2 - \left( 1 + {\cal R}^2/ r^2 \right)^{-1} d {\cal R}^2 - {\cal R}^2 d v^2  \;,
\end{equation}
so that it has negative cosmological constant $\Lambda = - 1 /r^2 < 0$, and Ricci scalar curvature, $- 6 /r^2 < 0$. As it is evident, this spacetime does not have an intrinsic horizon. We shall now show that our spacetimes of constant \textit{negative} curvature, are related to the dS rather than the AdS
spacetime\footnote{This result only apparently seems to contradict the discussion in \cite{deser}, in relation to the possibility to have a Hawking phenomenon through an embedding procedure into flat higher dimensional spacetimes. There it is shown that the spacetimes of constant negative curvature, AdS, cannot have an intrinsic Hawking phenomenon. It is necessary to include an acceleration, $a > 1/r$, in the higher dimensional Rindler spacetime. Here, instead, the spacetimes of negative curvature are related to dS.}.

A standard way to introduce both spacetimes is through the embedding into higher ((3+1) in this case) dimensional flat spacetimes. The two spacetimes are the solutions to these equations
\be
\eta_{A B} x^A x^B = + r^2 \quad  {\rm and} \quad \tilde{\eta}_{A B} x^A x^B = - r^2
\ee
the first for dS, the second for AdS. Here $A, B = 0, 1, 2, 3$, $\eta_{A B} = {\rm diag} (+1, -1, -1, -1)$, and  $\tilde{\eta}_{A B} =
{\rm diag} (+1, -1, -1, +1)$, so that ${\rm dS} \leftrightarrow {\rm AdS}$ when $r \leftrightarrow i r$ and $z \leftrightarrow i z$, where $x^3 \equiv z$. Usually, no physical meaning is ascribed to the higher dimensional embedding manifold, but only to the resultant spacetime, see, e.g., \cite{deser}. Thus, a signature like that of $\tilde{\eta}_{A B}$ is not a problem. For us this cannot be the case, as we do give physical meaning to the embedding spacetime, hence we cannot have the former signature, but only the one of $\eta_{A B}$. With this in mind, what we shall now do is to consider the well-known Weyl-equivalence of an AdS spacetime to an Einstein Static Universe (ESU) spacetime.

By defining
\be
\frac{1}{{\cal R}^2} \equiv \frac{1}{R^2} - \frac{1}{r^2} = \frac{1}{r^2 \cos^2(u/r)} - \frac{1}{r^2} \;,
\ee
and shifting the $u$ variable, $u \to u + r \pi/2$, the line element in (\ref{AdS1}) can be written as
\begin{equation}\label{AdS2}
ds_{{\rm AdS}_3}^2 = \frac{1}{\cos^2 (u/r)} \left[ d t^2 - d u^2 - r^2 \sin^2 (u/r) \; d v^2 \right] \;,
\end{equation}
where the line element in square brackets is what we have found for the spherically shaped graphene membrane, with $R (u) = r \sin (u/r)$. The first consideration here is that, as announced earlier, when graphene is shaped in a spherical fashion, since its line element is related to the AdS spacetime, we do not expect any horizon. The second consideration is more important, and what we are looking for here.

\begin{table}[h]
\caption{Quantification of how good is to approximate the Hilbert horizon of the elliptic pseudosphere spacetime with a cosmological event horizon. The closer $\zeta_{Ell} \equiv ( {\cal R}_{E h} - {\cal R}_{H h}) / r$ is to zero, the better is the approximation. In the table we indicate three values of $\ell / r$ comparable to those used in Table~\ref{tablebeltrami}, the corresponding values of $\zeta_{Ell}$, and of how close the Hilbert horizon of this spacetime ($R = r \cos \vartheta$) is to the Hilbert horizon of the Beltrami spacetime ($R = r$). The latter column, then, is also a measure of how well the elliptic pseudosphere spacetime can be identified with the Beltrami spacetime. The values are all approximate.}
\centering
\begin{tabular}{|l|c|l|}
  \hline
  $\vartheta \sim \ell/r$ & $ \zeta_{Ell} $ & $ (R_{H h} - r)/r$ \\ \hline
  0.1 & $5 \times 10^{-4}$ & $5 \times 10^{-3}  $ \\
  $10^{-4}$ & $5 \times 10^{-13}$ & $5 \times 10^{-9} $ \\
  $10^{-7}$ & $5 \times 10^{-22}$ & $5 \times 10^{-15} $ \\
    \hline
\end{tabular}
\label{tableelliptic}
\end{table}

Consider the line element obtained by substituting $r \to i r$ in (\ref{AdS2}), and including the factor $\sin \vartheta$ to take into account the geometry of the pseudosphere
\be\label{dSellip}
ds^2 = \frac{1}{\cosh^2 (u/r)} \left[ d t^2 - d u^2 - (r^2 \sin^2 \vartheta) \sinh^2 (u/r) \; d v^2 \right] \;.
\ee
This is Weyl related, through the time-independent conformal factor $1/\cosh^{2} (u/r)$, to the graphene spacetime for the elliptic pseudosphere, in square brackets, that we have already encountered. The radius is $R (u) = c \sinh (u/r)$, with the parametrization $c = r \sin \vartheta \leq r$. Substituting $R(u)$ for $u$ in (\ref{dSellip}), the line element becomes
\begin{eqnarray}
ds^2 = \left( 1 + \frac{R^2}{r^2 \sin^2 \vartheta} \right)^{-1} \left[ dt^2 - \frac{1}{\sin^2 \vartheta}
\left( 1 + \frac{R^2}{r^2 \sin^2 \vartheta} \right)^{-1} dR^2 - R^2 d^2 v \right] &  & \\
\equiv \left( 1 - \frac{{\cal R}^2}{r^2 \sin^2 \vartheta} \right) d t^2
- \frac{1}{\sin^2 \vartheta} \left( 1 - \frac{{\cal R}^2}{r^2 \sin^2 \vartheta} \right)^{-1} d {\cal R}^2
- {\cal R}^2 dv^2 &= ds_{{\rm dS}_3}^2& \;,
\end{eqnarray}
where
\be\label{defdsradial}
\frac{1}{{\cal R}^2} \equiv \frac{1}{R^2} + \frac{1}{r^2} = \frac{1}{(r \sin \vartheta)^2 \sinh^2(u/r)} + \frac{1}{(r \sin \vartheta)^2} \;.
\ee
That means that the graphene spacetime for an elliptic pseudosphere is Weyl related to a ${\rm dS}_3$ spacetime
\be\label{ellds}
ds^2_{\rm Ell} = \left( 1 + \frac{R^2}{r^2 \sin^2 \vartheta} \right) ds_{{\rm dS}_3}^2
= \left( 1 - \frac{{\cal R}^2}{r^2 \sin^2 \vartheta} \right)^{-1} ds_{{\rm dS}_3}^2 \;,
\ee
and an event horizon appears at
\be
{\cal R}_{E h} = r \sin \vartheta \;.
\ee
Let us clarify here that this dS spacetime is not the one produced by shaping a graphene membrane as an elliptic pseudosphere. It is only Weyl-related to it, through (\ref{ellds}). Hence, the fact that, e.g., the Ricci scalar curvature of this dS spacetime is $+ 6 / r^2$, should not create confusion. The latter is the Ricci curvature of the Weyl-related dS spacetime, while $- 2 / r^2$ is the Ricci curvature of the elliptic pseudosphere spacetime, as it must be.

What we need to do is to compare this horizon with the Hilbert horizon, and to use our \textit{Ansatz} $c = \ell$, that gives
\be\label{thetalr}
\vartheta = \arcsin (\ell / r)
\ee
The Hilbert horizon, in terms of the measurable radial variable $R$ is at $R_{H h} = r \cos \vartheta$, see previous discussions and Fig.~\ref{elliptic}. So that, from (\ref{defdsradial}), we see that in terms of the radial ${\rm dS}_3$ coordinate, is
\be
{\cal R}_{H h} = \frac{1}{2} \; r \sin (2 \vartheta) \;.
\ee
Clearly, for small $\vartheta$, that means small $\ell /r$, (see (\ref{thetalr})), the two horizons coincide, and have the value
\be
{\cal R}_{E h} \simeq {\cal R}_{H h} \simeq r \; \frac{\ell}{r} = \ell \;.
\ee
This holds for any value of $r$, provided this is big enough $r > \ell$. In the same limit, the elliptic spacetime tends to the Beltrami spacetime, and, in terms of the measurable radial coordinate $R$, the Hilbert horizon of the former, tends to the Hilbert horizon of the latter
\be
R^{Elliptic}_{H h} = r \cos \vartheta \to r = R^{Beltrami}_{H h}\;.
\ee
In other words, the very same Hilbert horizon we have seen earlier to be related to the Rindler event horizon, it is also related to a cosmological dS event horizon. In Table~\ref{tableelliptic} we show how good are these approximations for a graphene membrane.

\section{The hyperbolic pseudosphere spacetime: conformal to BTZ}

\begin{figure}[tbp]
\begin{center}
\includegraphics[width=0.8\textwidth]{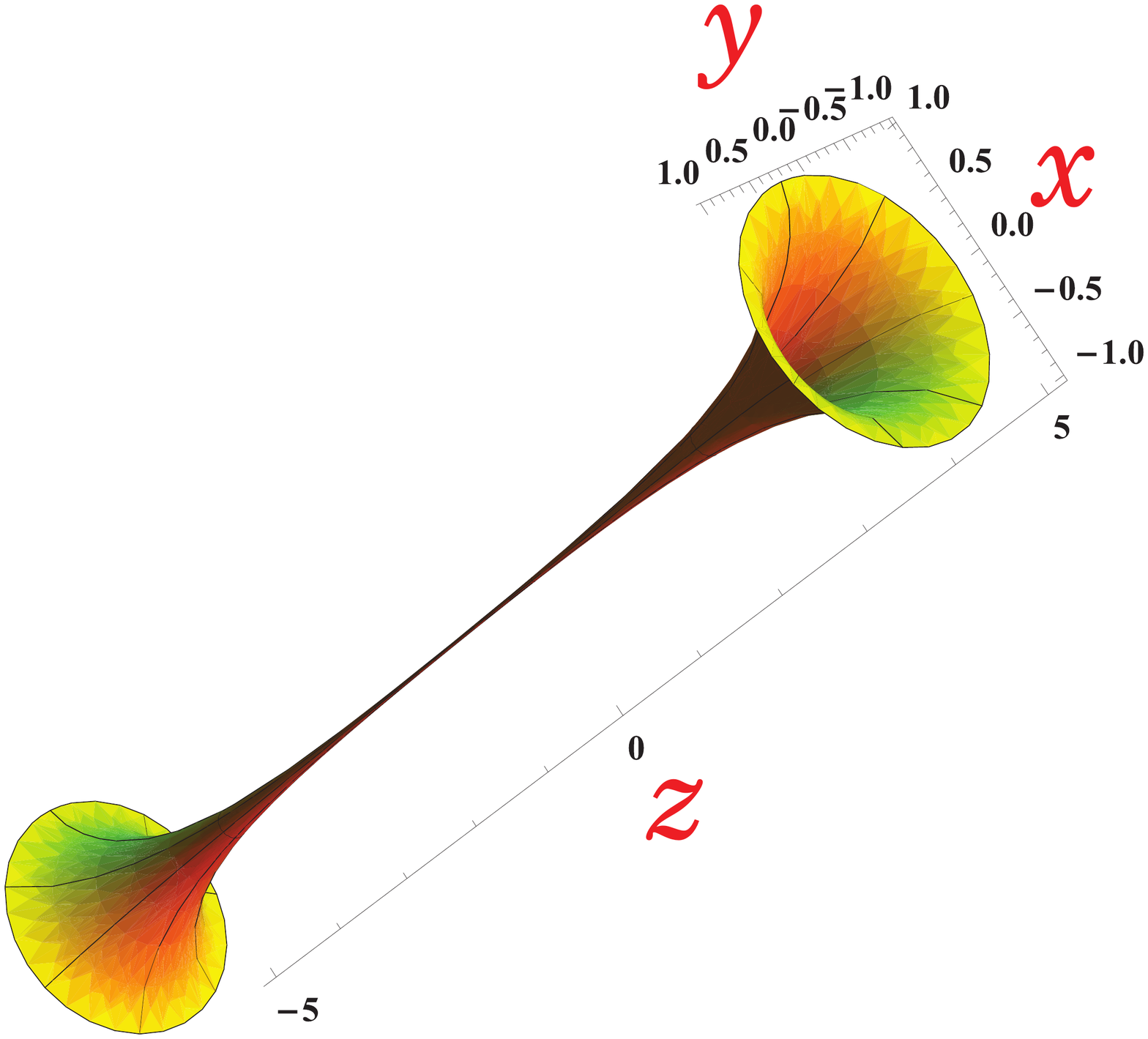}
\end{center}
\caption{\label{fig2} The hyperbolic pseudosphere for $c/r = 1/100$. The plot is for $u \in [- {\rm arccosh}(\sqrt{10001}), + {\rm arccosh}(\sqrt{10001})]$, and $v\in [0, 2\pi]$. The Hilbert horizons are located at the two maximal circles: $R_{max} \simeq 1.00005$. Figure taken from \cite{prd2014}.} \label{hyperbolicwormhole}
\end{figure}

It was shown in \cite{zermelo} that the line element of the non-rotating BTZ black hole is Weyl-related to the line element of the hyperbolic pseudosphere spacetime. There it was concluded that the Hilbert horizon and the event horizon could not match. For a non-extremal hyperbolic pseudosphere, strictly speaking, this is true. Nonetheless, when the geometrical/phyiscal role of the $c$ parameter is duly taken into account (the {\it Ansatz} $c = \ell$), the two horizons coincide, in the $\ell / r \to 0$ limit, although the mass of the hole goes to zero even faster. In that limit the hyperbolic pseudosphere tends to two Beltrami pseudospheres ``glued'' at the tails (see previous discussion and Fig.~\ref{hyperbolicwormhole}). Thus the correct statement here is that, the Hilbert horizon of the Beltrami spacetime (that, in terms of the measurable radial coordinate, is always given by $R = r$) is also a limiting case of zero mass of a BTZ event horizon (i.e., ${\cal R} \sim 0$, in terms of the BTZ radial coordinate). Let us show this here.

\begin{table}[h]
\caption{Quantification of how good is to approximate the Hilbert horizon of the hyperbolic pseudosphere spacetime with a BTZ black hole event horizon. The closer $\zeta_{Hyp} \equiv ( {\cal R}_{H h} - {\cal R}_{E h}) / {\cal R}_{E h}$ is to zero, the better is the approximation. In the table we indicate three values of $\ell / r$ comparable to those used in Tables~\ref{tablebeltrami} and \ref{tableelliptic}, the corresponding values of $\zeta_{Hyp}$, then how close the Hilbert horizon of this spacetime ($R = r \sqrt{1 + (\ell^2 / r^2)}$) is to the Hilbert horizon of the Beltrami spacetime ($R = r$) (that is also a measure of how well the hyperbolic pseudosphere spacetime can be identified with the Beltrami spacetime). In the last column are the values of the BTZ mass $M$ in terms of graphene parameters. All the values are approximate.}
\centering
\begin{tabular}{|l|c|l|l|}
  \hline
  $ \ell/r$ & $ \zeta_{Hyp} $ & $ (R_{H h} - r)/r$ & $M$ \\ \hline
  $0.1$ & $5 \times 10^{-3}$ & $5 \times 10^{-2} $ & $10^{-2}$\\
  $10^{-4}$ & $5 \times 10^{-9}$ & $5 \times 10^{-5}  $ & $10^{-8}$\\
  $10^{-7}$ & $5 \times 10^{-15}$ & $5 \times 10^{-8} $ & $10^{-14}$\\

    \hline
\end{tabular}
\label{tablehyperbolic}
\end{table}

The line element of the BTZ black hole, with zero angular momentum is \cite{btz}
\begin{eqnarray}
    ds^2_{BTZ} &= & \left(\frac{{\cal R}^2}{c^2} - M\right) dt^2
    - \frac{d{\cal R}^2}{\displaystyle{\frac{{\cal R}^2}{c^2} - M}}-{\cal R}^2 d\chi^2 \nonumber \\
    &\equiv &\left(\frac{{\cal R}^2}{c^2} - M \right) ds^2 \;, \label{eq1}
\end{eqnarray}
where $c$ and $M$ are two non negative real constants, the cosmological constant is negative, $\Lambda = - 1/c^2 < 0$, and
\begin{equation}\label{eq4}
    ds^2\equiv  dt^2-c^4\frac{d{\cal R}^2}{\displaystyle{\left({\cal R}^2-{\cal R}_{E h}^2\right)^2}} - c^2\,
    \frac{{\cal R}^2}{\displaystyle{\left({\cal R}^2-{\cal R}_{E h}^2\right)}} d\chi^2\,.
\end{equation}
Here
\be\label{hypevent}
{\cal R}_{E h} \equiv c \sqrt{M} \;,
\ee
is the event horizon of the black hole.

Let us define, $\chi \equiv v$ as the angular variable\footnote{This identification is particularly important to turn a standard $AdS_3$ spacetime into the BTZ black hole, see \cite{btz}, \cite{BHTZ}, \cite{carlipreview}. Here we do not touch upon this and other important issues.}, and
\begin{equation}\label{eq7}
    du \equiv - \frac{c^2}{{\cal R}^2-{\cal R}_{E h}^2}\, d{\cal R}\,, \qquad R({\cal R}) \equiv \frac{c {\cal R}}{{\cal R}^2-{\cal R}_{E h}^2}\,,
 \end{equation}
from which one easily obtains
\be\label{radiusBTZ}
{\cal R} (u ) = {\cal R}_{E h} \coth({\cal R}_{E h} u / c^2) \;,
\ee
that gives
\be
R({\cal R} (u)) \equiv R(u) = c \cosh({\cal R}_{E h} u / c^2)
\ee
i.e., the line element in (\ref{eq4}) is that of the hyperbolic pseudosphere spacetime
\be\label{btzhyp}
ds^2_{BTZ} = \left(\frac{{\cal R}^2}{c^2} - M \right) ds^2_{Hyp} \;,
\ee
with $r \equiv c^2 / {\cal R}_{E h} = c / \sqrt{M}$ (see (\ref{hypevent})), or $M = c^2 / r^2$. We now use our {\it Ansatz} for graphene, $c=\ell$, and write the relevant BTZ quantities after this identification
\be\label{identifBTZ}
\Lambda \equiv -\frac{1}{\ell^2} \quad , \quad M \equiv \frac{\ell^2}{r^2} \quad , \quad {\cal R}_{E h } \equiv \frac{\ell^2}{r} \;.
\ee

We need now to compare this event horizon to the Hilbert horizon of the hyperbolic pseudosphere spacetime, that, in terms of the radial coordinate of the pseudosphere, is at
\be\label{realBTZhorizon}
R_{H h} = \sqrt{r^2 + \ell^2} = r \, \sqrt{1 + \ell^2 / r^2} \;,
\ee
or, in terms of the meridian coordinate, $u_{H h} = r {\rm arccosh} \left( \sqrt{1 + r^2/\ell^2} \right)$. Substituting this value into (\ref{radiusBTZ}), and using (\ref{identifBTZ})
\be
{\cal R}_{H h} \equiv {\cal R} (u_{H h}) = {\cal R}_{E h} \coth\left( {\rm arccosh} \left( \sqrt{1 + r^2/\ell^2} \right) \right) \;.
\ee
For $r = 10^n \ell$ this formula approximates to
\begin{equation}\label{eqrn}
    {\cal R}_{H h} = {\cal R}_{E h} \times \frac{10^n}{(10^{2n}-1)^{1/2}}\simeq {\cal R}_{E h} \times \left( 1 + 5\times 10^{-(2n+1)} \right) \,.
\end{equation}
From the Table~\ref{tablehyperbolic}, it is clear that, again, in the small $\ell/r$ limit these two horizons coincide, but that is also the limit where $M \to 0$, and, accordingly ${\cal R}_{E h} \to 0$, i.e. the black-hole has disappeared, and we are left with what in \cite{btz} is called ``the vacuum state''. This means that, in order to have a proper BTZ black-hole something different needs to be done, but we shall not probe into that here, as our scope is to illustrate, yet from another perspective, that the Beltrami Hilbert horizon, $R=r$, is an event horizon, although, in this case, of a very limited nature. Indeed this happens. First, the spacetime here, in the limit, becomes two copies of the Beltrami spacetimes (see previous discussion, and Fig.~\ref{hyperbolicwormhole}). Second, although ${\cal R}_{E h} \to {\cal R}_{H h} \to 0$, this corresponds to a nonzero Hilbert horizon for the hyperbolic pseudosphere spacetime, $R_{H h} \to r$, that in turn coincides with the Hilbert horizon of the Beltrami spacetime. Here we have two such horizons (see Fig.~\ref{hyperbolicwormhole}), a situation that evokes a wormhole.

Some last comments are in order. The definition (\ref{identifBTZ}) of the cosmological constant gives a very large negative value, $\Lambda = - 1 / \ell^2 \simeq - 2.5 \times 10^{19} {\rm m}^{-2}$. This makes it less appealing for cosmological considerations than the definition $\Lambda = + 1/r^2$, used in the de Sitter/elliptic pseudosphere case. On the other hand, that definition makes justice of our {\it Ansatz} that relates $c$ to what sets the length scale of the given spacetime, especially in this case where the mass is a dimensionless parameter. It must be clear, though, that the BTZ spacetime we have briefly evoked here, is not what we have when we shape graphene as an hyperbolic pseudosphere, but it is only related to it through (\ref{btzhyp}). Hence, the identification $\Lambda = - 1/ \ell^2$, that points to a Ricci scalar curvature of $- 6 / \ell^2$, should not create confusion. The latter is the Ricci curvature of the Weyl-related BTZ spacetime, while the Ricci curvature of the hyperbolic pseudosphere spacetime is $- 2 / r^2$.

\chapter{Choice of the quantum vacuum}

Let us now come to the key issue of which quantum vacuum we need to refer to when computing our Green functions. For this Section only, with $c$ we indicate the speed of light in vacuum.

The first thing to consider is that this is a very peculiar situation, with two spacetimes of different nature that come into contact. On the one hand, we have the spacetime of the laboratory, that is (3+1)-dimensional, and \textit{non-relativistic in the sense of $c$ as limiting speed}
\be\label{dslabc}
ds^2_{lab} = c^2 dt^2 - dx^2 - dy^2 - dz^2 \;,
\ee
so, here the $0^{\rm th}$ component of the position vector is $X^0 \equiv c t$. Non-relativistic means that the transformations associated to this line element are ``small'' SO(3,1) transformations\footnote{We do not consider here the translations, an issue that for graphene deserves further study. About the Lorentz group in graphene, instead, we have elaborated earlier here.}, i.e., for instance, for a boost along the first axis
\be
\Lambda_{full} = \left(
                    \begin{array}{cccc}
                      \gamma & -\beta \gamma & 0 & 0 \\
                      - \beta \gamma & \gamma & 0 & 0 \\
                      0 & 0 & 1 & 0 \\
                      0 & 0 & 0 & 1 \\
                    \end{array}
                  \right) \in {\rm SO(3,1)}_c
                  \quad \Rightarrow \quad
\Lambda_{small} \simeq \left(
                    \begin{array}{cccc}
                      1 & -\beta & 0 & 0 \\
                      - \beta & 1 & 0 & 0 \\
                      0 & 0 & 1 & 0 \\
                      0 & 0 & 0 & 1 \\
                    \end{array}
                  \right) \in {\rm SO(3,1)}_{c}^{small}
\ee
where $\beta \equiv v / c$, and $\gamma \equiv (1 - \beta^2)^{-1/2}$, so that, at the $O(\beta^2)$ approximation, the line element (\ref{dslabc}) is invariant under $\Lambda_{small}$, and one sees that
\begin{eqnarray}
  c t' & = & ct - \beta x \Leftrightarrow t' = t + O(\beta^2) \simeq t \\
  x' & = & x - \beta ct = x - v t \;,
\end{eqnarray}
the transformations reduce to Galilei's (far right side), hence time is untouched. The notation ${\rm SO(3,1)}_c$ is just to remind that the elements of the group have $c$ , but group-wise the object is the standard SO(3,1). Similar considerations hold for ${\rm SO(3,1)}_{c}^{small}$.

We call this spacetime $\mathbf{R}^{(3,1)}_{c \; \; small}$, where ``small'' refers to the associated non-relativistic transformations. This is an abuse of notation, as the spacetime is, once and for all, $\mathbf{R}^{(3,1)}_c$, but this way we emphasize the fact that, at small velocities compared to $c$, time decouples entirely from space ($t' = t$), and the very same concept of spacetime has no meaning. Thus a non-relativistic spacetime is not a Euclidean spacetime (that would amount to have SO(4) as invariance group, hence a like-signs signature, e.g., $(+,+,+,+)$), but a spacetime for which the light-cone is so far away from the worldlines, that the effects of linking together space and time are negligible, and they are effectively separated entities. The $\psi$-electrons of graphene, that move at the Fermi speed $v_F \sim c/300$, {\it when considered from the laboratory frame}, fit this non-relativistic scenario very well, since for them $O(\beta^2) \sim 10^{-8}$.

On the other hand, we have the effective spacetime of planar graphene, that is (2+1)-dimensional, and \textit{relativistic in the sense of $v_F$, Fermi velocity, as limiting speed}
\be\label{dsvacuumsect}
ds^2_{graphene} = v_F^2 dt^2 - dx^2 - dy^2 \;,
\ee
hence, here the $0^{\rm th}$ component of the position vector is $x^0 \equiv v_F t$. We choose the planar graphene case, as that is the important case for our considerations. This line element is invariant under ${\rm SO(2,1)}_{v_F}$, with the same notation as before, hence, for the boost along $x$, the matrix is
\be
\Lambda_{full} = \left(
                    \begin{array}{ccc}
                      \gamma & -\beta \gamma & 0 \\
                      - \beta \gamma & \gamma & 0 \\
                      0 & 0 & 1 \\
                    \end{array}
                  \right) \in {\rm SO(2,1)}_{v_F} \;,
\ee
but, now $\beta \equiv v / v_F$. We call this spacetime $\mathbf{R}^{(2,1)}_{v_F}$, with the notation explained earlier.

Thus, the same time label $t$, enters two dramatically different spacetimes
\be
X^0 \equiv c t \in \mathbf{R}^{(3,1)}_{c \; \; small} \quad {\rm and} \quad  x^0 \equiv v_F t \in \mathbf{R}^{(2,1)}_{v_F} \;.
\ee
From the point of view of the $\psi$-electrons of graphene, $t$ enters the variable $x^0$ which is the time part of a proper spacetime distance. While, the same variable $t$, for the laboratory observer, enters a different variable $X^0$ and, being part of a line element that transforms under Galilei transformations, does not contribute to a \textit{spacetime}, but only to a time distance, as there space and time are decoupled.

This shows that the inner time variable $x^0$, although numerically given by $v_F$ times the same parameter of the outside clocks, it is intrinsically different from the external time variable $X^0$, from a relativistic point of view. Nonetheless, we have to account for the nowdays numerous experimental observations of the $v_F$-relativistic effects of the $\psi$-electrons of graphene. Within the picture illustrated above, the simplest way is to \textit{describe the external observer/lab spacetime as $\mathbf{R}^{(2,1)}_{v_F}$, and this must hold for no matter which spacetime is effectively reached by the $\psi$-electrons of graphene, including curved spacetimes.} Notice that, when the spacetime curvature effects occur on graphene (even only of spatial kind), it makes no sense to insist in identifying the two spacetimes, the inner and the outer. This would imply a curved laboratory spacetime.
Evidently, the only issue is with the time variable, as one can easily envisage an observer constrained to be on a two-dimensional spatial slice. The previous discussion shows that indeed the $0^{\rm th}$ components on the two sides (graphene and lab) have a different interpretation, hence, when the $0^{\rm th}$ component on the graphene side can be seen, e.g., as related to a Rindler time (see previous Section and \cite{ioriolambiase,prd2014}), the $0^{\rm th}$ component on the laboratory is, once and for all, a (2+1)-dimensional $v_F$-Minkowski time variable.

The role of the third spatial dimension is not completely gone in this picture, so we do reproduce some physics of the extra dimension. Indeed, in all the previous discussions about the embedding, we have considered $\mathbf{R}^{(3,1)}_{v_F}$. In fact, the role of this larger space is only seen in the effects of embedding the surfaces in spatial $\mathbf{R}^3$, (Hilbert horizons, dS vs AdS, etc). Once the surface is obtained, and the peculiarities of the embedding are taken into account in the resultant (2+1)-dimensional curved spacetime, the external observer/lab spacetime is modeled as $\mathbf{R}^{(2,1)}_{v_F}$.

We could use $\mathbf{R}^{(3,1)}_{c}$, although this choice is less natural for non-relativistic (in the sense of $c$) electrons. This choice would evoke quantum gravity scenarios, where indeed universes with different constants of nature are contemplated. Here we would have two such universes, with different ``speeds of light'', that get in contact\footnote{Quantum gravity scenarios also might enter due to the nature of these Dirac fields here, generated by a more elementary (and discrete) structure of the spacetime itself}. We shall not proceed this way here, and shall use instead the operationally-valid {\it Ansatz} illustrated above.

It is perhaps worth clarifying that this procedure should in no way be taken seriously from a general point of view. For a generic phenomenon, e.g., the dynamics of a classic non-relativistic marble rolling on the graphene sheet, the graphene surface is just a surface in a non-relativistic spacetime. What matters to us is that, the procedure works well for the case in point of the $\psi$-electrons of our concern. It is only for them that the spacetime of graphene is $v_F$-relativistic, and not for the classic marble. Furthermore, the $\psi$-electrons here, within the limits of the model illustrated earlier, are the quanta of a quantum field. The procedure then is satisfactory implemented when we prescribe that the structure of the n-point functions is always of the following kind
\be\label{generagreen}
S^{\rm any}(q_1, ..., q_n) \equiv \langle 0_M|\psi (q_1)... {\bar \psi}(q_n)|0_M\rangle \;,
\ee
where $\psi(q)$, with $q^\mu = (v_F t, u, v)$, is the Dirac quantum field associated to {\it any} graphene surface, while $|0_M\rangle$ is demanded to be always the quantum vacuum for the Dirac field of the flat spacetime, $\mathbf{R}^{(2,1)}_{v_F}$, that mimics the laboratory frame.

This choice for the quantum vacuum of reference in the computations of the Green functions also takes into account other issues arising in the process of measure. In the standard relativistic scenarios, say that of the Rindler spacetime, we have to measure on the Minkowski vacuum to see the Unruh effect. Indeed, say that we are traveling in a space-rocket at constant acceleration. If we bring on board a ``counter'' (thermometer), whose zero has been set on the ground (at zero acceleration, i.e. in the inertial frame), it is this device that, on the accelerated rocket, will count particles (and see a nonzero temperature). If we, on the contrary, had build the ``counter'' {\it on the accelerated rocket}, the value zero (of temperature, of particle number, etc.), would be associated to the given acceleration, and the device would not see anything. This is formalized in the ``cross related'' observables, that is to say, Green functions of the kind (\ref{generagreen}), with fields of one frame and vacuum of the other frame..

In the experimental set-up for the measurement of the Hawking-Unruh effect of graphene on the Beltrami pseudosphere, recalled later here, we are in a very similar situation as the one depicted for the accelerated rocket. On the one hand, the measuring device shares the same frame of the $\psi$-electrons, hence the spacetime seen is (Weyl-related to) a Rindler spacetime. On the other hand, the device has a ``counter'' that has been build in an inertial frame (the lab). This completes the parallel.

Another issue that is possible to face with the choice (\ref{generagreen}) is that of the inequivalent quantization schemes for fields in presence of curvature. When we use the continuum field approximation to describe the dynamics of the electrons of the $\pi$ orbitals of graphene, we not only need to demand the wavelengths to be bigger than the lattice length, $\lambda > 2 \pi \ell$, we also (and, perhaps, most importantly) need to have the conditions for the existence of unitarily inequivalent vacua, the most distinctive feature of QFT (on the general issue see, e.g., \cite{uirs,habilitation}, and, for an application to supersymmetry breaking, see \cite{susymix}). Only then we can be confident to have re-obtained the necessary conditions for typical QFT phenomena to take place on a ``simulating device'', as we propose graphene to be.

\begin{figure}
 \centering
  \includegraphics[height=.4\textheight]{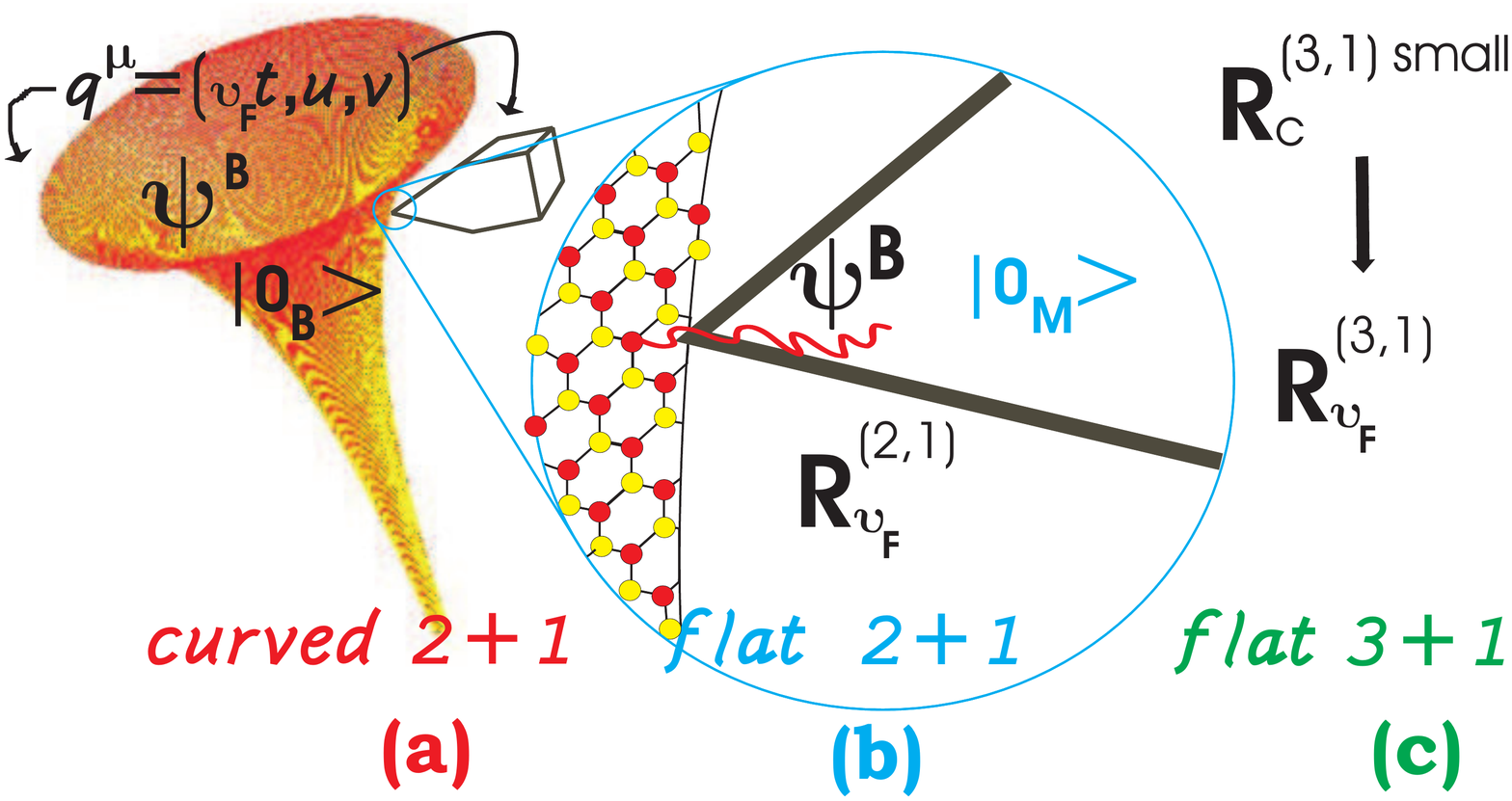}
  \caption{Schematic representation of the relativistic-like features of the measuring process. Figure taken from \cite{prd2014}.} \label{vacua}
\end{figure}

The matter deserves a thoroughly investigation, nonetheless, for the moment, we can take advantage from the results of \cite{siddhartha}. There it is proved that, when the singularity associated to a conical defect is properly taken into account, through the self-adjoint extension of the Hamiltonian operator, inequivalent quantization schemes naturally emerge in graphene. This inequivalence is of the same topological nature as the one arising in the quantization of a particle constrained to move on a circle \cite{kastrup}. Thus, although the system in point has a finite number of degrees of freedom, the Stone-von Neumann theorem of quantum mechanics is violated much in the same way as for a system with infinite number of degrees of freedom \cite{strocchi}. We shall not make direct use of those results here, but our logic, in this respect, is as follows.

We are interested in reproducing the conditions for a standard QFT in curved spacetime description of the electronic properties of graphene. As clarified before, this means that we shall focus on the effects of the intrinsic curvature, so that the action to consider can be taken to be the standard action (\ref{actionAcurvedpaper}). For the hexagonal lattice of graphene, intrinsic curvature means disclination defects, five-folded for positive curvature, seven-folded for negative curvature. Those are topological defects, carrying a singularity of a similar nature as the one associated to the conical defects (see, e.g., \cite{siddhartha}, and our discussion around Eq.~(\ref{a11})). Hence, intrinsic curvature here is tightly linked to the unitarily inequivalent representations necessary for a proper QFT in curved spacetime. With this in mind, we take for granted here that quantum vacua associated to the curved graphene spacetime in point, e.g., the Beltrami spacetime, are unitarily inequivalent to the quantum vacuum associated to the flat graphene spacetime of interest. Furthermore, we assume that the Rindler vacuum (emerging in the way illustrated earlier, when negative curvature is present), and the Minkowski vacuum are too. This last assumption relies on the fact that this fictitious Rindler spacetime emerges also from the curvature of the graphene sheet, hence the topological inequivalence above mentioned applies here too.

The assumptions described in this Section are summarized in Fig.~\ref{vacua} for the case of the Beltrami spacetime, and for a Scanning Tunneling Microscope (STM) measurement. The STM closely follows the profile of a Beltrami pseudosphere, hence the spatial coordinates are the same for both, the $\psi$-electrons on the surface, and the tip of the STM. The time label, $t$, is also the same for both, but it enters a ``Beltrami time'' (related to the Rindler time, see earlier) when considered from the $\psi$-electron point of view, and it enters a Minkowski time, when considered from the laboratory point of view. This hybrid situation is taken care of by the choice of the vacuum. The ``curved'' electron, $\psi^B$ (the wavy line), is supposed to tunnel into the measuring device, and indicated in the zoomed part of the figure (the circle in the middle). The final stage is indicated at the far right, the part (c). There the ``$\psi$-description'' ceases to be valid, and we are left with standard electrons. The core of the assumptions is in the part (b) of the figure, and, as explained, it consists in modeling the measuring process as an hybrid (i) of an operation happening in the graphene curved spacetime (same $q^\mu$ for graphene and for the device), and (ii) of setting a Minkowski vacuum $|0\rangle_M$ (relative to $\mathbf{R}^{(2,1)}_{v_F}$) as the vacuum of reference during the measuring process. The latter vacuum is assumed to be non-equivalent to $|0\rangle_B$ (and to $|0\rangle_R$).

\chapter{Visible signatures: Hawking effect as a geometry-dependent density of states}\label{hawking}

As shown, we can reproduce, on suitably curved graphene sheets, conditions for the existence of event horizons. These horizons coincide, within experimental limits, with the ``end of the world'' represented by the Hilbert horizon. The appearance of the cosmological type of horizon, and even the fact that a BTZ black-hole horizon might be in sight (provided other actions are undertaken), together with the fact that the physical end of (any) surface (always) comes with a potential barrier, indicate that, when the QFT description of action (\ref{actionAcurvedpaper}) and of the quantum vacuum of the previous Section holds, the mechanisms of pair creation and quantum tunneling through the horizon should take place here too, giving raise to Hawking type of effects, interpreted in the spirit of Parikh and Wilczek \cite{pariwilc}. In this approach, the other side of the horizon (the ``out'' region) is beyond where the surface has ended. The entanglement, necessary for the effect to take place, is between the particle that has left the graphene membrane, and the hole/antiparticle that it has left behind, and viceversa.

Although we evoked also other types of horizons, the cleanest horizon we have found is of the Rindler type, hence we shall focus on that one. The entanglement now is between particles (antiparticles) of one wedge and the corresponding antiparticles (particles) of the other wedge. The above picture holds all the way. One needs to consider that, after a long enough Rindler time $\eta$ (future or past null infinity, for particles and antiparticles, respectively) the particles/antiparticles reach the Hilbert/event horizon (see discussion about the geodesics of Beltrami spacetime given earlier), and, through quantum tunneling, leave the surface, giving raise to the same mechanism described above. Recall that, in the mathematical limit $c \to 0$, the future/past null infinity is reached always, $\eta = r/c \; t$. For the physical case $c = \ell$, the lab time $t$ it takes to reach the horizon is still short, see \cite{ioriolambiase}, but the best interpretation of this fact is to say, yet from another perspective, that the effect takes place for particles and antiparticles of very small energy, $E \sim 1/ \eta$, i.e. of very large wavelength, namely, large enough to feel the curvature effects $\lambda > r$.

In what follows, we shall focus on the two point Green function for the Beltrami spacetime, that is related to the important measurable quantity LDOS.

\section{Thermal power spectrum and graphene density of states}

We shall focus on the one particle Green function that contains all the information on the single particle properties of the system such as the LDOS, life time of the quasi-particles and thermodynamic properties (specific heat). For the reasons illustrated above, for us this function is defined as
\be\label{greenBel}
S^{(B)}(q_1, q_2)\equiv \langle 0_M|\psi^{(B)}(q_1){\bar \psi}^{(B)}(q_2)|0_M\rangle \;,
\ee
where with $B$ we indicate the reference to the Beltrami spacetime, and $q^\mu = (t,u,v)$. That is the {\it positive frequency Wightman function}, in the language of QFT in curved spacetimes \cite{birrellanddavies, israel, takagi}). To obtain this function, we use local Weyl symmetry of the action (\ref{actionAcurvedpaper}), as this case is a perfect match for its implementation \cite{iorio}
\be
g_{\mu\nu}^{(B)} = \phi^2(u)  g_{\mu\nu}^{(R)} \;,  \quad \psi^{(B)} = \phi^{-1}(u) \psi^{(R)} \;,
\ee
with $\phi (u) = \ell / r \; e^{u/r}$ and the Rindler type of metric
\be\label{rindlermetric}
g^{(R)}_{\mu \nu} (q) = {\rm diag} \left( \frac{r^2}{\ell^2} \; e^{-2u/r}, - \frac{r^2}{\ell^2} \; e^{-2u/r}, - r^2 \right) \;,
\ee
that was studied in detail earlier. The point we want to make here is that, local Weyl symmetry allows to translate the problem of computing (\ref{greenBel}) to the much easier task of computing
\be\label{greenRind}
S^{(R)} (q_1, q_2) \equiv \langle 0_M|\psi^{(R)}(q_1) {\bar \psi}^{(R)}(q_2)|0_M\rangle \;,
\ee
because
\be\label{sbel}
S^{(B)}(q_1, q_2) = \phi^{-1}(q_1) \phi^{-1}(q_2)S^{(R)}(q_1, q_2) \;.
\ee
Thus, our goal now is to compute (\ref{greenRind}).

First let us recall once more the peculiarity of the Rindler like spacetime (\ref{rindlermetric}). The time coordinate, in both wedges, ranges as usual,
$\eta \equiv (r / \ell) \; t \in [- \infty, + \infty]$, while the relevant space coordinate, {\it taken for curvatures that give a good $\xi_{Hh} \simeq \xi_{Eh}$} (see Table~\ref{tablebeltrami}), ranges as follows
\be\label{xiranges}
\xi \equiv - \frac{r}{\ell} \; u \in [\sim - \infty, 0] \quad {\rm and} \quad \xi \equiv \frac{r}{\ell} \; u \in [0, \sim + \infty] \;,
\ee
in the right wedge and in the left wedge, respectively. Of course, everywhere, $v \in [0, 2 \pi]$. In both cases, $\alpha (\xi) = a e^{- a \xi}$, but, in the right wedge,
$a \equiv \ell / r^2 > 0$, whereas in the left wedge, $a = - \ell / r^2 < 0$. The proper time is
\be\label{propertime}
\tau = e^{a \xi} \; \eta = \frac{r}{\ell} \; e^{- u/r} \; t \;.
\ee
The ranges of $\xi$ in (\ref{xiranges}) indicate that we are in a case where a boundary is present at $\xi = 0$, and when computing the Green function (\ref{greenRind}) we need to take into account that the degrees of freedom of the quantum field $\psi$, beyond that value of $\xi$, are absent. It is worth recalling that $\xi = 0 = u$ corresponds here to the smallest possible value of the radius of the pseudosphere (see Fig.~\ref{Beltrami}), that we set $R (0) = \ell$.

As explained earlier, at $\xi= 0$, the proper acceleration is well approximated with $\alpha \simeq 0$, i.e. it corresponds to the inertial observer. Now we require that the measuring procedure on the Beltrami spacetime reproduces, on the associated Rindler spacetime just recalled, the conditions for a worldline of constant acceleration. That is simply
\be\label{worldlinexiconst}
\xi = {\rm const} \;,
\ee
that means to keep the tip of the Scanning Tunneling Microscope (STM) at a fixed value of the meridian coordinate $u$, as explained also in Fig.~\ref{vacua} and around there. So, at any given measurement, the distance to which one has to compare how far is the boundary $b$ is $d(\xi) = \alpha^{-1}(\xi) - r$, see (\ref{dbeltrind}). Thus $b$, measured in units of $d(\xi)$, is
\be
b(0) \simeq 1 < b (\xi) < + \infty \simeq b (\xi_{Eh}) \;.
\ee
We expect that the effects of the boundary are negligible (the boundary is too far away) when the measurements are taken near the Hilbert/event horizon $\xi_{Eh}$, and when $b$ is located at $\xi = 0$, that is the ideal case of a non-truncated surface. On the other hand, the boundary term, also takes into account the practical issue that the Beltrami surface, when realized with the monolayer graphene, might be truncated before $\xi = 0$. It must be clear that all the computations are done for the worldline of constant acceleration, so that the conditions for the Unruh effect on the Rindler-like spacetime are fulfilled. Hence $\xi$ is going to be constant all over.

With this in mind, the Green function $S^{(R)}$ that we have to compute needs be evaluated at {\it the same point in space and at two different times},
$S^{(R)} (t, {\bf  q}, {\bf q}) \equiv \langle 0_M | \psi^{(R)} (t_1 = 0; {\bf q}) {\bar \psi}^{(R)} (t_2 = t; {\bf q}) |0_M \rangle$, where the dependence on $t_2 - t_1 \equiv t$ is a result of the stationarity of the worldline in point, and we have set the initial time to zero. Eventually, what we have to consider is
\be
S^{(R)}(\tau, {\bf q}, {\bf q}) \;,
\ee
where the proper time is related to $t$ through the relation (\ref{propertime}), and, for the Green function to be a proper positive frequency Wightman function, see \cite{birrellanddavies, takagi}, we need to evaluated it at $\tau \to \tau - i \varepsilon$, with $\varepsilon$ an infinitesimal positive parameter. This also takes into account the nonzero size of the detector\footnote{For the STM experiment we have in mind, $\varepsilon$ is the size, in ``natural units'', of the STM needle or tip. For a tungsten needle $\varepsilon \sim 0.25{\rm mm} \times v_F^{-1} \sim 10^{-10}$s, while for a typical tip $\varepsilon \sim 10 {\rm {\AA}} \times v_F^{-1} \sim 10^{-15}$s (see, e.g.,\cite{stmtip}). Those values of $\varepsilon$ are indeed infinitesimal.}.

The power spectrum one obtains from $S^{(R)}$ is \cite{birrellanddavies,takagi}
\begin{equation}\label{fourierF}
    F^{(R)}(\omega, {\bf q})\equiv \frac{1}{2} {\rm Tr}\left[\gamma^0\int_{-\infty}^{+\infty}d\tau e^{-i\omega \tau} S^{(R)}(\tau, {\bf q}, {\bf q}) \right]\,,
\end{equation}
and, for graphene, besides inessential constants, it coincides with the definition of the electronic LDOS \cite{altland,vozmediano},
$\rho^{(R)}(\omega, {\bf q}) \equiv \frac{2}{\pi} F^{(R)}(\omega, {\bf q})$. This is not yet the {\it physical LDOS}, as the latter is only obtained once we move to the Beltrami spacetime. Nonetheless, due to Weyl symmetry, the latter step is very simple since the Weyl factor in (\ref{sbel}) is time-independent, it goes through the Fourier transform, i.e. $F^{(B)} (\omega, {\bf q}) = \phi^{-2}({\bf q}) F^{(R)} (\omega, {\bf q})$, with obvious notation, hence the physical LDOS is
\be
\rho^{({\rm B})}(\omega, {\bf q}) = \phi^{-2}({\bf q}) \rho^{(R)}(\omega, {\bf q}) \;.
\ee
Thus, the only necessary computation is that of $F^{(R)}$ in (\ref{fourierF}). In order to do so, we can resort to the exact results obtained in this massless case, see \cite{takagi}. First one recalls that, in general, and for any spacetime dimension $n$, the Dirac ($S_n$) and scalar ($G_n$) Green functions are related as: $S_n = i \not\!\partial \, G_n$ (here $m=0$). With our choice of the worldline (i.e., for us, of the measuring procedure) we then have the exact expression
\be\label{fermivsbose}
S_n^{(R)}(\tau) =  \gamma^0 \partial_z G_n^{(R)}(\tau) = \gamma^0 \lambda_n G_{n+1}^{(R)}(\tau) \;,
\ee
where $z = \varepsilon + 2 i \alpha^{-1} \sinh(\alpha \tau/2)$ and $\lambda_n = 2 \sqrt{\pi} \, \Gamma(n/2)/\Gamma((n-1)/2)$, see \cite{takagi}. Thus, we see here that to compute our 3-dimensional Dirac Green functions we need a 4-dimensional scalar field. By taking in (\ref{fermivsbose}) $n=3$, the Fourier transform and the trace, as in (\ref{fourierF}), one easily obtains
\be\label{FlambdaD}
F^{(R)}(\omega) \equiv F_3^{(R)}(\omega) = \lambda_3 B_4^{(R)}(\omega) \;,
\ee
where $B_4^{(R)}(\omega)$ is the power spectrum of the 4-dimensional scalar field. Thus what is left to compute is
\be
B_4^{(R)}(\omega) \equiv B_{thermal}^{(R)} (\omega) + B_{boundary}^{(R)} (\omega) \;,
\ee
that is made of two parts, one showing thermal features, one due to the presence of the boundary in $b$, and this splits in two all relevant quantities, $F^{(R)}$, $F^{(B)}$, $\rho^{(R)}$, and the most important $\rho^{(B)}$.

The expression for $B_{thermal}^{(R)} (\omega)$ has been obtained in many places. Here we use the notation of \cite{takagi} that gives (see also \cite{ioriolambiase,prd2014})
$B_{thermal}^{(R)} (\omega) = ( \omega / 2 \pi ) /  (e^{2 \pi \omega / \alpha} - 1)$. Using this in (\ref{FlambdaD}), one immediately obtains
\be
F^{(R)}_{thermal} (\omega, {\bf q}) = \frac{\omega / 2}{e^{2 \pi \omega/\alpha}-1} \;,
\ee
where we used $\lambda_3 = \pi$, and a Unruh type of temperature appears \cite{unruh}, and $\alpha$ is positive \cite{takagi} (see also discussion after (\ref{ldosbound}) below)
\begin{equation}\label{36}
    {\cal T} \equiv \frac{\hbar v_F}{k_B} \; \frac{\alpha}{2\pi} = \frac{\hbar v_F}{k_B} \; \frac{\ell }{2 \pi r^2} \; e^{u/r} \;,
\end{equation}
with $u \in [0, r \ln (r / \ell)]$, and where the proper dimensional units were reintroduced, and a Tolman factor \cite{wald} $e^{u/r} = e^{-a \xi}$ appears, as required by local measurements. The expression for the thermal part of the physical LDOS is then immediate (recall that $\phi (u) = \ell / r \; \exp(u/r)$)
\begin{equation} \label{37}
    \rho^{({\rm B})}_{thermal} (E, u, r)= \frac{4}{\pi} \frac{1}{(\hbar v_F)^2} \frac{r^2}{\ell^2} e^{-2 u/r} \frac{E}{\exp{\left[ E / (k_B {\cal T}(u,r)) \right]}-1} \;,
\end{equation}
where we included the $g=4$ degeneracy, and the proper dimensions, e.g., $\omega \equiv \omega / v_F$, $E \equiv \hbar \omega$. This is the LDOS when boundary effects are absent.


From (\ref{37}) one sees that the largest temperature $\cal T$ is reached at the Hilbert horizon, and the value is \cite{ioriolambiase,prd2014}
\be
 {\cal T} (r \ln(r / \ell)) = \frac{\hbar v_F}{k_B} \; \frac{1 }{2 \pi r} \;,
\ee
and the Hilbert and event horizons coincide. Notice also that in (\ref{37}) the factor $r^2/\ell^2 \sim + \infty$ is fully balanced by the exponential factor next to it, $e^{-2 u/r}$, only on the horizon $u = r \ln(r/\ell)$,
\be
(r^2/\ell^2) \; e^{- 2 u/r}|_{u = r \ln(r/\ell)} = 1 \;,
\ee
as could have been guessed by the fact that $\phi|_{horizon} = 1$, see discussion after (\ref{rightchoice}). These facts clarify that the interesting phenomena happen near the horizon, and it is there that we should look for the Hawking effect on graphene. More indications of this come from the considerations of the effects of the boundary, that we shall consider next.

\section{Effects of a Minkowski-static truncation}

A complete calculation of the effects of the boundary would need the full knowledge of how the surface is truncated on the thin side, and of how that can be described in terms of our Rindler spacetime. There is an entire literature on the effects of various shapes, and locations of boundaries and mirrors on the Unruh effect, see, e.g., \cite{nistor} and references therein. Here we shall follow \cite{ohnishi}, and shall consider the case of the static boundary in $b$. The formula we shall obtain must not be trusted in all details, but will serve well the scope of showing how non-thermal features of the LDOS do not necessarily mean that our approach, based on QFT in curved spacetime, is not working. In fact, those non-thermal features can be understood within this model.

The positive frequency Wightman function, for a 4-dimensional scalar field, evaluated along a worldline of constant acceleration (that is the one obtained by measuring at a fixed meridian coordinate on the surface), in presence of one static boundary, located at a (dimensionless) distance $b$, in units of the distance of the point of measurement from the horizon (see earlier discussion), is given by \cite{ohnishi}
\be\label{ldosbound}
G^{(R)}_{boundary} (\tau, b) = - \frac{1}{4 \pi^2} \frac{\alpha^2}{4} \frac{1}{[\cosh(\alpha \tau - i \varepsilon) - b\,]^2} \;.
\ee
This result, as it stands, is for one boundary in one sector (wedge) only. To adapt the results of \cite{ohnishi} to our case, we need also to consider another boundary in a symmetric position, in the other sector (wedge). As explained earlier, our picture here is that the other wedge is obtained by the existence of antiparticles, for which the time flows in opposite directions, hence the meaning of positive and negative frequency swap. Thus what we have to do is consider both, the positive frequency and the negative frequency Wightman functions, and keep both $\alpha$ and $b$ positive.

Here we compute the LDOS associated to the existence of a static boundary at $b$. The Green function for the 4-dimensional scalar field is \cite{ohnishi} (hereafter we remove the label $boundary$ from all quantities)
\be
G^{(R)}_+ (\tau, b) = - \frac{1}{4 \pi^2} \frac{\alpha^2}{4} \frac{1}{[\cosh(\alpha \tau - i \varepsilon) - b\,]^2} \;.
\ee
where the prescription $- i \varepsilon$ is necessary to have the {\it positive frequency} Wightman function, that in turn is necessary to compute the power spectrum and the density of states. With this method we do not need to take the imaginary part of the Green function.
The power spectrum is then
\be
B^{(R)}_+ (\omega, b) = - \frac{\alpha^2}{16 \pi^2}\int^{+ \infty}_{-\infty} d\tau
\frac{e^{- i \omega \tau}}{[\cosh(\alpha \tau - i \varepsilon) - b\,]^2} \equiv - \frac{\alpha^2}{16 \pi^2} \; I_+(\omega, b)\;.
\ee
We now compute the integral $I_+(\omega,b)$
\be
I_+(\omega,b) = \alpha^{-1} \; \int^{+ \infty}_{-\infty} dx
\frac{e^{- i \omega/\alpha x}}{[\cosh(x - i \varepsilon) - b\,]^2} \;,
\ee
where $x = \alpha \tau$. The poles of
\be
f_+(z) = \alpha^{-1}  \; \frac{e^{- i \omega/\alpha z}}{[\cosh(z - i \varepsilon) - b\,]^2}
\ee
with $z = x + i y$, are
\be
z_{1,2} = \pm \tilde{b} + i \varepsilon
\ee
with $\tilde{b} = {\rm arcosh} b$. The residues are
\bea
Res(f_+, z_1= \tilde{b} + i \varepsilon) & = & - i \; \; \frac{\frac{\omega}{\alpha^2} (b^2 - 1)^{1/2} - i \frac{b}{\alpha}}{(b^2 - 1)^{3/2}} \; e^{\frac{\omega}{\alpha}(\varepsilon - i \tilde{b})} \;, \\
Res(f_+, z_2= - \tilde{b} + i \varepsilon) & = & - i \; \; \frac{\frac{\omega}{\alpha^2} (b^2 - 1)^{1/2} + i \frac{b}{\alpha}}{(b^2 - 1)^{3/2}} \; e^{\frac{\omega}{\alpha}(\varepsilon + i \tilde{b})} \;.
\eea
We have
\be
\oint dz f_+(z) = \alpha^{-1} \; \int^{+ R}_{- R} dx
\frac{e^{- i \frac{\omega}{\alpha} \; x}}{[\cosh(x - i \varepsilon) - b\,]^2} +
\alpha^{-1} \; \int_\Gamma dz \frac{e^{- i \frac{\omega}{\alpha} \; x + \frac{\omega}{\alpha} \; y}}{[\cosh(z - i \varepsilon) - b\,]^2} \;,
\ee
where $\Gamma$ is the upper or the lower half of a circle of radius $R$ centered at the origin of the $z$-plane. Since the poles are in the upper half $z$-plane ($ y > 0$), to have contribution we need $\Gamma$ there, and to have convergence in the $R \to + \infty$ limit, it must be
\be\label{cond+}
\omega/\alpha < 0 \Leftrightarrow \omega > 0 \; {\rm and} \; \alpha < 0 \quad {\rm OR} \quad \omega < 0 \; {\rm and} \; \alpha > 0 \;.
\ee
When this happens
\be
I_+(\omega, b) = 2 \pi i \; (Res(f_+, z_1) + Res(f_+, z_2)) \;.
\ee
As explained in the text, we need now to consider also the {\it negative frequency} Wightman function, again with $b > 0$, that is
\be
G_-^{(R)} (\tau, b) = - \frac{1}{4 \pi^2} \frac{\alpha^2}{4} \frac{1}{[\cosh(\alpha \tau + i \varepsilon) - b\,]^2} \;.
\ee
The power spectrum is then
\be
B^{(R)}_- (\omega, b) = - \frac{\alpha^2}{16 \pi^2}\int^{+ \infty}_{-\infty} d\tau
\frac{e^{- i \omega \tau}}{[\cosh(\alpha \tau + i \varepsilon) - b\,]^2} \equiv - \frac{\alpha^2}{16 \pi^2} \; I_-(\omega, b)\;.
\ee
We now compute the integral $I_-(\omega,b)$
\be
I_-(\omega,b) = \alpha^{-1} \; \int^{+ \infty}_{-\infty} dx
\frac{e^{- i \omega/\alpha x}}{[\cosh(x + i \varepsilon) - b\,]^2} \;,
\ee
where $x = \alpha \tau$. The poles of
\be
f_-(z) = \alpha^{-1}  \; \frac{e^{- i \omega/\alpha z}}{[\cosh(z + i \varepsilon) - b\,]^2}
\ee
with $z = x + i y$, are
\be
z_{1,2} = \pm \tilde{b} - i \varepsilon \;.
\ee
The residues are
\bea
Res(f_-, z_1= \tilde{b} - i \varepsilon) & = &  i \; \; \frac{\frac{\omega}{\alpha^2} (b^2 - 1)^{1/2} + i \frac{b}{\alpha}}{(b^2 - 1)^{3/2}} \; e^{\frac{\omega}{\alpha}(\varepsilon + i \tilde{b})} \;, \\
Res(f_-, z_2= - \tilde{b} - i \varepsilon) & = & i \; \; \frac{\frac{\omega}{\alpha^2} (b^2 - 1)^{1/2} - i \frac{b}{\alpha}}{(b^2 - 1)^{3/2}} \; e^{\frac{\omega}{\alpha}(\varepsilon - i \tilde{b})} \;.
\eea
We have
\be
\oint dz f_-(z) = \alpha^{-1} \; \int^{+ R}_{- R} dx
\frac{e^{- i \frac{\omega}{\alpha} \; x}}{[\cosh(x + i \varepsilon) - b\,]^2} +
\alpha^{-1} \; \int_\Gamma dz \frac{e^{- i \frac{\omega}{\alpha} \; x + \frac{\omega}{\alpha} \; y}}{[\cosh(z + i \varepsilon) - b\,]^2} \;,
\ee
where $\Gamma$ is the upper or the lower half of a circle of radius $R$ centered at the origin of the $z$-plane. Since the poles are in the lower half $z$-plane ($ y < 0$), to have contribution we need $\Gamma$ there, and to have convergence in the $R \to + \infty$ limit, it must be
\be\label{cond-}
\omega/\alpha > 0 \Leftrightarrow \omega > 0 \; {\rm and} \; \alpha > 0 \quad {\rm OR} \quad \omega < 0 \; {\rm and} \; \alpha < 0 \;.
\ee
When this happens
\be
I_-(\omega, b) = 2 \pi i \; (Res(f_-, z_1) + Res(f_-, z_2)) \;.
\ee
If we call $Res(f_+, z_1) \equiv - i w$, with $w$ complex, we wee that $Res(f_+, z_2) = - i w^*$, $Res(f_-, z_1) =  i w^*$, $Res(f_-, z_2) = i w$, so
\be
I_+ (\omega, b) = 4 \pi {\rm Re}(w) \quad {\rm and} \quad I_- (\omega, b) = - 4 \pi {\rm Re}(w) \;,
\ee
with Re = real part.

The result we are looking for then, is obtained by adding all the contributions with $\alpha > 0$, that means that we are picking one sector, and the other comes in as antiparticles
\bea
B^{(R)} (\omega, b) & = & - \frac{\alpha^2}{16 \pi^2}\; \;  4 \pi {\rm Re}(w)|_{\varepsilon = 0} \quad {\rm when} \; \omega < 0 \nonumber \\
& + & \left( - \frac{\alpha^2}{16 \pi^2} \right) \left( - 4 \pi {\rm Re}(w)|_{\varepsilon = 0}\right)  \quad {\rm when} \; \omega > 0 \label{middle} \;,
\eea
where we used the second condition in (\ref{cond+}), and the first condition in (\ref{cond-}), for the first and second line above, respectively. Since
\be
- \frac{\alpha^2}{16 \pi^2}\; \;  4 \pi {\rm Re}(w)|_{\varepsilon = 0} =
- \frac{1}{4 \pi} (b^2 - 1)^{-3/2} [\omega (b^2 - 1)^{1/2} \cos(\tilde{b} \, \omega / \alpha) - b \, \alpha \sin(\tilde{b} \, \omega / \alpha)] \;,
\ee
that inserted in (\ref{middle}) gives
\be
B^{(R)} (\omega, b) = \frac{1}{4 \pi} \; \frac{|\omega|}{b^2 - 1} \; \cos\left( \tilde{b} \; \frac{\omega}{\alpha} \right) \;,
\ee
hence the wanted power spectrum for the Dirac 3-dimensional Dirac field, is easily obtained through $F^{(R)} (\omega, b) = \pi B^{(R)} (\omega, b)$ (recall that $\lambda_3 = \pi$). With this, the physical LDOS is immediately obtained with the help of Weyl symmetry
\bea
\rho^{(B)}(\omega, b) & = & \frac{2}{\pi} \; \phi^{-2} (u) F^{(R)} (\omega, b) \;, \nonumber \\
& = & \frac{1}{2 \pi} \; \frac{r^2}{\ell^2} \; e^{- 2 u / r} \; \frac{|\omega|}{b^2 - 1} \; \cos\left( \tilde{b} \; \frac{\omega}{\alpha} \right) \;,
\eea
and by including the degeneracy $g = 4$, and the proper units, like for the thermal part, one obtains
\be\label{boundaryldos}
\rho^{({\rm B})}_{boundary} (E, u, r)= \frac{2}{\pi} \frac{1}{(\hbar v_F)^2} \frac{r^2}{\ell^2} e^{-2 u/r} \frac{|E|}{b^2 - 1}
\cos\left( \tilde{b} \frac{E}{\hbar v_F \alpha(u,r)}\right) \;,
\ee
where $\tilde{b} = {\rm arcosh} b$, and we reintroduce the label ''boundary'' from now on.

The behavior of the boundary term is as expected
\be
\rho^{({\rm B})}_{boundary} \rightarrow 0 \; {\rm for} \; b \to + \infty \quad {\rm and} \quad
\rho^{({\rm B})}_{boundary} \rightarrow \pm \infty \; {\rm for} \; b \to 1 \;.
\ee
Indeed, the first limit describes the case of infinite distance, in units of $\alpha^{-1}$, between the point $u$ where one measures, and the value of $u$ where the surface ends, that is a measurement taken near the Hilbert horizon, $R=r$, and the thin end of the surface ending at $R=\ell$. The second limit refers to a measurement taken on/near the boundary itself. There our approximations for the boundary do not hold fully, nonetheless we can trust that the boundary there will of course dominate. Let us stress again that, the boundary term (\ref{boundaryldos}) takes into account the fact that the infinities here are only approximated. This has two meanings. First, even in the ideal case of a Beltrami that ends at $R = \ell$ (and supposing that our QFT approximations work till there), the range of $u$ is not really infinite. Second, the real graphene surface will end before that ideal value of $R$, anyway. What is important, though, is to see how strong are the non-thermal corrections over the thermal spectrum.

Therefore the total LDOS for a graphene membrane shaped as a Beltrami pseudosphere, within the limits of validity of the Weyl symmetry, is then
\bea\label{totalLDOS}
 \rho^{({\rm B})} (E, u, r)  & = & \frac{4}{\pi (\hbar v_F)^2} \frac{r^2}{\ell^2} e^{-2 u / r}
 {\Big[} \frac{E}{\exp \left[ (2 \pi E)/ (\hbar v_F \alpha (u,r)) \right]-1} \nonumber \\
 & + & \frac{1}{2} \frac{|E|}{b^2 - 1} \cos\left( \frac{\tilde{b} \; E}{\hbar v_F \alpha(u,r)}\right) {\Big]} \;.
\eea

\begin{figure}
\centering
\begin{tabular}{cc}
\epsfig{file=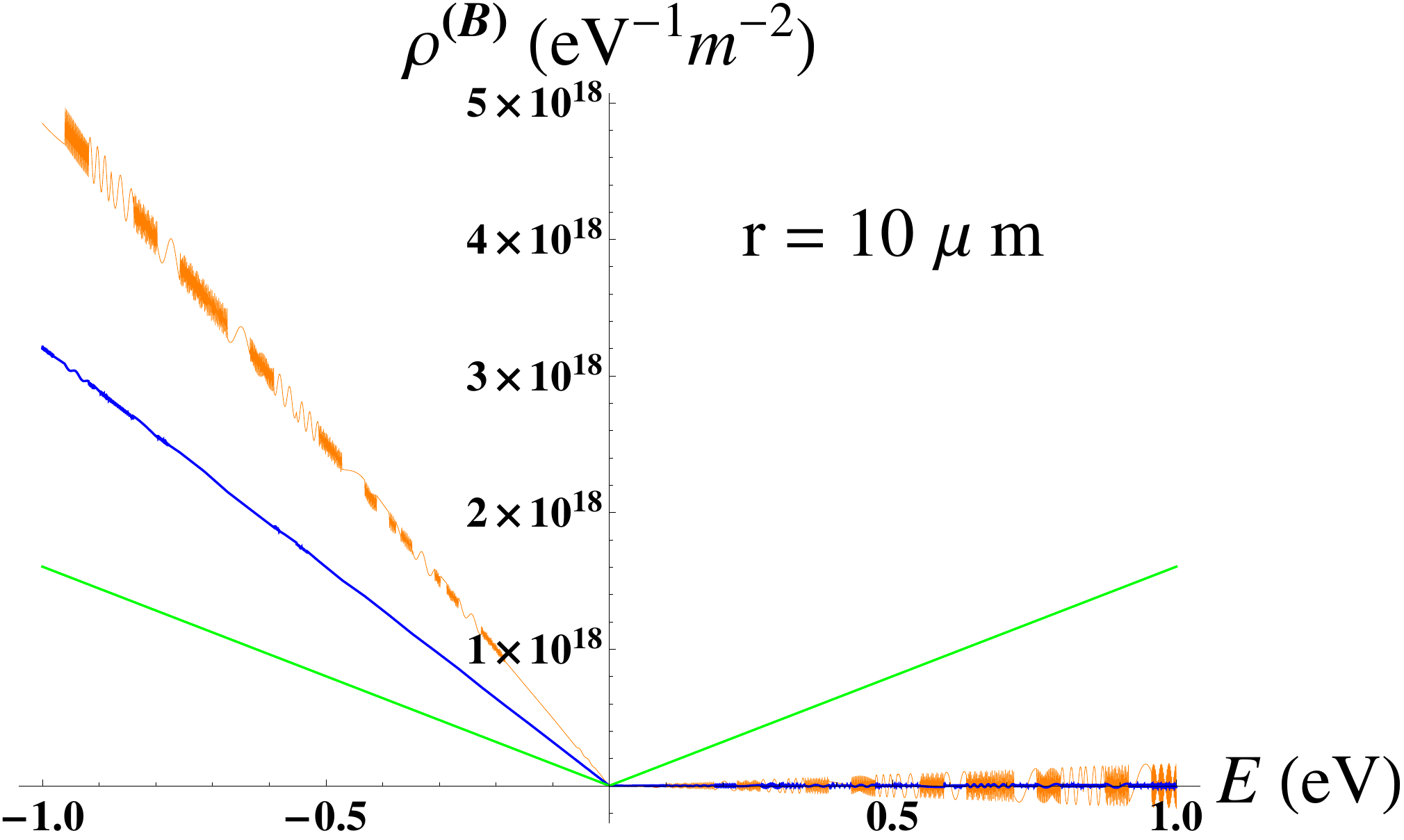, width=0.85\linewidth, clip=} \\
\epsfig{file=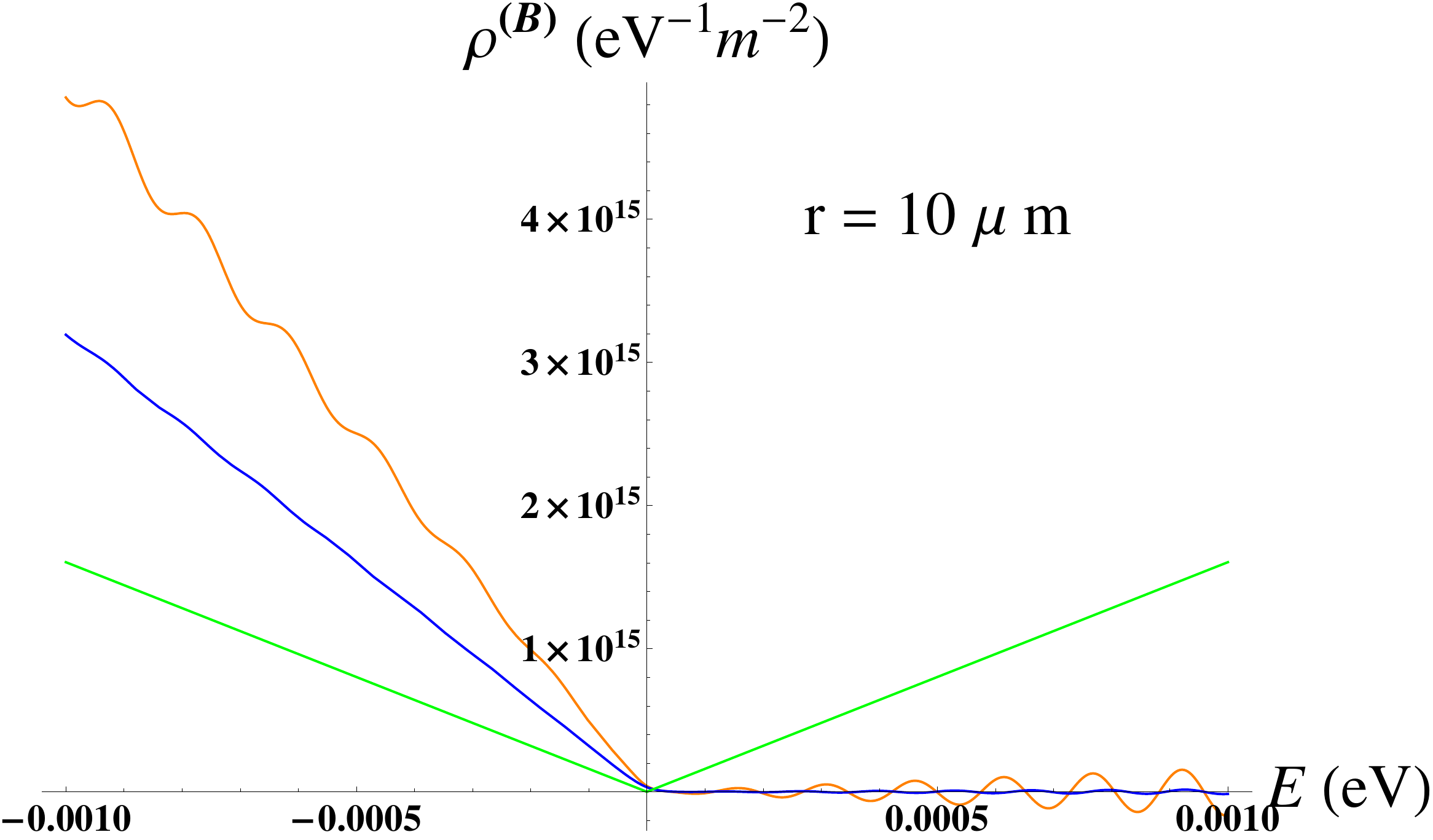, width=0.85\linewidth, clip=}
\end{tabular}
\caption{Plots of the total LDOS, including the effects of truncation, against the Energy. The radius of curvature is $r= 10\mu$m. The plots in blue are for measurements taken very near/at the horizon ($u = r \ln (r /\ell)$). The plots in orange are for measurements taken just below the horizon ($u = 0.98 r \ln (r /\ell)$). The figure below is a zooming-in ($E \in [-1 {\rm meV}, + 1 {\rm meV}]$) of the figure above ($E \in [-1 {\rm eV}, + 1 {\rm eV}]$), something that shows how thermal effect are shadowed when the energies are too much above $E_r$. The pictures also show that, no matter the range, the smoking gun of the Hawking effect here is a depletion of the positive energy LDOS in favor of the negative energy. This can be appreciated by comparing with the flat LDOS (green).}
\label{plottotal}
\end{figure}

In Fig.~\ref{plottotal} we plot $\rho^{({\rm B})}$ vs $E$, for radius of curvature $r = 10 \mu$m, and for two values of $b$, $b = 10$ for $u = r \ln(r/ \ell)$ (in blue), and $b = 4$ for $u = 0.98 \, r \ln(r/ \ell)$ (in orange). We also plot in green the flat LDOS. When the boundary term dominates, the thermal nature is gone (negative values of $\rho^{(B)}(E)$ need not be taken too seriously, as our approximations do not allow to trust the formula in all details too near the extremal values of $b = 1$). What is important here is that, for relatively small values of $b$, that is for measurements taken close to the horizon, the thermal character is still there, although shadowed by the oscillatory term (noise).

\section{General considerations for any surface of constant negative $\cal K$}

We can also have some sort of Hawking effect on a generic surface $\Sigma$ of constant negative ${\cal K}$. That might happen if on $\Sigma$ the conditions for an event horizon are realized. Then a Hawking effect, of the kind described here can take place through a LDOS whose structure is the same of the LDOS (\ref{totalLDOS}) of the Beltrami surface. Nonetheless, it might be quite complicated to have control of the procedure, especially of the all important occurrence of the event horizon. Let us now show why this is so, what are the issues, and a possible strategy to see whether the effect is there.

The line element of the spacetime is (\ref{lobspacetime})
\begin{eqnarray}
ds^2 =  \frac{r^2}{{\tilde y}^2}\left[\frac{{\tilde y}^2}{r^2}dt^2 - d{\tilde y}^2 - d{\tilde x}^2\right] \;, \nonumber
\end{eqnarray}
where the abstract Lobachevsky coordinates need be specified for $\Sigma$: $\tilde{x}_\Sigma (u_\Sigma, v_\Sigma)$, and $\tilde{y}_\Sigma (u_\Sigma, v_\Sigma)$, including the ranges of $u_\Sigma$, and $v_\Sigma$. Now, as recalled earlier (see \cite{eisenhart}) {\it any} $\Sigma$ is locally reducible to one of the three pseudospheres, i.e., its line element can be reduced to the line element of one of the three pseudospheres. In \cite{prd2014} it was shown that, in the limit for $c \to 0$, the three pseudospheres all become the Beltrami\footnote{Notice that, since we are always referring to the Beltrami spacetime, the type of horizon we are considering is of the Rindler type. One might as well use, say, the elliptic pseudosphere as reference, hence the horizon would be of the cosmological kind. Although interesting, though, this case is less appealing for our purposes as the spacetimes would be Weyl related to non-flat spacetimes, de Sitter, hence the formula (\ref{totalLDOS}) would not be of direct use.}, with some global differences that can become important for the existence of a well defined Hilbert horizon on $\Sigma$. At any rate, {\it locally}, by considering Beltrami in the $c \to 0$ limit, we are dealing also with $\Sigma$, {\it in the same limit}. Indeed, after the reduction of the line element of $\Sigma$, $c$ is in $dl^2_\Sigma (c)$ too, hence we can obtain the shape of $\Sigma$ in that limit, by knowing how the ranges of $u_\Sigma (c)$, and $v_\Sigma (c)$ are affected.

When, on $\Sigma$, a Hilbert horizon is well defined by the coordinates $(u^{Hh}_\Sigma, v^{Hh}_\Sigma)$, and when, for $c \to 0$, $\tilde{y}_\Sigma (u^{Hh}_\Sigma, v^{Hh}_\Sigma) \simeq 0$, within physically reasonable approximations, then the event horizon is present on $\Sigma$, and it coincides, within the same approximations, with its Hilbert horizon $u^{Eh}_\Sigma \simeq u^{Hh}_\Sigma$, and $v^{Eh}_\Sigma \simeq v^{Hh}_\Sigma$. What we cannot know a priori is whether indeed there is a good Hilbert horizon on $\Sigma$. The embedding in ${\mathbf R}^3$ that gives $\Sigma$ can be so intricate that the Hilbert horizon might be a meaningless concept there (see, e.g., the expressions for the embedding of Fig.~\ref{genericpseudosphere}), even though it might be mapped onto meaningful ones, $R = r$, or $R = r \cos \vartheta$, or $R = \sqrt{r^2 + c^2}$, and then, eventually, to the $R = r$ of Beltrami.

Summarizing, if we know that a Hilbert horizon exists on $\Sigma$, and we know the mapping from $\Sigma$ to ``its pseudosphere''
\be \label{mapgeneralcase}
u_\Sigma (u_p, v_p), \quad {\rm and} \quad u_\Sigma (u_p, v_p) \;,
\ee
where $p$ stands for any one of the three pseudospheres, we know how $c$ enters the line element, and the ranges of $u_\Sigma$ and $v_\Sigma$, so that we can perform the limit $c \to 0$ and we shall know the resulting shape of $\Sigma$, and the location of its Hilbert horizon, that will coincide with an event horizon. Furthermore, in that limit, the pseudosphere of reference has become the Beltrami, hence the formula (\ref{totalLDOS}) can be used
\be \label{ldosall}
 \rho^{(\Sigma)} (E, u_\Sigma, v_\Sigma) \simeq  \rho^{({\rm B})} (E, u_B (u_\Sigma, v_\Sigma)) \;,
\ee
where $u_B (u_\Sigma, v_\Sigma)$ is obtained by inverting (\ref{mapgeneralcase}), after the limit $c \to 0$ has been performed. But there is a {\it crucial warning} for the correct use of formula (\ref{ldosall}): {\it The formula is valid only locally}. That is why we use ``$\simeq$''. Due to the local nature of the geometric reduction of $\Sigma$ to the pseudosphere, the mapping (\ref{mapgeneralcase}) might be multivalued (hence not a true map). That means that, if we insist in using the formula for a closed path on $\Sigma$, the formula might give different values for the same point at each full turn, an instance that has no physical meaning. Hence, in general, we can only use (\ref{ldosall}) in a small neighbor for the given point of measurement $(u_\Sigma, v_\Sigma)$. Recall that we have encountered already problems of multivaluedness, due to the choice of coordinates we have constrained ourselves to use. That problem, and this too, in principle might be solved by a clever choice of new coordinates (see, e.g., (\ref{newcoord}) and (\ref{newconffact})), but then one needs to explain the physics of their realization in the laboratory.

Another issue with the use of (\ref{ldosall}) is more practical. Due to the nature of the mapping, it might happen that measuring at a give point, $(u^1_\Sigma, v^1_\Sigma)$, one sees thermal effects, while measuring at a close point, $(u^2_\Sigma, v^2_\Sigma)$, the thermal effects are gone. Indeed, close points on $\Sigma$ might correspond to far points on Beltrami, hence the effects of the boundary term might unexpectedly play the role of masking the Hawking effect that is, in fact, present.

\chapter{Conclusions and Outlook}

Let us conclude by pointing to the potential developments of the use of graphene, as a laboratory for testing fundamental issues of physics. As well known, the research on fundamental physics becomes nearly completely speculative, as soon as we depart from QFT, and from classical GR. Nonetheless, we do have a reliable starting point to merge QFT and GR, that is the theoretical discovery of Stephen Hawking \cite{hawking} that, when the special features of quantum fields (quantum vacuum) in a classical curved spacetime are taken into account, a black hole should radiate with a thermal spectrum, and a temperature emerging from the geometry of the spacetime. Any serious attempt to understand quantum gravity has to start from there. That is why black holes are at the crossroad of many of the speculations about the physics at the Planck scale.

From the results reported about here, and from those to come, a goal that seems in sight is the realization of reliable set-ups, where graphene well reproduces the black-hole thermodynamics scenarios, with the analogue gravity of the appropriate kind to emerge from the description of graphene's membrane, and with the Dirac QFT coupled to this background. This work should be done, primarily through theoretical research of the kind reported here, but needs to be supported, at least, by numerical computer simulations based on the real material. Work is in progress in that direction \cite{ComingTrumpet}. The lattice structure of graphene's membrane, the possibility to move through energy regimes where discrete and continuum descriptions coexist, and the unique features of matter fields whose relativistic structure is induced by the spacetime itself, are all issues that can be explored.

The first, and urgent, task is a test of the predictions of \cite{ioriolambiase, prd2014} on the possibility for a Hawking effect on graphene. It is only when we shall have this under full control, that the following steps towards a black-hole thermodynamics will be on a firm basis. As extensively discussed here, those predictions are based on the possibility to obtain very specific shapes, that should recreate, for the pseudoparticles of graphene, conditions (conformally) related to those of a spacetime with an horizon. The case we mostly focused here is the one of a Rindler type of horizon. Nonetheless, as seen, spacetimes related to the de Sitter spacetime are in hands, hence cosmologically relevant issues could also be addressed (see, e.g., \cite{siddhartha} for cosmological considerations in graphene). And many more things can be addressed from the theoretical perspective. For instance, there are many intriguing issues about closed time-like curves, or specific boundary conditions for the underlying anti de Sitter spacetime of a BTZ black hole, that, when the proper set up will be found, can all be addressed.

One can also depart from the free Dirac model. For instance, one could consider also external (3+1) dimensional electromagnetic fields, with the double scope of probing extra dimensions physics in this context, and also of manipulating the system in order to have more elaborate metrics than those obtained by spatial curvature only (this could be a way to reproduce the conditions for the BTZ). We can include scalars fields, as done for instance to describe vortices (see, e.g., \cite{jackiwpi}), we can move to interacting Dirac fields, for instance by mixing the two Dirac points, or by including four Fermi interaction terms \cite{gusynin}, just to mention a few cases. This would enrich the scenario of the type of QFT one can have on graphene.

Nonetheless, the main step ahead is the construction of the other side of the model, namely of the proper gravity theory that effectively describes the elastic theory of deformations of graphene. Although phrased in different words, this is a major challenge for the condensed matter community, and much work is currently devoted to this task. What we are looking for here is a gravity model that can well describe the energy regimes from the ``very low energy'' of the $\mu$eV--meV, where our approximations are expected to work, to the ``intermediate energy'' regime of 1-2 eV, where elastic effects become dominant, till the ``very high energy'' regimes, where the linear approximation no longer works. In these last regimes the (pseudo-)relativistic structure of the Dirac field will be deformed, and the discrete nature of the space(time) (the hexagonal lattice with spacing $\ell$) will become so important that the continuum description, in terms of smooth metrics, will no longer be valid. This is a region of the graphene spectrum, that is not accessible to our present QFT model description, that will become an important point to enforce analogy with quantum gravity scenarios. Indeed, even though it is just to early to foresee the results of this research, it is reasonable to expect that a natural analogue of the Planck length in this context is going to be the lattice spacing $\ell$, hence the quantum gravity models that we should be able to face are those that evoke discrete spacetime structures \cite{worldcrystal} (a problem to face will be the continuous nature of time in our context, but ``time crystals`` are not entirely exotic objects by now, see, e.g., \cite{wilczek}).

What we expect, though, is that the type of theory that will result will have to take into account both curvature (spin-connection) and torsion (Vielbein), in the spirit of the generalized (Cartan) theory of gravity, especially in the presence of both kinds of topological defects, of the disclination, as well as of the dislocation type (see, e.g., \cite{KatVol92}). The most natural approach will be the gauge-field approach to gravity in (2+1) dimensiona, based on the famous results of Edward Witten, of gravity as a gauge theory of ISO(2,1), the Poincar\`e group in (2+1) dimensiona, and including also a cosmological constant $\Lambda$, hence changing the group according to the sign of $\Lambda$ \cite{witten3dgravity}. Of course, these are just guidelines, but we expect (as also announced above) that lattice effects have to play a role, even from the very beginning, see, e.g. \cite{dejuannonabelian}, where indications that the effective gauge field that well describes at least some deformations of graphene is of a non-abelian nature. Important, in this respect, is also the work of \cite{guillo}, on the low energy electron-phonon effective actions emerging from symmetries of the lattice.

Many other issues are also in sight. Let us briefly mention only a few. There are results of \cite{mauriciosusy}, and of \cite{jakubsky}, that point towards the use of graphene in supersymmetric contexts. There are models of the early universe, based on (2+1) dimensional gravity \cite{vanderbij}, where graphene might play a role. Relativistic quantum computing, a field in rapid development, see e.g. \cite{fuentes1, fuentes2}, is another direction to probe. The work of \cite{marzuoli, palumbothesis} points to the use in graphene of the rich scenarios of topological field theories, such as the BF-theory.

\chapter*{Acknowledgements} The author is grateful to Pablo Pais, who substantially contributed to add clarity to key issues. The financial support from the Czech Science Foundation (GA\v{C}R), under contract no. 14-07983S, is acknowledged.

\end{document}